# New Fe II energy levels from stellar spectra*

F. Castelli[1] and R.L. Kurucz[2]

[1] Istituto Nazionale di Astrofisica– Osservatorio Astronomico di Trieste, Via Tiepolo 11, I-34131 Trieste, Italy
 e-mail: castelli@oats.inaf.it
[2] Harvard-Smithsonian Center for Astrophysics, 60 Garden Street, Cambridge, MA 02138, USA

**ABSTRACT**

*Aims.* The spectra of B-type and early A-type stars show numerous unidentified lines in the whole optical range, especially in the 5100-5400 Å interval. Because Fe II transitions to high energy levels should be observed in this region, we used semiempirical predicted wavelengths and gf-values of Fe II to identify unknown lines.
*Methods.* Semiempirical line data for Fe II computed by Kurucz are used to synthesize the spectrum of the slow-rotating, Fe -overabundant CP star HR 6000.
*Results.* We determined a total of 109 new 4f levels for Fe II with energies ranging from 122 324 cm$^{-1}$ to 128 110 cm$^{-1}$. They belong to the Fe II subconfigurations 3d$^6$($^3$P)4f (10 levels), 3d$^6$($^3$H)4f (36 levels), 3d$^6$($^3$F)4f (37 levels), and 3d$^6$($^3$G)4f (26 levels). We also found 14 even levels from 4d (3 levels), 5d (7 levels), and 6d (4 levels) configurations. The new levels have allowed us to identify more than 50 % of the previously unidentified lines of HR 6000 in the wavelength region 3800-8000 Å. Tables listing the new energy levels are given in the paper; tables listing the spectral lines with log $gf \geq -1.5$ that are transitions to the 4f energy levels are given in the Online Material. These new levels produce 18000 lines throughout the spectrum from the ultraviolet to the infrared.

**Key words.** line:identification-atomic data-stars:atmospheres-stars:chemically peculiar- stars:individual:HR 6000

## 1. Introduction

In a previous paper (Castelli, Kurucz & Hubrig, 2009) (Paper I) we have determined 21 new 3d$^6$($^3$H)4f high energy levels of Fe II on the basis of predicted energy levels, computed log $gf$ values for Fe II, and unidentified lines in UVES high resolution, high signal-to-noise spectra of HR 6000 and 46 Aql. Both stars are iron overabundant CP stars and have rotational velocity v*sini* of the order of 1.5 km s$^{-1}$ and 1.0 km s$^{-1}$, respectively.

In this paper we continue the effort to determine new highenergy levels of Fe II. We used the same spectra and models for HR 6000 that we adopted in Paper I, together with Fe II line lists which include transitions between observed-observed, observed-predicted, and predicted-predicted energy levels. In this paper we increase the number of the new energy levels from the 21 listed in Paper I, to a total of 109 energy levels, which belong to the Fe II subconfigutations: 3d$^6$($^3$P)4f (10 levels), 3d$^6$($^3$H)4f (36 levels), 3d$^6$($^3$F)4f (37 levels), and 3d$^6$($^3$G)4f (26 levels), and 14 levels from the even configurations 4d (3 levels), 5d (7 levels), and 6d (4 levels). The new levels have allowed us to identify more than the 50 % of the previously unidentified lines in the wavelength region 3800-8000 Å of HR 6000 (Castelli & Hubrig, 2007). The method that we adopted to determine the new energy levels is the same as described in Paper I. It is recalled here in Sect. 3. The comparison of the observed spectrum of HR 6000 with the synthetic spectrum which includes the new Fe II lines is available on the Castelli web site[1].

## 2. The star HR 6000

According to Paper I, the CP star HR 6000 (HD 144667) has an estimated rotational velocity of 1.5 km sec$^{-1}$. The model stellar parameters for an individual abundance ATLAS12 (Kurucz 2005) model are $T_{\rm eff}$=13450 K, log $g$=4.3. In addition to the large iron overabundance [+0.9], overabundances of Xe ([+4.6]), P (>[+1.5]), Ti ([+0.55]), Cr ([+0.2]), Mn ([+1.5]), Y ([+1.2]), and Hg ([+2.7]) were observed. This peculiar chemical composition, together with the underabundances of He, C, N, O, Al, Mg, Si, S, Cl, Sc, V, Co, Ni, and Sr gives rise to an optical line spectrum very rich in Fe II lines, with transitions involving upper energy levels close to the ionization limit (Johansson 2009). Also numerous Fe I and Fe III lines are observable in the spectrum.

---

*Send offprint requests to*: F. Castelli
\* Tables 6, 7, 8, and 9 are only available in electronic form at the CDS via anonymous ftp to cdsarc.u-strasbg.fr (130.79.128.5) or via http://cdsweb.u-strasbg.fr/cgi-bin/qcat?J/A+A/

[1] http://wwwuser.oat.ts.astro.it/castelli/hr6000new/hr6000.html

## 3. The method

To determine the new energy levels we used high-resolution UVES spectra of HR 6000 (see Paper I), the corresponding synthetic spectrum, and the list of the computed transitions with predicted values for levels with no experimentally available energies. Predicted energy levels and log $gf$ values were computed by Kurucz with his version of the Cowan (1981) code (Kurucz 2009). The calculation included 46 even configurations $d^7$, $d^64s-9s$, $d^64d-9d$, $d^65g-9g$, $d^67i-9i$, $d^69l$, $d^54s^2$, $d^54s5s-9s$, $d^54s4d-9d$, $d^54s5g-9g$, $d^54s7i-9i$, $d^54s9l$, $d^44s^24d$, and $d^54p^2$ with 19771 levels least-squares fitted to 418 known levels. The 39 odd configurations included $d^64p-9p$, $d^64f-9f$, $d^66h-9h$, $d^68k-9k$, $d^54s4p-9p$, $d^54s4f-9f$, $d^54s6h-9h$, $d^54s8k-9k$, $d^44s^24p-5p$, and $d^44s^24f$ with 19652 levels least-squares fitted to 596 known levels. The calculations were done in LS coupling with all configuration interactions included, with scaled Hartree-Fock starting guesses, and with Hartree-Fock transition integrals. A total of 7080169 lines were saved from the transition array of which 102833 lines are between known levels and have good wavelengths.

The computed line list was sorted into tables of all the strong lines connected to every predicted level. When a given predicted level gives rise to at least two Fe II lines having log $gf \geq -1.0$, we selected one of these transitions and searched in the spectrum for those unidentified lines which have wavelength within ±50 Å and residual flux within about ± 5% of those of the selected predicted line. From the observed wavelength of one of these unidentified lines and from the known energy of the lower or upper level of the predicted transition, we derived a possible energy for the predicted level. If most of transitions obtained with this energy correspond to lines observed in the spectrum, we kept the tentative energy value as a real value, otherwise we repeated the procedure using another line taken from the unidentified ones, and continued the searching until we found that energy for which most of the predicted lines correspond to the observed lines. Whenever one or more new levels were found, the whole semiempirical calculation was repeated to produce improved predicted wavelengths and log $gf$-values. Because all configuration interactions were included, and because the mixing is exceptionally strong in the 4d and 5d configurations, every new level changed the predictions. Mixing between close levels can produce large uncertainties in the log $gf$ values for lines that involve those levels.

This procedure is very successfull for levels which produce two or more transitions with log $gf > 0.0$, but becomes more and more difficult as the intensity of the predicted lines decreases. In fact, weak lines are usually blended with stronger components, so that the method may fail in these cases.

## 4. The new energy levels

The new energy levels of the $3d^6(^3P)4f$, $3d^6(^3H)4f$, $3d^6(^3F)4f$, and $3d^6(^3G)4f$ subconfigurations and from the even configurations $3d^64d$, $3d^65d$, and $3d^66d$ are listed in Tables 1–5. Because the $3d^64f$ states of Fe II tend to appear in pairs we have used the $j_c[K]_j$ notation of jK coupling for them, where $\mathbf{j}_c$ is the total angular momentum of the core and $\mathbf{K}=\mathbf{J}_c+\mathbf{l}$ is the coupling of $\mathbf{J}_c$ with the orbital angular momentum $\mathbf{l}$ of the active electron. The level pairs correspond to the two separate values of the total angular momentum $\mathbf{J}$ obtained when the spin s=±1/2 of the active electron is added to $\mathbf{K}$. The positive energies are those obtained by comparing observed and predicted line profiles, as described in Sect. 3 and shown in Fig. 2. The energies between parentheses in Tables 1−4 are predicted values for which we have been not able to find the corresponding observed level. The reason for the failure is that either all the lines from the energy level are weak or, even if some of the transitions are predicted as moderately strong (log $gf > 0.0$), they are blended with other stronger components, so that their identification is uncertain. The columns with label "c−o" in Tables 1-5 show the difference between the predicted and observed energy levels.

The 4d even energy levels listed in Table 5 give rise to some of the transitions listed in the Online Material. The strongest transitions related with the 5d, and 6d even energy levels occur in the 6000-8000 Å region and in the 4000-5000 Å region, respectively. The transitions to the odd energy levels are discussed in Sect. 5

The observed energy levels, the least squares fits, the predicted energy levels, and the line lists can be found on the Kurucz web site[2]. The observed levels come from the following sources: Johansson (1978), Sugar & Corliss (1985), Adam et al. (1987), Johansson & Baschek (1988), Johansson (1988, private communication), Rosberg & Johansson (1992), Castelli, Johansson & Hubrig (2008), Castelli, Kurucz, Hubrig (2009), and this work. The calculations on the web site are updated whenever there are improvements to the energy levels.

## 5. The new Fe II lines

The new Fe II lines in the 3800-8000 Å region, produced by transitions to the Fe II subconfigurations $(^3P)4f$, $(^3H)4f$, $(^3F)4f$, and $(^3G)4f$, are shown in Tables 6−9, respectively. Only lines with log $gf \geq -1.50$ are listed, because lines with lower log $gf$ values are not observable in this wavelength region of HR 6000. The new Fe II lines are mostly concentrated in the 5100-5400 Å interval. The upper energy levels (cols. 1−4) were derived as described in Sect. 3, the lower energy levels (cols. 5−6) are those described in Sect. 4, the calculated wavelength (col. 7) is the Ritz wavelength in air, the log $gf$ values (col. 8) were computed by Kurucz, the observed wavelengths (col. 9) are the wavelengths of lines well observable in the HR 6000 spectrum. Most of them were listed as unidentified lines in Castelli & Hubrig (2007)[3]. In the last column, comments derived from the comparison of the observed and computed spectra are added for most lines. In a few cases, both computed and observed stellar lines correspond to lines measured by Johansson in laboratory works (Johansson 1978; Castelli, Johansson, & Hubrig 2008). The notes "J78" and "lab" are added for these lines. When lines are computed weaker than the observed ones the disagreement can be due either to a too low log $gf$ value or to some unknown component which increases the line intensity. When lines are computed much stronger than the observed ones, some problem with the energy levels or/and log $gf$ computations is very probably present. When we observed a very good agreement between the observed and computed lines, either isolated or blends, we added the note "good agreement".

---

[2] http://kurucz.harvard.edu/atoms/2601
[3] http://wwwuser.oat.ts.astro.it/castelli/hr6000/unidentified.txt

**Table 1.** Fe II energy levels for the $3d^6$ ($^3P$)4f subconfiguration. Energies between parentheses are predicted values.

| Design-ation | J | Energy cm$^{-1}$ | c−o cm$^{-1}$ | Design-ation | J | Energy cm$^{-1}$ | c−o cm$^{-1}$ | Design-ation | J | Energy cm$^{-1}$ | c−o cm$^{-1}$ |
|---|---|---|---|---|---|---|---|---|---|---|---|
| 2[5] | 11/2 | 122351.810 | −20.236 | | | | | | | | |
|      | 9/2  | 122324.142 | −18.980 | | | | | | | | |
| 2[4] | 9/2  | 122355.116 | −6.685  | 1[4] | 9/2 | 123629.520 | −4.606 | | | | |
|      | 7/2  | 122355.553 | −6.801  |      | 7/2 | 123637.833 | −6.417 | | | | |
| 2[3] | 7/2  | 122351.488 | −18.489 | 1[3] | 7/2 | 123615.875 | −2.642 | 0[3] | 7/2 | (124167.229) | |
|      | 5/2  | (122353.541) |        |      | 5/2 | 123649.493 | −5.687 |      | 5/2 | 124157.060   | +15.841 |
| 2[2] | 5/2  | (122342.921) |        | 1[2] | 5/2 | (123637.063) | | | | | |
|      | 3/2  | (122336.098) |        |      | 3/2 | (123646.360) | | | | | |
| 2[1] | 3/2  | (122358.405) |        | | | | | | | | |
|      | 1/2  | (122332.608) |        | | | | | | | | |

**Table 2.** Fe II energy levels for the $3d^6$ ($^3H$)4f subconfiguration. Energies between parentheses are predicted values.

| Design-ation | J | Energy cm$^{-1}$ | c−o cm$^{-1}$ | Design-ation | J | Energy cm$^{-1}$ | c−o cm$^{-1}$ | Design-ation | J | Energy cm$^{-1}$ | c−o cm$^{-1}$ |
|---|---|---|---|---|---|---|---|---|---|---|---|
| 6[9] | 19/2 | 122954.180 | +14.465 | | | | | | | | |
|      | 17/2 | 122952.730 | +20.251 | | | | | | | | |
| 6[8] | 17/2 | 123007.910 | +26.752 | 5[8] | 17/2 | 123219.200 | −10.198 | | | | |
|      | 15/2 | 122910.920 | −16.531 |      | 15/2 | 123193.090 | −17.864 | | | | |
| 6[7] | 15/2 | 123018.430 | +34.439 | 5[7] | 15/2 | 123238.440 | −6.653  | 4[7] | 15/2 | 123396.250 | −33.027 |
|      | 13/2 | 123015.400 | +40.333 |      | 13/2 | 123168.680 | −33.645 |      | 13/2 | 123355.490 | −36.436 |
| 6[6] | 13/2 | 122990.620 | −2.720  | 5[6] | 13/2 | 123249.650 | −6.519  | 4[6] | 13/2 | 123414.730 | −32.244 |
|      | 11/2 | 123037.430 | +26.878 |      | 11/2 | 123270.340 | +0.899  |      | 11/2 | 123427.119 | −33.418 |
| 6[5] | 11/2 | 123002.288 | +33.455 | 5[5] | 11/2 | 123251.470 | −1.320  | 4[5] | 11/2 | 123441.100 | −26.889 |
|      | 9/2  | 123026.350 | +18.587 |      | 9/2  | 123269.378 | +2.937  |      | 9/2  | 123435.468 | −17.705 |
| 6[4] | 9/2  | 122988.215 | +30.836 | 5[4] | 9/2  | 123258.994 | −1.556  | 4[4] | 9/2  | 123460.690 | −26.898 |
|      | 7/2  | 122980.408 | +26.752 |      | 7/2  | 123258.021 | −1.362  |      | 7/2  | 123435.277 | −16.103 |
| 6[3] | 7/2  | 122946.419 | +21.403 | 5[3] | 7/2  | 123235.165 | +3.471  | 4[3] | 7/2  | 123451.449 | −21.115 |
|      | 5/2  | (122967.896) |        |      | 5/2  | (123248.017) |        |      | 5/2  | 123430.181 | −16.906 |
|      |      |            |         | 5[2] | 5/2  | 123211.159 | −1.017  | 4[2] | 5/2  | (123401.927) | |
|      |      |            |         |      | 3/2  | 123213.323 | −12.585 |      | 3/2  | (123384.857) | |
|      |      |            |         |      |      |            |         | 4[1] | 3/2  | (123356.410) | |
|      |      |            |         |      |      |            |         |      | 1/2  | (123343.705) | |

Figure 1 shows the Fe II spectrum in the 5185-5196 Å interval, computed before and after the determination of the new energy levels. Figure 2 compares the observed spectrum of HR 6000 with the synthetic spectrum computed with the line list including the new Fe II lines. When the two figures are considered together, the improvement in the comparison between the observed and computed spectra is evident.

**Table 3.** Fe II energy levels for the 3d$^6$ ($^3$F)4f subconfiguration. Energies between parentheses are predicted values.

| Design-ation | J | Energy cm$^{-1}$ | c−o cm$^{-1}$ | Design-ation | J | Energy cm$^{-1}$ | c−o cm$^{-1}$ | Design-ation | J | Energy cm$^{-1}$ | c−o cm$^{-1}$ |
|---|---|---|---|---|---|---|---|---|---|---|---|
| 4[7] | 15/2 | 124421.468 | +12.238 | | | | | | | | |
| | 13/2 | 124436.436 | +36.895 | | | | | | | | |
| 4[6] | 13/2 | 124400.107 | +4.567 | 3[6] | 13/2 | 124661.274 | +15.827 | | | | |
| | 11/2 | 124402.557 | −3.593 | | 11/2 | 124656.535 | +7.092 | | | | |
| 4[5] | 11/2 | 124388.840 | +3.174 | 3[5] | 11/2 | 124626.900 | +3.179 | 2[5] | 11/2 | 124803.873 | +20.054 |
| | 9/2 | 124385.706 | +2.938 | | 9/2 | 124636.116 | +3.120 | | 9/2 | 124809.727 | +15.721 |
| 4[4] | 9/2 | 124401.939 | +4.674 | 3[4] | 9/2 | 124623.120 | +3.085 | 2[4] | 9/2 | 124793.905 | +12.624 |
| | 7/2 | 124385.010 | +0.698 | | 7/2 | 124620.914 | +7.289 | | 7/2 | 124783.748 | +15.272 |
| 4[3] | 7/2 | 124416.110 | +13.187 | 3[3] | 7/2 | 124641.989 | +9.092 | 2[3] | 7/2 | (124814.025) | |
| | 5/2 | 124403.474 | +1.243 | | 5/2 | 124653.022 | −8.651 | | 5/2 | (124808.178) | |
| 4[2] | 5/2 | 124434.563 | +23.142 | 3[2] | 5/2 | (124670.316) | | 2[2] | 5/2 | (124835.676) | |
| | 3/2 | 124460.410 | −11.802 | | 3/2 | (124678.325) | | | 3/2 | (124833.418) | |
| 4[1] | 3/2 | (124487.989) | | 3[1] | 3/2 | (124697.077) | | 2[1] | 3/2 | (124876.972) | |
| | 1/2 | (124484.721) | | | 1/2 | (124708.453) | | | 1/2 | (124874.375) | |
| | | | | 3[0] | 1/2 | 124731.762 | −4.875 | | | | |

**Table 4.** Fe II energy levels for the 3d$^6$ ($^3$G)4f subconfiguration. Energies between parentheses are predicted values.

| Design-ation | J | Energy cm$^{-1}$ | c−o cm$^{-1}$ | Design-ation | J | Energy cm$^{-1}$ | c−o cm$^{-1}$ | Design-ation | J | Energy cm$^{-1}$ | c−o cm$^{-1}$ |
|---|---|---|---|---|---|---|---|---|---|---|---|
| 5[8] | 17/2 | 127507.241 | −5.657 | | | | | | | | |
| | 15/2 | 127524.1227 | +14.501 | | | | | | | | |
| 5[7] | 15/2 | 127484.653 | −1.445 | 4[7] | 15/2 | 127892.981 | +4.313 | | | | |
| | 13/2 | 127515.235 | +2.816 | | 13/2 | 127895.260 | +3.367 | | | | |
| 5[6] | 13/2 | 127489.429 | −4.823 | 4[6] | 13/2 | 127875.000 | +2.236 | 3[6] | 13/2 | 128110.214 | −2.182 |
| | 11/2 | 127489.977 | −0.294 | | 11/2 | 127880.436 | +1.216 | | 11/2 | (128076.012) | |
| 5[5] | 11/2 | 127482.748 | +3.147 | 4[5] | 11/2 | 127869.158 | +0.993 | 3[5] | 11/2 | 128071.171 | −10.517 |
| | 9/2 | (127484.561) | | | 9/2 | 127855.952 | −16.898 | | 9/2 | 128055.658 | −16.898 |
| 5[4] | 9/2 | 127485.362 | −15.194 | 4[4] | 9/2 | 127869.892 | −4.920 | 3[4] | 9/2 | 128062.710 | −15.669 |
| | 7/2 | 127485.699 | +9.404 | | 7/2 | (127871.098) | | | 7/2 | 128066.823 | −22.228 |
| 5[3] | 7/2 | (127476.624) | | 4[3] | 7/2 | (127877.776) | | 3[3] | 7/2 | (128047.849) | |
| | 5/2 | 127510.913 | +9.552 | | 5/2 | 127874.745 | +5.549 | | 5/2 | 128063.103 | −8.192 |
| 5[2] | 5/2 | (127499.343) | | 4[2] | 5/2 | (127868.807) | | 3[2] | 5/2 | 128089.313 | +10.032 |
| | 3/2 | 127487.681 | −0.341 | | 3/2 | (127895.930) | | | 3/2 | (128069.044) | |
| | | | | 4[1] | 3/2 | (127876.787) | | 3[1] | 3/2 | (128099.051) | |
| | | | | | 1/2 | (127898.510) | | | 1/2 | (128099.237) | |
| | | | | | | | | 3[0] | 1/2 | (128161.312) | |

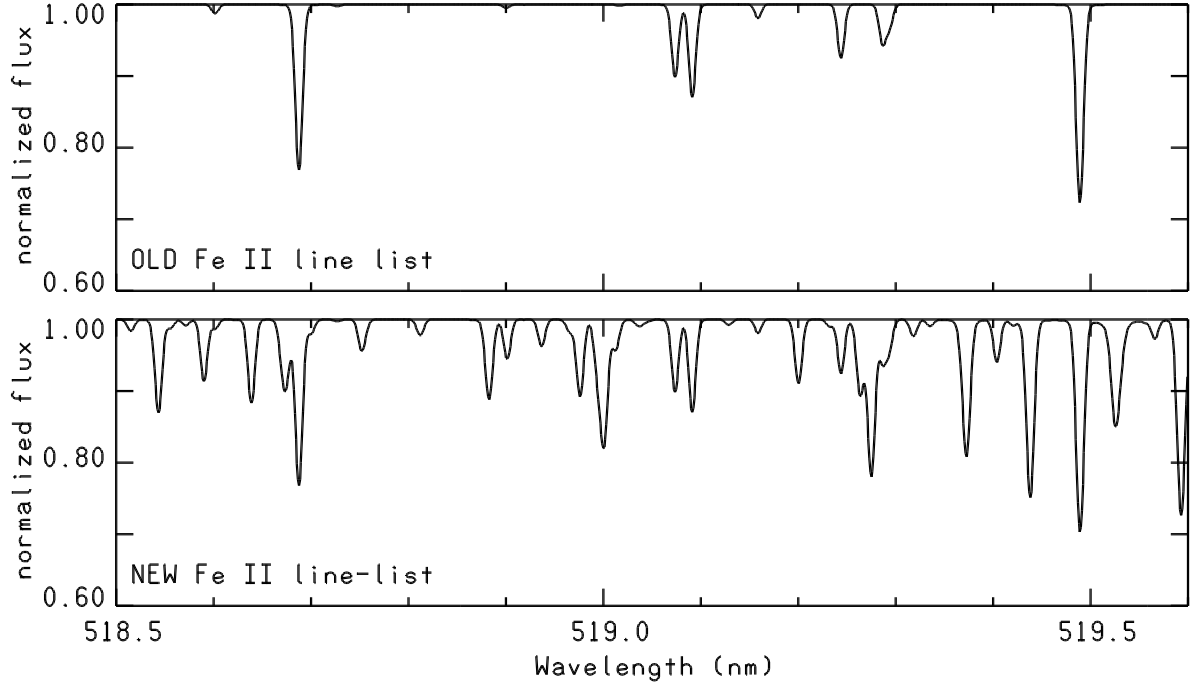

**Fig. 1.** Upper panel shows the Fe II synthetic spectrum for the parameters of HR 6000 ($T_{\mathrm{eff}}$=13450 K, log $g$=4.3, v$sini$=1.5 km$^{-1}$, [Fe/H]]=+0.9) computed with the line list availble before this work. The lower panel is the same, but with the new Fe II lines added in the line list.

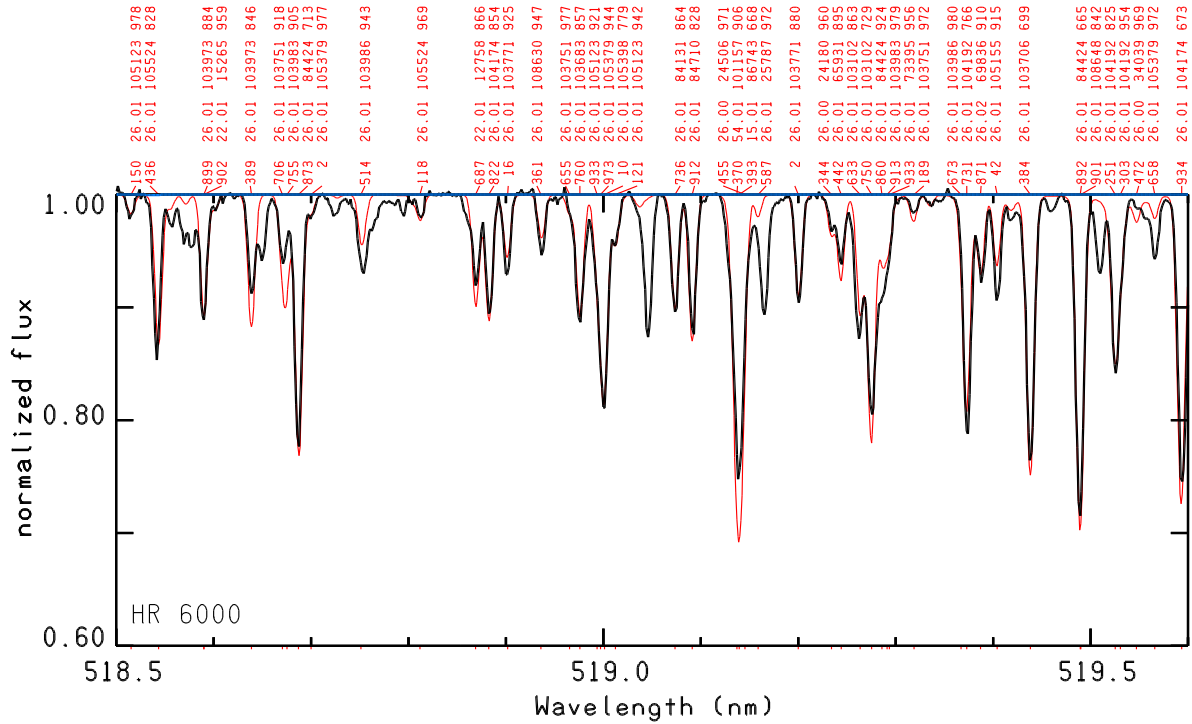

**Fig. 2.** Comparison of the UVES spectrum of HR 6000 (black line) with a synthetic spectrum (red line) computed with a line list including the new Fe II lines. The line identification can be decoded as follows: for the first line, 150 last 3 digits of wavelength 518.5150 nm; 26 atomic number of iron; .01 charge/100, i.e. 26.01 identifies the line as Fe II; 105123 is the energy of the lower level in cm$^{-1}$; 970 is the residual central intensity in per mil.

**Table 5.** Fe II new levels from $3d^64d$, $3d^65d$, and $3d^66d$ configurations.

| Designation | | J | Energy cm$^{-1}$ | c−o cm$^{-1}$ |
|---|---|---|---|---|
| $3d^6(^3P)4d$ | $^2F$ | 7/2 | 103191.917 | +27.014 |
| $3d^6(^3P)4d$ | $^2D$ | 5/2 | 103597.402 | −5.701 |
| $3d^6(^3F)4d$ | $^2F$ | 7/2 | 105775.491 | −42.697 |
| $3d^6(^3H)5d$ | $^4H$ | 13/2 | 124208.725 | +47.495 |
| $3d^6(^3H)5d$ | $^4G$ | 11/2 | 124251.805 | +44.041 |
| $3d^6(^3H)5d$ | $^4K$ | 15/2 | 124297.017 | −5.220 |
| $3d^6(^3H)5d$ | $^4I$ | 15/2 | 124357.304 | +12.292 |
| $3d^6(^3H)5d$ | $^4K$ | 13/2 | 124415.353 | −14.256 |
| $3d^6(^3H)5d$ | $^2I$ | 11/2 | 124976.008 | −38.096 |
| $3d^6(^3F)5d$ | $^4H$ | 13/2 | 125732.991 | +9.243 |
| $3d^6(^5D)6d$ | $^6D$ | 5/2 | 113934.466 | −58.836 |
| $3d^6(^5D)6d$ | $^4D$ | 7/2 | 114009.934 | −3.477 |
| $3d^6(^5D)6d$ | $^6G$ | 7/2 | 114428.399 | +51.787 |
| $3d^6(^5D)6d$ | $^6G$ | 5/2 | 114619.007 | +22.415 |

## 6. Conclusions

Computed atomic data and stellar spectra observed at high resolution and high signal-to-noise ratio of the iron−overabundant, slow−rotating star HR 6000 were used to extend laboratory studies on Fe II energy levels and line transitions. We identified as Fe II about 500 unidentified spectral lines in the 3800−8000 Å region. A few of these lines were already identified as iron from laboratory analyses (Johansson 2007, private communication), but they were never classified. Because numerous other new lines are components of blends they contribute to improve the agreement between observed and computed spectra. On the other hand, there is a small number of new lines which are not observed in the spectrum. We believe that they are due to computational problems related with the mixing of the even energy levels rather than to incorrect energy values for the new 4f odd levels.

In spite of the large number of the new identified lines, several medium-strong lines and a conspicuous number of weak lines remain still unidentified in the spectral region we analyzed. If we examine the list of the Fe II lines which correspond to transitions from predicted energy levels, we can count about 4600 lines with $\log gf \geq -1.0$, where about 400 of them have $\log gf \geq 0.0$. Because the transitions producing these lines occur between high-excitation energy levels that are not strongly populated, most of the lines are weak in a star like HR 6000. This large number of weak predicted lines could explain the spectrum of HR 6000 longward of about 5800 Å. The spectrum looks like it is affected by a noise larger than that due to the instrumental effects. Castelli & Hubrig (2007) explained this "noise" with the presence of a T-Tauri star affecting the HR 6000 spectrum. After this study, we prefer to state that the spectrum shows the presence of numerous weak Fe II lines from high-excitation levels, probably 4d, 5d, 6d − 4f, 5f, 6f transitions, which still have to be identified. The hypothesis of the presence of the T-Tauri star affecting the HR 6000 spectrum is an example of an incorrect conclusion that can be drawn owing to the use of incomplete line lists. We will extend this study of the Fe II spectrum to the near infrared region in the near future using CRIRES (CRyogenic high-resolution InfraRed Echelle Spectrograph) observations of HR 6000 and 46 Aql. The observations are scheduled in summer 2010 (ESO proposal 41380, P. I. S. Hubrig).

## References

Adam, J., Baschek, B., Johansson, S., Nilsson, A. E., & Brage, T. 1987, ApJ, 312, 337
Biémont, E., Johansson, S., & Palmeri, P. 1997, Phys. Scr, 55, 559
Castelli, F., & Hubrig, S. 2007, A&A, 475, 1041
Castelli, F., Johansson, S., & Hubrig, S. 2008, Journal of Physics Conference Series, 130, 012003
Castelli, F., Kurucz, R., & Hubrig, S. 2009, A&A, 508, 401 (Paper I)
Cowan, R. D. 1981, The Theory of Atomic Structure and Spectra (Berkeley: Univ. California Press)
Johansson, S. 1978, Phys. Scr, 18, 217
Johansson, S. 2009, Phys. Scr, T134, 014013
Johansson, S., & Baschek, B. 1988, Nuclear Instruments and Methods in Physics Research B, 31, 222
Kurucz, R. L. 2005, Memorie della Societa' Astronomica Italiana, Supplementi, 8, 14
Kurucz, R. L. 2009, American Institute of Physics Conference Series, 1171, 43
Rosberg, M., & Johansson, S. 1992, Phys. Scr, 45, 590
Sugar, J., & Corliss, C. 1985, J. Phys. Chem. Ref. Data, 14, Supp. 2

# Online Material

**Table 6.** Fe II lines in the 3800-8000 Å region with log $gf \geq -1.5$ and $3d^6(^3P)4f$ energy levels as upper levels

| Upper level | | | | Lower level | | λ(calc) | log gf | λ(obs) | Notes |
|---|---|---|---|---|---|---|---|---|---|
| cm$^{-1}$ | | | J | cm$^{-1}$ | | Å | KUR | Å | |
| 122351.810 | ($^3$P)4f | 2[5] | 11/2 | 103165.320 | ($^3$P)4d $^4$F$_{9/2}$ | 5210.550 | +0.795 | 5210.55 | good agreement |
| | | | | 103683.070 | ($^5$D)5d $^4$F$_{9/2}$ | 5355.059 | +0.164 | 5355.06 | computed too strong |
| | | | | 103771.320 | ($^3$H)4d $^4$G$_{9/2}$ | 5380.493 | −1.047 | | at the noise level |
| | | | | 104807.210 | ($^3$H)4d $^2$G$_{9/2}$ | 5698.178 | −0.539 | | blend with a telluric line |
| | | | | 104916.550 | ($^3$H)4d $^4$F$_{9/2}$ | 5733.913 | −0.635 | 5733.90 | computed too weak |
| | | | | 106722.170 | ($^3$F)4d $^4$F$_{9/2}$ | 6396.332 | −0.741 | 6396.32 | computed too weak |
| | | | | 109811.920 | ($^3$G)4d $^4$F$_{9/2}$ | 7972.359 | −0.985 | | at the noise level |
| 122324.142 | ($^3$P)4f | 2[5] | 9/2 | 103102.860 | ($^3$P)4d $^4$D$_{7/2}$ | 5201.118 | −0.056 | | wrong, not observed |
| | | | | 103191.917 | ($^3$P)4d $^2$F$_{7/2}$ | 5225.329 | +0.634 | | blend, good agreement |
| | | | | 103986.330 | ($^3$H)4d $^4$H$_{7/2}$ | 5451.698 | −1.133 | | blend, good agreement |
| | | | | 104107.950 | ($^3$P)4d $^4$F$_{7/2}$ | 5488.097 | −0.362 | | blend, good agreement |
| | | | | 104481.590 | ($^3$H)4d $^2$F$_{7/2}$ | 5603.024 | −0.170 | 5603.05 | |
| | | | | 105123.000 | ($^3$H)4d $^2$G$_{7/2}$ | 5811.956 | −1.441 | | blend, good agreement |
| | | | | 105775.491 | ($^3$F)4d $^2$F$_{7/2}$ | 6041.116 | −0.837 | 6041.1 | weak, good agreement |
| 122355.116 | ($^3$P)4f | 2[4] | 9/2 | 102394.718 | ($^5$D)6s $^4$D$_{7/2}$ | 5008.523 | −0.809 | | weak, computed too strong |
| | | | | 103102.860 | ($^3$P)4d $^4$D$_{7/2}$ | 5192.750 | +0.657 | 5192.75 | lab, good agreement |
| | | | | 103165.320 | ($^3$P)4d $^4$F$_{9/2}$ | 5209.652 | −0.035 | 5209.66 | lab, good agreement |
| | | | | 103191.917 | ($^3$P)4d $^2$F$_{7/2}$ | 5216.883 | −0.404 | | blend |
| | | | | 103683.070 | ($^5$D)5d $^4$F$_{9/2}$ | 5354.110 | −0.637 | 5354.1 | weak |
| | | | | 104107.950 | ($^3$P)4d $^4$F$_{7/2}$ | 5478.781 | −1.319 | | at the continuum level |
| | | | | 104807.210 | ($^3$H)4d $^2$G$_{9/2}$ | 5697.105 | −1.443 | | at the continuum level |
| | | | | 106767.210 | ($^3$F)4d $^4$F$_{7/2}$ | 6413.457 | −1.407 | | blend |
| 122355.550 | ($^3$P)4f | 2[4] | 7/2 | 102394.718 | ($^5$D)6s $^4$D$_{7/2}$ | 5008.414 | −1.258 | | good agreement |
| | | | | 102802.312 | ($^5$D)6s $^4$D$_{5/2}$ | 5112.818 | −0.959 | 5112.82 | computed too weak |
| | | | | 103002.670 | ($^3$P)4d $^4$D$_{5/2}$ | 5165.751 | +0.441 | 5165.75 | lab, good agreement |
| | | | | 103102.860 | ($^3$P)4d $^4$D$_{7/2}$ | 5192.633 | +0.155 | 5192.62 | lab, computed too weak |
| | | | | 103165.320 | ($^3$P)4d $^4$F$_{9/2}$ | 5209.534 | −1.105 | | blend, good agreement |
| | | | | 103191.917 | ($^3$P)4d $^2$F$_{7/2}$ | 5216.765 | −0.764 | | blend |
| | | | | 106796.660 | ($^3$F)4d $^4$P$_{5/2}$ | 6425.418 | −1.436 | | at the continuum level |
| 122351.488 | ($^3$P)4f | 2[3] | 7/2 | 103102.860 | ($^3$P)4d $^4$D$_{7/2}$ | 5193.729 | −1.320 | | blend |
| | | | | 103191.917 | ($^3$P)4d $^2$F$_{7/2}$ | 5217.871 | −0.250 | 5217.870 | lab |
| | | | | 103597.402 | ($^3$P)4d $^2$D$_{5/2}$ | 5330.689 | +0.525 | 5330.680 | lab |
| | | | | 104023.910 | ($^3$H)4d $^4$G$_{5/2}$ | 5454.742 | −1.327 | | at the continuum level |
| | | | | 104107.950 | ($^3$P)4d $^4$F$_{7/2}$ | 5479.870 | −1.320 | | at the continuum level |
| | | | | 104481.590 | ($^3$H)4d $^2$F$_{7/2}$ | 5594.450 | −1.116 | 5594.42 | computed too weak ? |
| | | | | 104569.230 | ($^3$P)4d $^4$F$_{5/2}$ | 5622.022 | −0.573 | 5622.02 | computed too weak ? |
| | | | | 105234.237 | ($^3$H)4d $^4$F$_{5/2}$ | 5840.440 | −1.282 | | at the continuum level |
| | | | | 107407.800 | ($^3$F)4d $^2$D$_{5/2}$ | 6689.941 | −0.330 | 6689.91 | |
| 123629.520 | ($^3$P)4f | 1[4] | 9/2 | 103102.860 | ($^3$P)4d $^4$D$_{7/2}$ | 4870.353 | −1.402 | | at the continuum level |
| | | | | 104000.810 | ($^5$D)5d $^6$P$_{7/2}$ | 5093.159 | −0.981 | | blend |
| | | | | 104107.950 | ($^3$P)4d $^4$F$_{7/2}$ | 5121.112 | +0.327 | 5121.1 | lab, good agreement |
| | | | | 104481.590 | ($^3$H)4d $^2$F$_{7/2}$ | 5221.043 | +0.408 | 5221.04 | lab, good agreement |
| | | | | 104873.230 | ($^5$D)5d $^4$D$_{7/2}$ | 5330.062 | −1.183 | | blend |
| | | | | 104993.860 | ($^3$F)4d $^4$D$_{7/2}$ | 5364.564 | −0.118 | 5364.55 | computed too strong |
| | | | | 105123.000 | ($^3$H)4d $^2$G$_{7/2}$ | 5401.999 | −0.418 | | blend |
| | | | | 105220.600 | ($^3$H)4d $^4$F$_{7/2}$ | 5430.640 | −1.066 | 5430.64 | computed too weak |
| | | | | 105775.491 | ($^3$F)4d $^2$F$_{7/2}$ | 5599.422 | −0.624 | 5599.42 | good agreement |
| | | | | 106767.210 | ($^3$F)4d $^4$F$_{7/2}$ | 5928.743 | −0.677 | 5928.72 | at the noise level |
| | | | | 110167.280 | ($^3$G)4d $^4$F$_{7/2}$ | 7426.139 | −1.173 | | |

**Table 6.** Fe II lines in the 3800-8000 Å region with log $gf \geq -1.5$ and 3d$^6$($^3$P)4f energy levels as upper levels

| Upper level | | | | Lower level | | $\lambda$(calc) | log $gf$ | $\lambda$(obs) | Notes |
|---|---|---|---|---|---|---|---|---|---|
| cm$^{-1}$ | | | J | cm$^{-1}$ | | Å | KUR | Å | |
| 123637.833 | ($^3$P)4f | 1[4] | 7/2 | 102802.312 | ($^5$D)6s $^4$D$_{5/2}$ | 4798.155 | −1.297 | | at the continuum level |
| | | | | 103002.670 | ($^3$P)4d $^4$D$_{5/2}$ | 4844.743 | −0.954 | | computed too strong |
| | | | | 103597.402 | ($^3$P)4d $^2$D$_{5/2}$ | 4988.521 | −0.339 | 4988.51 | lab |
| | | | | 104107.950 | ($^3$P)4d $^4$F$_{7/2}$ | 5118.932 | −0.819 | 5118.95 | lab, computed too weak |
| | | | | 104120.270 | ($^5$D)5d $^6$P$_{5/2}$ | 5122.163 | −1.282 | | |
| | | | | 104481.590 | ($^3$H)4d $^2$F$_{7/2}$ | 5218.777 | −0.644 | | blend |
| | | | | 104569.230 | ($^3$P)4d $^4$F$_{5/2}$ | 5242.763 | +0.180 | 5242.775 | lab |
| | | | | 104993.860 | ($^3$F)4d $^4$D$_{7/2}$ | 5362.172 | −1.268 | | at the continuum level |
| | | | | 105127.770 | ($^5$D)5d $^4$D$_{5/2}$ | 5400.965 | −1.143 | | at the continuum level |
| | | | | 105234.237 | ($^3$H)4d $^4$F$_{5/2}$ | 5432.211 | −0.531 | | wrong, not observed |
| | | | | 105379.430 | ($^3$F)4d $^4$D$_{5/2}$ | 5475.409 | −0.552 | 5475.42 | computed too strong |
| | | | | 105711.730 | ($^5$D)5d $^6$S$_{5/2}$ | 5576.909 | −1.432 | | at the continuum level |
| | | | | 106208.560 | ($^3$F)4d $^2$F$_{5/2}$ | 5735.883 | −1.221 | | at the continuum level |
| | | | | 106796.660 | ($^3$F)4d $^4$P$_{5/2}$ | 5936.184 | −1.317 | | at the level of the noise |
| | | | | 106866.760 | ($^3$F)4d $^4$F$_{5/2}$ | 5960.996 | −0.565 | 5961.00 | |
| | | | | 107407.800 | ($^3$F)4d $^2$D$_{5/2}$ | 6159.712 | −0.665 | 6179.75 | blend ? |
| | | | | 110428.280 | ($^3$G)4d $^4$F$_{5/2}$ | 7568.195 | −1.229 | | no spectrum |
| 123615.875 | ($^3$P)4f | 1[3] | 7/2 | 103597.402 | ($^3$P)4d $^2$D$_{5/2}$ | 4993.993 | −1.435 | | |
| | | | | 104023.910 | ($^3$H)4d $^4$G$_{5/2}$ | 5102.711 | −0.526 | 5102.7 | lab, good agreement |
| | | | | 104107.950 | ($^3$P)4d $^4$F$_{7/2}$ | 5124.694 | −1.046 | 5124.69 | good agreement |
| | | | | 104120.270 | ($^5$D)5d $^6$P$_{5/2}$ | 5127.932 | −0.244 | | wrong, not obs |
| | | | | 104209.610 | ($^3$H)4d $^2$F$_{5/2}$ | 5151.540 | −0.081 | 5151.52 | J78, lab, computed too weak |
| | | | | 104481.590 | ($^3$H)4d $^2$F$_{7/2}$ | 5224.766 | −0.973 | 5227.77 | good agreement |
| | | | | 104569.230 | ($^3$P)4d $^4$F$_{5/2}$ | 5248.807 | −0.232 | 5248.801 | computed too strong |
| | | | | 105127.770 | ($^5$D)5d $^4$D$_{5/2}$ | 5407.380 | −1.391 | 5407.37 | computed too weak |
| | | | | 105234.237 | ($^3$H)4d $^4$F$_{5/2}$ | 5438.700 | −0.416 | 5438.70 | computed too strong |
| | | | | 106208.560 | ($^3$F)4d $^2$F$_{5/2}$ | 5743.118 | −0.454 | 5743.10 | good agreement |
| 123649.493 | ($^3$P)4f | 1[3] | 5/2 | 104209.610 | ($^3$H)4d $^2$F$_{5/2}$ | 5142.631 | −1.288 | | at the continuum level |
| | | | | 104569.230 | ($^3$P)4d $^4$F$_{5/2}$ | 5239.559 | −1.150 | 5239.56 | good agreement |
| | | | | 104572.920 | ($^3$P)4d $^4$F$_{3/2}$ | 5240.573 | +0.071 | 5240.587 | lab, good agreement |
| | | | | 104588.710 | ($^5$D)5d $^6$D$_{3/2}$ | 5244.914 | −1.288 | | blend |
| | | | | 104839.998 | ($^3$P)4d $^2$D$_{3/2}$ | 5314.985 | −0.441 | | blend, computed too strong |
| | | | | 105234.237 | ($^3$H)4d $^4$F$_{5/2}$ | 5428.771 | −1.471 | | blend |
| | | | | 105317.440 | ($^3$P)4d $^2$P$_{3/2}$ | 5453.411 | +0.082 | 5453.42 | lab, computed too strong |
| | | | | 105518.140 | ($^3$H)4d $^4$F$_{3/2}$ | 5513.777 | −0.591 | | wrong, not observed |
| | | | | 106846.650 | ($^3$F)4d $^4$F$_{3/2}$ | 5949.725 | −1.358 | | at the continuum level |
| | | | | 107430.250 | ($^3$F)4d $^2$D$_{3/2}$ | 6163.810 | −0.253 | | wrong, not observed |
| | | | | 108105.900 | ($^3$F)4d $^2$P$_{3/2}$ | 6431.741 | −0.724 | | blend |
| 124157.060 | ($^3$P)4f | 0[3] | 5/2 | 104569.230 | ($^3$P)4d $^4$F$_{5/2}$ | 5103.788 | −1.191 | 5103.8 | good agreement |
| | | | | 104572.920 | ($^3$P)4d $^4$F$_{3/2}$ | 5104.750 | +0.094 | 5104.75 | lab, good agreement |
| | | | | 104588.710 | ($^5$D)5d $^6$D$_{3/2}$ | 5108.869 | −1.369 | | |
| | | | | 104839.998 | ($^3$P)4d $^2$D$_{3/2}$ | 5175.329 | −1.125 | | blend |
| | | | | 105234.237 | ($^3$H)4d $^4$F$_{5/2}$ | 5283.154 | −0.937 | | blend |
| | | | | 105317.440 | ($^3$P)4d $^2$P$_{3/2}$ | 5306.486 | −1.020 | 5306.49 | computed too weak |
| | | | | 105460.230 | ($^3$F)4d $^4$D$_{3/2}$ | 5347.013 | −0.482 | 5347.05 | blend |
| | | | | 105518.140 | ($^3$H)4d $^4$F$_{3/2}$ | 5363.626 | +0.082 | 5363.61 | computed too strong |
| | | | | 106846.650 | ($^3$F)4d $^4$F$_{3/2}$ | 5775.269 | −0.286 | 5775.25 | good agreement |
| | | | | 107430.250 | ($^3$F)4d $^2$D$_{3/2}$ | 5976.771 | −0.922 | | blend |
| | | | | 108105.900 | ($^3$F)4d $^2$P$_{3/2}$ | 6228.356 | −0.686 | 6228.34 | good agreement |
| | | | | 110609.540 | ($^3$G)4d $^4$F$_{3/2}$ | 7379.392 | −1.370 | | at the continuum level |

**Table 7.** Fe II lines in the 3800-8000 Å region with log $gf \geq -1.5$ and $3d^6(^3H)4f$ energy level as upper levels

| Upper level | | | | Lower level | | $\lambda$(calc) | log $gf$ | $\lambda$(obs) | Notes |
|---|---|---|---|---|---|---|---|---|---|
| cm$^{-1}$ | | | J | cm$^{-1}$ | | Å | KUR | Å | |
| 122954.180 | ($^3$H)4f | 6[9] | 19/2 | 103644.800 | ($^3$H)4d $^4$K$_{17/2}$ | 5177.388 | +1.169 | 5177.394 | J78, lab, good agreement |
| 122952.730 | ($^3$H)4f | 6[9] | 17/2 | 103644.800 | ($^3$H)4d $^4$K$_{17/2}$ | 5177.777 | −0.930 | | blend |
| | | | | 103706.530 | ($^3$H)4d $^4$K$_{15/2}$ | 5194.384 | +0.798 | 5194.387 | lab, good agreement |
| | | | | 103878.370 | ($^3$H)4d $^4$I$_{15/2}$ | 5241.181 | +0.558 | 5241.183 | J78, lab, good agreement |
| | | | | 104119.710 | ($^3$H)4d $^2$K$_{15/2}$ | 5308.346 | +0.518 | 5308.350 | J78, lab, good agreement |
| 123007.910 | ($^3$H)4f | 6[8] | 17/2 | 103644.800 | ($^3$H)4d $^4$K$_{17/2}$ | 5163.021 | +0.498 | 5163.018 | J78, lab, good agreement |
| | | | | 103706.530 | ($^3$H)4d $^4$K$_{15/2}$ | 5179.534 | +0.534 | 5179.540 | J78, lab, good agreement |
| | | | | 103878.370 | ($^3$H)4d $^4$I$_{15/2}$ | 5226.062 | +0.820 | 5226.070 | lab, good agreement |
| | | | | 104119.710 | ($^3$H)4d $^2$K$_{15/2}$ | 5292.838 | −1.419 | | |
| | | | | 108337.860 | ($^3$G)4d $^4$I$_{15/2}$ | 6814.729 | −1.183 | | at the noise level |
| 122910.920 | ($^3$H)4f | 6[8] | 15/2 | 103706.530 | ($^3$H)4d $^4$K$_{15/2}$ | 5205.693 | −0.207 | 5205.70 | blend |
| | | | | 103832.050 | ($^3$H)4d $^4$K$_{13/2}$ | 5239.942 | +0.015 | 5239.948 | J78, lab, computed too weak |
| | | | | 103878.370 | ($^3$H)4d $^4$I$_{15/2}$ | 5252.695 | −0.107 | 5252.702 | lab, computed too weak |
| | | | | 104064.670 | ($^3$H)4d $^4$I$_{13/2}$ | 5304.620 | −0.357 | 5304.60 | lab, computed too weak |
| | | | | 104119.710 | ($^3$H)4d $^2$K$_{15/2}$ | 5320.157 | +0.082 | 5320.18 | lab, good agreement |
| | | | | 104315.370 | ($^3$H)4d $^2$K$_{13/2}$ | 5376.136 | +0.132 | 5376.12 | lab, computed too weak |
| | | | | 104622.300 | ($^3$H)4d $^2$I$_{13/2}$ | 5466.362 | +0.698 | 5466.38 | good agreement |
| | | | | 108463.910 | ($^3$G)4d $^4$I$_{13/2}$ | 6919.939 | −0.887 | | at the continuum level |
| | | | | 108648.695 | ($^1$I)5s e$^2$I$_{13/2}$ | 7009.596 | −1.436 | 7009.6 ? | computed too weak ? |
| | | | | 109049.600 | ($^3$G)4d $^2$I$_{13/2}$ | 7212.332 | −1.456 | 7212.33 ? | computed too weak ? |
| 123018.430 | ($^3$H)4f | 6[7] | 15/2 | 103617.580 | ($^3$H)4d $^4$H$_{13/2}$ | 5152.978 | +0.761 | 5152.985 | lab, good agreement |
| | | | | 103644.800 | ($^3$H)4d $^4$K$_{17/2}$ | 5160.218 | −0.354 | 5160.213 | lab, good agreement |
| | | | | 103706.530 | ($^3$H)4d $^4$K$_{15/2}$ | 5176.713 | +0.364 | 5176.722 | J78, lab, good agreement |
| | | | | 103832.050 | ($^3$H)4d $^4$K$_{13/2}$ | 5210.580 | −1.104 | 5210.65 ? | computed too weak ? |
| | | | | 103878.370 | ($^3$H)4d $^4$I$_{15/2}$ | 5223.190 | +0.447 | 5223.25 | blend, good agreement |
| | | | | 104064.670 | ($^3$H)4d $^4$I$_{13/2}$ | 5274.530 | −1.138 | 5274.53 | good agreement |
| | | | | 104119.710 | ($^3$H)4d $^2$K$_{15/2}$ | 5289.892 | −0.894 | 5289.899 | lab, good agreement |
| | | | | 104622.300 | ($^3$H)4d $^2$I$_{13/2}$ | 5434.415 | −1.378 | | at the noise level |
| | | | | 108337.860 | ($^3$G)4d $^4$I$_{15/2}$ | 6809.845 | −1.228 | | at the noise level |
| 123015.400 | ($^3$H)4f | 6[7] | 13/2 | 103600.430 | ($^3$H)4d $^4$G$_{11/2}$ | 5149.230 | +0.424 | 5149.243 | lab, good agreement |
| | | | | 103617.580 | ($^3$H)4d $^4$H$_{13/2}$ | 5153.783 | +0.761 | 5153.786 | lab, good agreement |
| | | | | 103706.530 | ($^3$H)4d $^4$K$_{15/2}$ | 5177.525 | −0.341 | | blend |
| | | | | 103751.660 | ($^3$H)4d $^4$H$_{11/2}$ | 5189.655 | −0.783 | | blend, good agreement |
| | | | | 103878.370 | ($^3$H)4d $^4$I$_{15/2}$ | 5224.017 | −0.132 | 5224.025 | lab, good agreement |
| | | | | 104119.710 | ($^3$H)4d $^2$K$_{15/2}$ | 5290.740 | −1.258 | 5290.730 | computed too weak |
| | | | | 104765.450 | ($^3$H)4d $^2$I$_{11/2}$ | 5477.945 | −1.275 | 5477.95 | good agreement |
| | | | | 105063.550 | ($^3$F)4d $^4$G$_{11/2}$ | 5568.910 | −1.164 | 5568.92 | good agreement |
| | | | | 105288.850 | ($^3$F)4d $^4$H$_{13/2}$ | 5639.690 | −1.357 | | blend |
| | | | | 106045.690 | ($^3$H)4d $^2$H$_{11/2}$ | 5891.220 | −1.302 | | blend |
| | | | | 108181.550 | ($^3$G)4d $^4$G$_{11/2}$ | 6739.478 | −1.459 | | at the noise level |
| 122990.620 | ($^3$H)4f | 6[6] | 13/2 | 103706.530 | ($^3$H)4d $^4$K$_{15/2}$ | 5184.178 | −0.976 | | blend |
| | | | | 103751.660 | ($^3$H)4d $^4$H$_{11/2}$ | 5196.339 | −0.126 | 5196.32 | computed too weak |
| | | | | 103832.050 | ($^3$H)4d $^4$K$_{13/2}$ | 5218.143 | −0.028 | 5218.149 | lab, good agreement |
| | | | | 103878.370 | ($^3$H)4d $^4$I$_{15/2}$ | 5230.790 | −1.208 | 5230.80 | good agreement |
| | | | | 103973.780 | ($^3$H)4d $^4$K$_{11/2}$ | 5257.034 | −0.940 | | blend |
| | | | | 104064.670 | ($^3$H)4d $^4$I$_{13/2}$ | 5282.281 | −1.039 | 5282.29 | blend, computed too weak |
| | | | | 104119.710 | ($^3$H)4d $^2$K$_{15/2}$ | 5297.687 | −1.010 | 5297.7 | blend |
| | | | | 104174.270 | ($^3$H)4d $^4$I$_{11/2}$ | 5313.049 | −0.954 | | blend |
| | | | | 104315.370 | ($^3$H)4d $^2$K$_{13/2}$ | 5353.192 | +0.205 | 5353.22 | blend, computed too strong |
| | | | | 104622.300 | ($^3$H)4d $^2$I$_{13/2}$ | 5442.643 | +0.049 | 5442.65 | J78, lab, good agreement |
| | | | | 104765.450 | ($^3$H)4d $^2$I$_{11/2}$ | 5485.393 | +0.141 | 5485.40 | computed too strong |

**Table 7.** Fe II lines in the 3800-8000 Å region with log $gf \geq -1.5$ and $3d^6(^3H)4f$ energy level as upper levels

| Upper level | | | | Lower level | | λ(calc) | log gf | λ(obs) | Notes |
|---|---|---|---|---|---|---|---|---|---|
| cm$^{-1}$ | | | J | cm$^{-1}$ | | Å | KUR | Å | |
| 122990.620 | cont. | | | 105063.550 | $(^3F)4d\ ^4G_{11/2}$ | 5576.608 | −0.487 | 5576.60 | computed too strong |
| | | | | 105763.270 | $(^3F)4d\ ^2H_{11/2}$ | 5803.114 | −0.380 | 5803.12 | computed too weak |
| | | | | 106045.690 | $(^3H)4d\ ^2H_{11/2}$ | 5899.835 | +0.277 | 5899.82 | good agreement |
| | | | | 108630.429 | $(^1I)5s\ e^2I_{11/2}$ | 6961.775 | −1.168 | | at the continuum level |
| | | | | 109049.600 | $(^3G)4d\ ^2I_{13/2}$ | 7171.100 | −1.477 | | at the continuum level |
| | | | | 109389.880 | $(^3G)4d\ ^2I_{11/2}$ | 7350.516 | −1.297 | 7350.49 ? | computed too weak ? |
| | | | | 109683.280 | $(^3G)4d\ ^2H_{11/2}$ | 7512.581 | −0.706 | | blend, computed too weak ? |
| | | | | | | | | | |
| 123037.430 | $(^3H)4f$ | 6[6] | 11/2 | 103751.660 | $(^3H)4d\ ^4H_{11/2}$ | 5183.727 | +0.242 | 5183.713 | J78, lab, blend |
| | | | | 103771.320 | $(^3H)4d\ ^4G_{9/2}$ | 5189.016 | −0.187 | 5189.013 | lab |
| | | | | 103832.050 | $(^3H)4d\ ^4K_{13/2}$ | 5205.425 | −0.558 | 5205.427 | lab, blend |
| | | | | 103874.260 | $(^3H)4d\ ^4H_{9/2}$ | 5216.891 | −0.503 | | blend |
| | | | | 104064.670 | $(^3H)4d\ ^4I_{13/2}$ | 5269.248 | −0.797 | 5269.235 | |
| | | | | 104315.370 | $(^3H)4d\ ^2K_{13/2}$ | 5339.807 | −0.759 | | |
| | | | | 104622.300 | $(^3H)4d\ ^2I_{13/2}$ | 5428.808 | −0.405 | 5428.80 | lab |
| | | | | 104765.450 | $(^3H)4d\ ^2I_{11/2}$ | 5471.340 | −0.934 | | |
| | | | | 104807.210 | $(^3H)4d\ ^2G_{9/2}$ | 5483.874 | −0.019 | 5483.85 | lab |
| | | | | 104916.550 | $(^3H)4d\ ^4F_{9/2}$ | 5516.963 | −0.234 | | wrong, not obs |
| | | | | 105063.550 | $(^3F)4d\ ^4G_{11/2}$ | 5562.084 | −1.223 | | |
| | | | | 105398.850 | $(^3F)4d\ ^4H_{11/2}$ | 5667.818 | −1.176 | | |
| | | | | 105763.270 | $(^3F)4d\ ^2H_{11/2}$ | 5787.389 | −0.146 | 5787.35 | |
| | | | | 106045.690 | $(^3H)4d\ ^2H_{11/2}$ | 5883.582 | +0.287 | 5883.58 | J78 |
| | | | | 106097.520 | $(^3H)4d\ ^2H_{9/2}$ | 5901.584 | −0.581 | | blend |
| | | | | 106924.430 | $(^3F)4d\ ^2G_{9/2}$ | 6204.452 | −1.391 | | |
| | | | | 109683.280 | $(^3G)4d\ ^2H_{11/2}$ | 7486.247 | −0.596 | | |
| | | | | | | | | | |
| 123002.288 | $(^3H)4f$ | 6[5] | 11/2 | 103165.320 | $(^3P)4d\ ^4F_{9/2}$ | 5039.690 | −0.526 | | |
| | | | | 103600.430 | $(^3H)4d\ ^4G_{11/2}$ | 5152.712 | +0.662 | 5152.70 | lab |
| | | | | 103617.580 | $(^3H)4d\ ^4H_{13/2}$ | 5157.271 | +0.380 | | blend |
| | | | | 103683.070 | $(^3H)4d\ ^4F_{9/2}$ | 5174.754 | −0.491 | 5174.75 | lab |
| | | | | 103751.660 | $(^3H)4d\ ^4H_{11/2}$ | 5193.192 | −0.719 | 5193.191 | blend |
| | | | | 103771.320 | $(^3H)4d\ ^4G_{9/2}$ | 5198.501 | −1.338 | | |
| | | | | 104765.450 | $(^3H)4d\ ^2I_{11/2}$ | 5481.886 | −1.256 | | |
| | | | | 104807.210 | $(^3H)4d\ ^2G_{9/2}$ | 5494.468 | −0.835 | | |
| | | | | 104916.550 | $(^3H)4d\ ^4F_{9/2}$ | 5527.686 | −1.221 | 5527.68 | computed too weak |
| | | | | 105063.550 | $(^3F)4d\ ^4G_{11/2}$ | 5572.983 | −0.697 | 5572.98 | |
| | | | | 106045.690 | $(^3H)4d\ ^2H_{11/2}$ | 5895.778 | −1.407 | | |
| | | | | 106722.170 | $(^3F)4d\ ^4F_{9/2}$ | 6140.765 | −0.940 | | |
| | | | | 108181.550 | $(^3G)4d\ ^4G_{11/2}$ | 6745.444 | −1.310 | | |
| | | | | 109811.920 | $(^3G)4d\ ^4F_{9/2}$ | 7579.208 | −1.201 | | |
| | | | | | | | | | |
| 123026.350 | $^3H)4f$ | 6[5] | 9/2 | 103102.860 | $(^3P)4d\ ^4D_{7/2}$ | 5017.801 | −1.092 | | |
| | | | | 103751.660 | $(^3H)4d\ ^4H_{11/2}$ | 5186.706 | −0.152 | 5186.722 | lab |
| | | | | 103771.320 | $(^3H)4d\ ^4G_{9/2}$ | 5192.002 | +0.073 | 5192.010 | lab |
| | | | | 103874.260 | $(^3H)4d\ ^4H_{9/2}$ | 5219.909 | −0.488 | | blend |
| | | | | 104107.950 | $(^3P)4d\ ^4F_{7/2}$ | 5284.389 | −0.355 | | |
| | | | | 104481.590 | $(^3H)4d\ ^2F_{7/2}$ | 5390.860 | −1.184 | | |
| | | | | 104807.210 | $(^3H)4d\ ^2G_{9/2}$ | 5487.209 | +0.186 | 5487.21 | lab |
| | | | | 104916.550 | $(^3H)4d\ ^4F_{9/2}$ | 5520.339 | −0.063 | | wrong, not observed |
| | | | | 104993.860 | $(^3F)4d\ ^4D_{7/2}$ | 5544.006 | −1.091 | | |
| | | | | 105763.270 | $(^3F)4d\ ^2H_{11/2}$ | 5791.103 | −0.522 | 5791.05 | |
| | | | | 106045.690 | $(^3H)4d\ ^2H_{11/2}$ | 5887.421 | −0.109 | 5887.42 | |
| | | | | 106097.520 | $(^3H)4d\ ^2H_{9/2}$ | 5905.446 | −0.710 | | |
| | | | | 106722.170 | $(^3F)4d\ ^4F_{9/2}$ | 6131.699 | −1.253 | | |
| | | | | 106767.210 | $(^3F)4d\ ^4F_{7/2}$ | 6148.685 | −1.351 | | |
| | | | | 106924.430 | $(^3F)4d\ ^2G_{9/2}$ | 6208.722 | −0.916 | | |
| | | | | 109683.280 | $(^3G)4d\ ^2H_{11/2}$ | 7492.464 | −1.002 | | |

**Table 7.** Fe II lines in the 3800-8000 Å region with log $gf \geq -1.5$ and $3d^6(^3H)4f$ energy levels as upper levels

| Upper level | | | | Lower level | | $\lambda$(calc) | log $gf$ | $\lambda$(obs) | Notes |
|---|---|---|---|---|---|---|---|---|---|
| cm$^{-1}$ | | | J | cm$^{-1}$ | | Å | KUR | Å | |
| 122988.215 | ($^3$H)4f | 6[4] | 9/2 | 103165.320 | ($^3$P)4d $^4$F$_{9/2}$ | 5043.266 | −0.030 | | |
| | | | | 103600.430 | ($^3$H)4d $^4$G$_{11/2}$ | 5156.450 | +0.529 | 5156.45 | lab |
| | | | | 103683.070 | ($^3$H)4d $^4$F$_{9/2}$ | 5178.524 | −0.018 | 5178.53 | lab |
| | | | | 103751.660 | ($^3$H)4d $^4$H$_{11/2}$ | 5196.989 | −0.773 | | |
| | | | | 103771.320 | ($^3$H)4d $^4$G$_{9/2}$ | 5202.306 | −0.787 | | |
| | | | | 104765.450 | ($^3$H)4d $^2$I$_{11/2}$ | 5486.117 | −1.286 | | |
| | | | | 104807.210 | ($^3$H)4d $^2$G$_{9/2}$ | 5498.718 | −0.382 | 5498.72 | |
| | | | | 104916.550 | ($^3$H)4d $^4$F$_{9/2}$ | 5531.988 | −1.028 | | |
| | | | | 105063.550 | ($^3$F)4d $^4$G$_{11/2}$ | 5577.356 | −0.785 | 5577.35 | |
| | | | | 106045.690 | ($^3$H)4d $^2$H$_{11/2}$ | 5900.673 | −1.342 | | |
| | | | | 106722.170 | ($^3$F)4d $^4$F$_{9/2}$ | 6146.075 | −0.412 | 6146.08 | |
| | | | | 106924.430 | ($^3$F)4d $^2$G$_{9/2}$ | 6223.461 | −1.178 | | |
| | | | | 108181.550 | ($^3$G)4d $^4$G$_{11/2}$ | 6751.852 | −1.421 | | |
| | | | | 109811.920 | ($^3$G)4d $^4$F$_{9/2}$ | 7587.298 | −0.695 | | |
| 122980.408 | ($^3$H)4f | 6[4] | 7/2 | 103102.860 | ($^3$P)4d $^4$D$_{7/2}$ | 5029.399 | −0.735 | | |
| | | | | 103165.320 | ($^3$P)4d $^4$F$_{9/2}$ | 5045.253 | −0.962 | | |
| | | | | 103683.070 | ($^3$H)4d $^4$F$_{9/2}$ | 5180.619 | −1.116 | | |
| | | | | 103771.320 | ($^3$H)4d $^4$G$_{9/2}$ | 5204.420 | −0.034 | 5204.419 | |
| | | | | 103874.260 | ($^3$H)4d $^4$H$_{9/2}$ | 5232.461 | −0.656 | | |
| | | | | 103921.630 | ($^3$H)4d $^4$G$_{7/2}$ | 5245.466 | −1.235 | | |
| | | | | 104107.950 | ($^3$P)4d $^4$F$_{7/2}$ | 5297.253 | +0.049 | 5297.26 | |
| | | | | 104481.590 | ($^3$H)4d $^2$F$_{7/2}$ | 5404.248 | −0.598 | | |
| | | | | 104807.210 | ($^3$H)4d $^2$G$_{9/2}$ | 5501.081 | −0.147 | | |
| | | | | 104916.550 | ($^3$H)4d $^4$F$_{9/2}$ | 5534.379 | −0.071 | | |
| | | | | 104993.860 | ($^3$F)4d $^4$D$_{7/2}$ | 5558.167 | −0.731 | | |
| | | | | 106097.520 | ($^3$H)4d $^2$H$_{9/2}$ | 5921.516 | −0.986 | | |
| | | | | 106722.170 | ($^3$F)4d $^4$F$_{9/2}$ | 6149.026 | −0.728 | | |
| | | | | 106767.210 | ($^3$F)4d $^4$F$_{7/2}$ | 6166.108 | −1.069 | | |
| | | | | 106924.430 | ($^3$F)4d $^2$G$_{9/2}$ | 6226.487 | −1.380 | | |
| 122946.419 | ($^3$H)4f | 6[3] | 7/2 | 103102.860 | ($^3$P)4d $^4$D$_{7/2}$ | 5038.014 | −1.413 | | |
| | | | | 103165.320 | ($^3$P)4d $^4$F$_{9/2}$ | 5053.922 | +0.160 | | |
| | | | | 103683.070 | ($^3$H)4d $^4$F$_{9/2}$ | 5189.760 | +0.167 | 5189.763 | lab. |
| | | | | 103771.320 | ($^3$H)4d $^4$G$_{9/2}$ | 5213.645 | −0.746 | | |
| | | | | 104107.950 | ($^3$P)4d $^4$F$_{7/2}$ | 5306.811 | −0.814 | | |
| | | | | 104807.210 | ($^3$H)4d $^2$G$_{9/2}$ | 5511.388 | −0.043 | 5511.40 | |
| | | | | 105155.090 | ($^3$F)4d $^4$G$_{9/2}$ | 5619.156 | −1.229 | | |
| | | | | 105211.062 | ($^5$D)5d $^4$G$_{9/2}$ | 5636.890 | −1.411 | | |
| | | | | 106097.520 | ($^3$H)4d $^2$H$_{9/2}$ | 5933.462 | −1.332 | | |
| | | | | 106722.170 | ($^3$F)4d $^4$F$_{9/2}$ | 6161.908 | −0.227 | 6161.90 | |
| | | | | 106924.430 | ($^3$F)4d $^2$G$_{9/2}$ | 6239.696 | −0.856 | | |
| | | | | 109811.920 | ($^3$G)4d $^4$F$_{9/2}$ | 7611.442 | −0.504 | | |
| 123219.200 | ($^3$H)4f | 5[8] | 17/2 | 103644.800 | ($^3$H)4d $^4$K$_{17/2}$ | 5107.290 | −0.983 | | |
| | | | | 103706.530 | ($^3$H)4d $^4$K$_{15/2}$ | 5123.448 | +0.347 | 5123.45 | lab |
| | | | | 103878.370 | ($^3$H)4d $^4$I$_{15/2}$ | 5168.969 | +0.064 | | blend |
| | | | | 104119.710 | ($^3$H)4d $^2$K$_{15/2}$ | 5234.285 | +0.991 | 5234.283 | lab |
| 123193.090 | ($^3$H)4f | 5[8] | 15/2 | 103706.530 | ($^3$H)4d $^4$K$_{15/2}$ | 5130.313 | −0.507 | | |
| | | | | 103832.050 | ($^3$H)4d $^4$K$_{13/2}$ | 5163.574 | +0.908 | 5163.55 | lab |
| | | | | 103878.370 | ($^3$H)4d $^4$I$_{15/2}$ | 5175.957 | −0.540 | 5175.95 | |
| | | | | 104064.670 | ($^3$H)4d $^4$I$_{13/2}$ | 5226.368 | −0.216 | | blend |
| | | | | 104119.710 | ($^3$H)4d $^2$K$_{15/2}$ | 5241.450 | −0.301 | 5241.465 | lab |
| | | | | 104315.370 | ($^3$H)4d $^2$K$_{13/2}$ | 5295.776 | −0.452 | 5295.773 | |
| | | | | 104622.300 | ($^3$H)4d $^2$I$_{13/2}$ | 5383.304 | +0.146 | 5383.32 | blend |

**Table 7.** Fe II lines in the 3800-8000 Å region with log $gf \geq -1.5$ and $3d^6(^3H)4f$ energy levels as upper levels

| Upper level | | | | Lower level | | λ(calc) | log gf | λ(obs) | Notes |
|---|---|---|---|---|---|---|---|---|---|
| cm$^{-1}$ | | | J | cm$^{-1}$ | | Å | KUR | Å | |
| 123238.440 | ($^3$H)4f | 5[7] | 15/2 | 103617.580 | ($^3$H)4d $^4$H$_{13/2}$ | 5095.196 | −0.836 | 5095.19 | |
| | | | | 103706.530 | ($^3$H)4d $^4$K$_{15/2}$ | 5118.401 | −0.254 | 5118.40 | lab |
| | | | | 103832.050 | ($^3$H)4d $^4$K$_{13/2}$ | 5151.507 | −0.716 | | blend |
| | | | | 103878.370 | ($^3$H)4d $^4$I$_{15/2}$ | 5163.831 | −0.599 | 5163.82 | lab |
| | | | | 104064.670 | ($^3$H)4d $^4$I$_{13/2}$ | 5214.007 | +0.873 | 5214.99 | blend |
| | | | | 104119.710 | ($^3$H)4d $^2$K$_{15/2}$ | 5229.017 | −0.045 | 5229.030 | lab |
| | | | | 104315.370 | ($^3$H)4d $^2$K$_{13/2}$ | 5283.085 | +0.323 | 5283.093 | lab |
| | | | | 105288.850 | ($^3$F)4d $^4$H$_{13/2}$ | 5569.611 | −1.005 | | blend |
| 123168.680 | ($^3$H)4f | 5[7] | 13/2 | 103600.430 | ($^3$H)4d $^4$G$_{11/2}$ | 5108.895 | −1.165 | | |
| | | | | 103706.530 | ($^3$H)4d $^4$K$_{15/2}$ | 5136.747 | −1.256 | | |
| | | | | 103751.660 | ($^3$H)4d $^4$H$_{11/2}$ | 5148.687 | +0.010 | 5148.7 | lab |
| | | | | 103832.050 | ($^3$H)4d $^4$K$_{11/2}$ | 5170.092 | −1.170 | | |
| | | | | 103973.780 | ($^3$H)4d $^4$K$_{11/2}$ | 5208.267 | −0.275 | 5208.268 | computed too weak |
| | | | | 104064.670 | ($^3$H)4d $^4$I$_{13/2}$ | 5233.046 | +0.138 | 5233.041 | |
| | | | | 104174.270 | ($^3$H)4d $^4$I$_{11/2}$ | 5263.242 | −0.600 | | |
| | | | | 104315.370 | ($^3$H)4d $^2$K$_{13/2}$ | 5302.633 | −0.581 | | |
| | | | | 104622.300 | ($^3$H)4d $^2$I$_{13/2}$ | 5390.389 | +0.010 | 5390.38 | computed too strong |
| | | | | 104765.450 | ($^3$H)4d $^2$I$_{11/2}$ | 5432.319 | +0.495 | 5432.31 | lab |
| | | | | 105063.550 | ($^3$F)4d $^4$G$_{11/2}$ | 5521.763 | −0.481 | 5521.78 | |
| | | | | 105398.850 | ($^3$F)4d $^4$H$_{11/2}$ | 5625.954 | −1.425 | | |
| | | | | 105763.270 | ($^3$F)4d $^2$H$_{11/2}$ | 5743.747 | −0.321 | 5743.75 | computed too strong |
| | | | | 106045.690 | ($^3$H)4d $^2$H$_{11/2}$ | 5838.483 | −0.311 | | |
| | | | | 108630.429 | ($^1$I)5s e$^2$I$_{11/2}$ | 6876.509 | −1.228 | | |
| | | | | 109683.280 | ($^3$G)4d $^2$H$_{11/2}$ | 7413.385 | −0.848 | | |
| 123249.650 | ($^3$H)4f | 5[6] | 13/2 | 103600.430 | ($^3$H)4d $^4$G$_{11/2}$ | 5087.842 | −0.510 | 5087.85 | lab |
| | | | | 103706.530 | ($^3$H)4d $^4$K$_{15/2}$ | 5115.465 | −1.027 | | |
| | | | | 103751.660 | ($^3$H)4d $^4$H$_{11/2}$ | 5127.305 | +0.392 | 5127.32 | lab, blend |
| | | | | 103832.050 | ($^3$H)4d $^4$K$_{13/2}$ | 5148.533 | +0.357 | 5148.52 | lab |
| | | | | 103973.780 | ($^3$H)4d $^4$K$_{11/2}$ | 5186.389 | +0.210 | 5186.396 | lab |
| | | | | 104064.670 | ($^3$H)4d $^4$I$_{13/2}$ | 5210.960 | −0.403 | 5210.964 | |
| | | | | 104119.710 | ($^3$H)4d $^2$K$_{15/2}$ | 5225.953 | −0.742 | | blend |
| | | | | 104174.270 | ($^3$H)4d $^4$I$_{11/2}$ | 5240.901 | −0.464 | 5240.911 | |
| | | | | 104315.370 | ($^3$H)4d $^2$K$_{13/2}$ | 5279.957 | −0.647 | | blend |
| | | | | 104622.300 | ($^3$H)4d $^2$I$_{13/2}$ | 5366.958 | +0.032 | 5366.95 | lab |
| | | | | 105063.550 | ($^3$F)4d $^4$G$_{11/2}$ | 5497.178 | −1.156 | | |
| | | | | 105288.850 | ($^3$F)4d $^4$H$_{13/2}$ | 5566.135 | −1.260 | | |
| | | | | 105763.270 | ($^3$F)4d $^2$H$_{11/2}$ | 5717.150 | −0.553 | 5717.18 | |
| | | | | 106045.690 | ($^3$H)4d $^2$H$_{11/2}$ | 5811.004 | −0.182 | 5811.00 | |
| | | | | 109049.600 | ($^3$G)4d $^2$I$_{13/2}$ | 7040.287 | −1.496 | | |
| | | | | 109683.280 | ($^3$G)4d $^2$H$_{11/2}$ | 7369.139 | −1.023 | | |
| 123270.340 | ($^3$H)4f | 5[6] | 11/2 | 103600.430 | ($^3$H)4d $^4$G$_{11/2}$ | 5082.491 | −0.827 | | blend |
| | | | | 103683.070 | ($^3$H)4d $^4$F$_{9/2}$ | 5103.934 | −1.365 | | |
| | | | | 103751.660 | ($^3$H)4d $^4$H$_{11/2}$ | 5121.871 | +0.373 | 5121.89 | lab |
| | | | | 103771.320 | ($^3$H)4d $^4$G$_{9/2}$ | 5127.035 | −0.542 | 5127.05 | |
| | | | | 103832.050 | ($^3$H)4d $^4$K$_{11/2}$ | 5143.054 | −0.456 | 5143.05 | |
| | | | | 103874.260 | ($^3$H)4d $^4$H$_{9/2}$ | 5154.246 | +0.127 | 5154.25 | lab |
| | | | | 103973.780 | ($^3$H)4d $^4$K$_{11/2}$ | 5180.829 | −0.529 | 5180.84 | lab |
| | | | | 104064.670 | ($^3$H)4d $^4$I$_{13/2}$ | 5205.347 | −0.844 | 5235.225 | |
| | | | | 104174.270 | ($^3$H)4d $^4$I$_{11/2}$ | 5235.223 | −0.536 | | |
| | | | | 104192.480 | ($^3$H)4d $^4$I$_{9/2}$ | 5240.220 | −1.229 | | |
| | | | | 104315.370 | ($^3$H)4d $^2$K$_{13/2}$ | 5274.195 | −1.310 | | |
| | | | | 104622.300 | ($^3$H)4d $^2$I$_{13/2}$ | 5361.004 | −0.422 | 5361.00 | lab |

**Table 7.** Fe II lines in the 3800-8000 Å region with log $gf \geq -1.5$ and $3d^6(^3H)4f$ energy levels as upper levels

| Upper level | | | | Lower level | | λ(calc) | log gf | λ(obs) | Notes |
|---|---|---|---|---|---|---|---|---|---|
| cm$^{-1}$ | | | J | cm$^{-1}$ | | Å | KUR | Å | |
| 123270.340 | cont. | | | 104807.210 | ($^3$H)4d $^2$G$_{9/2}$ | 5414.696 | −0.589 | 5414.7 | blend |
| | | | | 104916.550 | ($^3$H)4d $^4$F$_{9/2}$ | 5446.953 | −0.182 | 5446.95 | |
| | | | | 105063.550 | ($^3$F)4d $^4$G$_{11/2}$ | 5490.931 | −1.162 | | |
| | | | | 105155.090 | ($^3$F)4d $^4$G$_{9/2}$ | 5518.678 | −0.927 | | wrong, not observed |
| | | | | 105763.270 | ($^3$F)4d $^2$H$_{11/2}$ | 5710.394 | −0.287 | 5710.40 | |
| | | | | 106045.690 | ($^3$H)4d $^2$H$_{11/2}$ | 5804.025 | −0.029 | 5804.02 | |
| | | | | 106722.170 | ($^3$F)4d $^4$F$_{9/2}$ | 6041.291 | −1.018 | | |
| | | | | 106924.430 | ($^3$F)4d $^2$G$_{9/2}$ | 6116.045 | −1.092 | | |
| | | | | 109683.280 | ($^3$G)4d $^2$H$_{11/2}$ | 7357.917 | −0.867 | | |
| 123251.470 | ($^3$H)4f | 5[5] | 11/2 | 103751.660 | ($^3$H)4d $^4$H$_{11/2}$ | 5126.827 | −0.236 | | blend |
| | | | | 103771.320 | ($^3$H)4d $^4$G$_{9/2}$ | 5132.001 | +0.078 | 5132.0 | lab |
| | | | | 103874.260 | ($^3$H)4d $^4$H$_{9/2}$ | 5159.265 | +0.007 | 5159.29 | lab, blend |
| | | | | 103973.780 | ($^3$H)4d $^4$K$_{11/2}$ | 5185.899 | +0.058 | 5185.901 | lab |
| | | | | 104064.670 | ($^3$H)4d $^4$I$_{13/2}$ | 5210.466 | −0.583 | | |
| | | | | 104174.270 | ($^3$H)4d $^4$I$_{11/2}$ | 5240.401 | −0.177 | 5240.405 | lab |
| | | | | 104192.480 | ($^3$H)4d $^4$I$_{9/2}$ | 5245.408 | −1.139 | | blend |
| | | | | 104315.370 | ($^3$H)4d $^2$K$_{13/2}$ | 5279.449 | −1.308 | | |
| | | | | 104765.450 | ($^3$H)4d $^2$I$_{11/2}$ | 5407.990 | +0.040 | 5407.99 | lab |
| | | | | 104807.210 | ($^3$H)4d $^2$G$_{9/2}$ | 5420.234 | −1.131 | | |
| | | | | 104916.550 | ($^3$H)4d $^4$F$_{9/2}$ | 5452.558 | −0.967 | 5452.55 | |
| | | | | 105063.550 | ($^3$F)4d $^4$G$_{11/2}$ | 5496.628 | −0.739 | 5496.62 | |
| | | | | 105155.090 | ($^3$F)4d $^4$G$_{9/2}$ | 5524.433 | −1.032 | | |
| | | | | 105524.460 | ($^3$F)4d $^4$H$_{9/2}$ | 5639.544 | −1.347 | | |
| | | | | 106018.640 | ($^3$F)4d $^2$H$_{9/2}$ | 5801.269 | −0.770 | | computed too strong |
| | | | | 106045.690 | ($^3$H)4d $^2$H$_{11/2}$ | 5810.389 | −1.328 | | |
| | | | | 106097.520 | ($^3$H)4d $^2$H$_{9/2}$ | 5827.945 | −0.015 | 5827.95 | computed too weak |
| | | | | 106924.430 | ($^3$F)4d $^2$G$_{9/2}$ | 6123.114 | −0.236 | | |
| | | | | 109625.200 | ($^3$G)4d $^2$G$_{9/2}$ | 7336.744 | −1.064 | | |
| | | | | 110008.300 | ($^3$G)4d $^2$H$_{9/2}$ | 7548.984 | −1.185 | | |
| 123269.378 | ($^3$H)4f | 5[5] | 9/2 | 103751.660 | ($^3$H)4d $^4$H$_{11/2}$ | 5122.123 | −1.173 | | blend |
| | | | | 103771.320 | ($^3$H)4d $^4$G$_{9/2}$ | 5127.287 | −0.734 | | blend |
| | | | | 103874.260 | ($^3$H)4d $^4$H$_{9/2}$ | 5154.501 | +0.418 | 5154.50 | lab |
| | | | | 103921.630 | ($^3$H)4d $^4$G$_{7/2}$ | 5167.121 | −0.470 | 5167.1 | computed too weak |
| | | | | 103973.780 | ($^3$H)4d $^4$K$_{11/2}$ | 5181.086 | −0.545 | 5181.1 | blend, computed too weak |
| | | | | 103983.510 | ($^3$G)5s $^2$G$_{7/2}$ | 5183.700 | −0.079 | | blend |
| | | | | 103986.330 | ($^3$H)4d $^4$H$_{7/2}$ | 5184.458 | −0.485 | 5184.463 | computed too strong |
| | | | | 104107.950 | ($^3$P)4d $^4$F$_{7/2}$ | 5217.365 | −1.017 | | |
| | | | | 104174.270 | ($^3$H)4d $^4$I$_{11/2}$ | 5235.486 | −0.560 | | |
| | | | | 104765.450 | ($^3$H)4d $^2$I$_{11/2}$ | 5402.756 | −0.812 | | |
| | | | | 104807.210 | ($^3$H)4d $^2$G$_{9/2}$ | 5414.977 | −0.955 | | |
| | | | | 104993.860 | ($^3$F)4d $^4$D$_{7/2}$ | 5470.281 | −1.409 | | |
| | | | | 105123.000 | ($^3$H)4d $^2$G$_{7/2}$ | 5509.211 | −0.290 | 5509.2 | |
| | | | | 105220.600 | ($^3$H)4d $^4$F$_{7/2}$ | 5539.003 | −1.382 | | |
| | | | | 105524.460 | ($^3$F)4d $^4$H$_{9/2}$ | 5633.853 | −1.381 | | |
| | | | | 106018.640 | ($^3$F)4d $^2$H$_{9/2}$ | 5795.246 | −0.974 | | |
| | | | | 106097.520 | ($^3$H)4d $^2$H$_{9/2}$ | 5821.868 | −0.325 | 5821.88 | |
| | | | | 106722.170 | ($^3$F)4d $^4$F$_{9/2}$ | 6041.643 | −1.431 | | |
| | | | | 106900.370 | ($^3$F)4d $^2$G$_{7/2}$ | 6107.415 | −0.980 | | |
| | | | | 106924.430 | ($^3$F)4d $^2$G$_{9/2}$ | 6116.405 | −0.472 | | blend |
| | | | | 109625.200 | ($^3$G)4d $^2$G$_{9/2}$ | 7327.115 | −1.238 | | |
| 123258.994 | ($^3$H)4f | 5[4] | 9/2 | 103165.320 | ($^3$P)4d $^4$F$_{9/2}$ | 4975.303 | −1.479 | | |
| | | | | 103191.917 | ($^3$P)4d $^2$F$_{7/2}$ | 4981.898 | −0.587 | | |
| | | | | 103600.430 | ($^3$H)4d $^4$G$_{11/2}$ | 5085.425 | −1.404 | | |
| | | | | 103683.070 | ($^3$H)4d $^4$F$_{9/2}$ | 5106.894 | −0.960 | | |
| | | | | 103751.660 | ($^3$H)4d $^4$H$_{11/2}$ | 5124.850 | +0.047 | 5124.82 | lab |

**Table 7.** Fe II lines in the 3800-8000 Å region with log $gf \geq -1.5$ and $3d^6(^3H)4f$ energy level as upper levels

| Upper level | | | Lower level | | $\lambda$(calc) | log $gf$ | $\lambda$(obs) | Notes |
|---|---|---|---|---|---|---|---|---|
| cm$^{-1}$ | | J | cm$^{-1}$ | | Å | KUR | Å | |
| 123258.994 | cont. | | 103771.320 | ($^3$H)4d $^4$G$_{9/2}$ | 5130.020 | +0.269 | 5130.0 | lab |
| | | | 103874.260 | ($^3$H)4d $^4$H$_{9/2}$ | 5157.263 | −0.663 | | blend |
| | | | 104481.590 | ($^3$H)4d $^2$F$_{7/2}$ | 5324.070 | −0.506 | | blend |
| | | | 104807.210 | ($^3$H)4d $^2$G$_{9/2}$ | 5418.025 | −0.657 | 5418.02 | lab |
| | | | 104916.550 | ($^3$H)4d $^4$F$_{9/2}$ | 5450.323 | +0.051 | 5450.30 | wrong, computed too strong |
| | | | 105063.550 | ($^3$F)4d $^4$G$_{11/2}$ | 5494.356 | −1.301 | | |
| | | | 105123.000 | ($^3$H)4d $^2$G$_{7/2}$ | 5512.367 | −0.848 | | |
| | | | 105155.090 | ($^3$F)4d $^4$G$_{9/2}$ | 5522.138 | −0.450 | 5522.10 | computed too strong |
| | | | 105211.062 | ($^5$D)5d $^4$G$_{9/2}$ | 5539.264 | −1.434 | | |
| | | | 105763.270 | ($^3$F)4d $^2$H$_{11/2}$ | 5714.098 | −0.740 | 5714.10 | |
| | | | 106045.690 | ($^3$H)4d $^2$H$_{11/2}$ | 5807.851 | −0.440 | 5807.85 | blend |
| | | | 106097.520 | ($^3$H)4d $^2$H$_{9/2}$ | 5825.392 | −0.814 | | |
| | | | 106722.170 | ($^3$F)4d $^4$F$_{9/2}$ | 6045.483 | −0.970 | | |
| | | | 106767.210 | ($^3$F)4d $^4$F$_{7/2}$ | 6061.948 | −1.148 | | |
| | | | 106900.370 | ($^3$F)4d $^2$G$_{7/2}$ | 6111.293 | −1.488 | | |
| | | | 108391.500 | ($^3$G)4d $^4$G$_{9/2}$ | 6724.229 | −1.436 | | |
| | | | 109683.280 | ($^3$G)4d $^2$H$_{11/2}$ | 7364.069 | −1.370 | | |
| | | | 110167.280 | ($^3$G)4d $^4$F$_{7/2}$ | 7636.319 | −1.343 | | |
| 123258.021 | ($^3$H)4f | 5[4] 7/2 | 102802.312 | ($^5$D)6s $^4$D$_{5/2}$ | 4887.246 | −1.497 | | blend |
| | | | 103002.670 | ($^3$P)4d $^4$D$_{5/2}$ | 4935.589 | −1.223 | | blend |
| | | | 103102.860 | ($^3$P)4d $^4$D$_{7/2}$ | 4960.124 | −1.397 | | at the continuum level |
| | | | 103771.320 | ($^3$H)4d $^4$G$_{9/2}$ | 5130.276 | −0.633 | | blend |
| | | | 103874.260 | ($^3$H)4d $^4$H$_{9/2}$ | 5157.521 | −0.254 | | blend |
| | | | 103921.630 | ($^3$H)4d $^4$G$_{7/2}$ | 5170.156 | −0.375 | | blend |
| | | | 103983.510 | ($^3$G)5s $^2$G$_{7/2}$ | 5186.755 | −0.078 | | blend |
| | | | 103986.330 | ($^3$H)4d $^4$H$_{7/2}$ | 5187.514 | −0.396 | 5187.52 | |
| | | | 104107.950 | ($^3$P)4d $^4$F$_{7/2}$ | 5220.459 | −1.202 | | computed too strong |
| | | | 104120.270 | ($^5$D)5d $^6$P$_{5/2}$ | 5223.820 | −0.829 | | blend |
| | | | 104209.610 | ($^3$H)4d $^2$F$_{5/2}$ | 5248.321 | −0.898 | | blend |
| | | | 104569.230 | ($^3$P)4d $^4$F$_{5/2}$ | 5349.313 | −0.940 | | wrong, not observed |
| | | | 104916.550 | ($^3$H)4d $^4$F$_{9/2}$ | 5450.611 | −1.412 | | blend |
| | | | 104993.860 | ($^3$F)4d $^4$D$_{7/2}$ | 5473.683 | −0.926 | | blend |
| | | | 105123.000 | ($^3$H)4d $^2$G$_{7/2}$ | 5512.661 | +0.003 | 5512.65 | |
| | | | 105220.600 | ($^3$H)4d $^4$F$_{7/2}$ | 5542.490 | −1.205 | | blend |
| | | | 106018.640 | ($^3$F)4d $^2$H$_{9/2}$ | 5799.064 | −1.320 | | blend |
| | | | 106097.520 | ($^3$H)4d $^2$H$_{9/2}$ | 5825.721 | −0.559 | 5825.73 | |
| | | | 106866.760 | ($^3$F)4d $^4$F$_{5/2}$ | 6099.124 | −1.189 | | blend |
| | | | 106900.370 | ($^3$F)4d $^2$G$_{7/2}$ | 6111.655 | −0.698 | | blend |
| | | | 106924.430 | ($^3$F)4d $^2$G$_{9/2}$ | 6120.658 | −0.942 | | at the continuum level |
| | | | 110167.280 | ($^3$G)4d $^4$F$_{7/2}$ | 7636.885 | −1.434 | | no spectrum |
| 123235.165 | ($^3$H)4f | 5[3] 7/2 | 103191.917 | ($^3$P)4d $^2$F$_{7/2}$ | 4987.820 | −0.173 | | |
| | | | 103771.320 | ($^3$H)4d $^4$G$_{9/2}$ | 5136.300 | −0.037 | 5136.30 | |
| | | | 103874.260 | ($^3$H)4d $^4$H$_{9/2}$ | 5163.610 | −0.154 | | blend |
| | | | 103921.630 | ($^3$H)4d $^4$G$_{7/2}$ | 5176.274 | −0.716 | 5176.25 | |
| | | | 103983.510 | ($^3$G)5s $^2$G$_{7/2}$ | 5192.913 | −0.799 | | blend |
| | | | 103986.330 | ($^3$H)4d $^4$H$_{7/2}$ | 5193.673 | −0.887 | | blend |
| | | | 104107.950 | ($^3$P)4d $^4$F$_{7/2}$ | 5226.698 | −1.309 | | |
| | | | 104481.590 | ($^3$H)4d $^2$F$_{7/2}$ | 5330.834 | −0.226 | 5330.81 | computed too strong |
| | | | 104807.210 | ($^3$H)4d $^2$G$_{9/2}$ | 5425.030 | −0.825 | 5425.01 | |
| | | | 104916.550 | ($^3$H)4d $^4$F$_{9/2}$ | 5457.411 | −0.238 | 5457.40 | |
| | | | 105123.000 | ($^3$H)4d $^2$G$_{7/2}$ | 5519.618 | −1.438 | | |
| | | | 105155.090 | ($^3$F)4d $^4$G$_{9/2}$ | 5529.415 | −0.668 | 5529.40 | wrong, computed too strong |
| | | | 105220.600 | ($^3$H)4d $^4$F$_{7/2}$ | 5549.523 | −1.242 | | |
| | | | 105291.010 | ($^3$F)4d $^4$G$_{7/2}$ | 5571.298 | −1.482 | | |
| | | | 106722.170 | ($^3$F)4d $^4$F$_{9/2}$ | 6054.160 | −1.224 | | |

**Table 7.** Fe II lines in the 3800-8000 Å region with log $gf \geq -1.5$ and $3d^6(^3H)4f$ energy levels as upper levels

| Upper level | | | | Lower level | | λ(calc) | log gf | λ(obs) | Notes |
|---|---|---|---|---|---|---|---|---|---|
| cm$^{-1}$ | | | J | cm$^{-1}$ | | Å | KUR | Å | |
| 123235.165 | cont. | | | 106767.210 | ($^3$F)4d $^4$F$_{7/2}$ | 6070.719 | −0.626 | 6070.71 | |
| | | | | 110167.280 | ($^3$G)4d $^4$F$_{7/2}$ | 7650.242 | −0.970 | | |
| | | | | 110570.300 | ($^3$G)4d $^2$F$_{7/2}$ | 7893.688 | −1.448 | | |
| 123211.159 | ($^3$H)4f | 5[2] | 5/2 | 103193.917 | ($^3$P)4d $^2$F$_{7/2}$ | 4993.801 | −0.145 | 4993.80 | computed too strong |
| | | | | 103921.630 | ($^3$H)4d $^4$G$_{7/2}$ | 5182.716 | −1.163 | 5182.707 | good agreement |
| | | | | 103986.330 | ($^3$G)5s $^2$G$_{7/2}$ | 5200.159 | −1.442 | | |
| | | | | 104481.590 | ($^3$H)4d $^2$F$_{7/2}$ | 5337.666 | −0.236 | | blend |
| | | | | 104993.860 | ($^3$F)4d $^4$D$_{7/2}$ | 5487.763 | −1.396 | | blend |
| | | | | 105123.000 | ($^3$H)4d $^2$G$_{7/2}$ | 5526.943 | −0.560 | 5526.92 | computed too strong |
| | | | | 105291.010 | ($^3$F)4d $^4$G$_{7/2}$ | 5578.762 | −1.365 | | at the level of the noise |
| | | | | 106767.210 | ($^3$F)4d $^4$F$_{7/2}$ | 6079.581 | −0.532 | 6709.60 | good agreement |
| | | | | 106900.370 | ($^3$F)4d $^2$G$_{7/2}$ | 6129.215 | −1.126 | | blend |
| | | | | 110167.280 | ($^3$G)4d $^4$F$_{7/2}$ | 7664.321 | −0.703 | | in telluric |
| | | | | 110570.300 | ($^3$G)4d $^2$F$_{7/2}$ | 7908.679 | −1.384 | | in telluric |
| 123213.323 | ($^3$H)4f | 5[2] | 3/2 | 102802.312 | ($^5$D)6s $^4$D$_{5/2}$ | 4897.949 | −1.090 | 4897.90 | at the level of the noise |
| | | | | 103597.402 | ($^3$P)4d $^2$D$_{5/2}$ | 5096.480 | −1.325 | | at the level of the noise |
| | | | | 104120.270 | ($^5$D)5d $^6$P$_{5/2}$ | 5236.050 | −0.269 | 5236.046 | computed too strong |
| | | | | 104209.610 | ($^3$H)4d $^2$F$_{5/2}$ | 5260.666 | −0.338 | 5260.682 | lab, good agreement |
| | | | | 104569.230 | ($^3$P)4d $^4$F$_{5/2}$ | 5362.139 | −0.684 | | wrong, not observed |
| | | | | 105234.237 | ($^3$H)4d $^4$F$_{5/2}$ | 5560.475 | −1.142 | | |
| | | | | 105414.180 | ($^3$F)4d $^4$G$_{5/2}$ | 5616.690 | −1.055 | | blend |
| | | | | 106796.660 | ($^3$F)4d $^4$P$_{5/2}$ | 6089.687 | −1.322 | | blend |
| | | | | 106866.760 | ($^3$F)4d $^4$F$_{5/2}$ | 6115.802 | −0.758 | 6115.80 | good agreement |
| | | | | 110428.280 | ($^3$G)4d $^4$F$_{5/2}$ | 7819.490 | −1.269 | | at the continuum level |
| 123396.250 | ($^3$H)4f | 4[7] | 15/2 | 103706.530 | ($^3$H)4d $^4$K$_{15/2}$ | 5077.377 | −1.404 | | |
| | | | | 103832.050 | ($^3$H)4d $^4$K$_{13/2}$ | 5109.953 | −0.102 | 5109.95 | lab |
| | | | | 104064.670 | ($^3$H)4d $^4$I$_{13/2}$ | 5171.443 | +0.259 | 5171.45 | lab |
| | | | | 104315.370 | ($^3$H)4d $^2$K$_{13/2}$ | 5239.390 | +0.861 | 5239.394 | J78 |
| | | | | 104622.300 | ($^3$H)4d $^2$I$_{13/2}$ | 5325.048 | +0.257 | 5325.05 | J78, lab |
| 123355.490 | ($^3$H)4f | 4[7] | 13/2 | 103600.430 | ($^3$H)4d $^4$G$_{11/2}$ | 5060.583 | −1.409 | | |
| | | | | 103751.660 | ($^3$H)4d $^4$H$_{11/2}$ | 5099.623 | −0.221 | 5099.6 | lab |
| | | | | 103832.050 | ($^3$H)4d $^4$K$_{13/2}$ | 5120.621 | −1.170 | 5120.62 | lab, computed too weak |
| | | | | 103973.780 | ($^3$H)4d $^4$K$_{11/2}$ | 5158.067 | +0.788 | 5158.05 | J78, lab |
| | | | | 104064.670 | ($^3$H)4d $^4$I$_{13/2}$ | 5182.370 | +0.034 | 5182.371 | lab |
| | | | | 104119.710 | ($^3$H)4d $^2$K$_{15/2}$ | 5197.198 | −1.475 | | |
| | | | | 104315.370 | ($^3$H)4d $^2$K$_{13/2}$ | 5250.606 | −0.778 | 5250.609 | computed too weak |
| | | | | 104622.300 | ($^3$H)4d $^2$I$_{13/2}$ | 5336.635 | −0.215 | 5336.62 | |
| | | | | 104765.450 | ($^3$H)4d $^2$I$_{11/2}$ | 5377.729 | −0.165 | 5377.71 | J78, lab, computed too weak |
| | | | | 105763.270 | ($^3$F)4d $^2$H$_{11/2}$ | 5682.754 | −0.574 | 5682.75 | |
| | | | | 106045.690 | ($^3$H)4d $^2$H$_{11/2}$ | 5775.473 | −0.674 | | |
| | | | | 109683.280 | ($^3$G)4d $^2$H$_{11/2}$ | 7312.092 | −1.277 | | |
| 123414.730 | ($^3$H)4f | 4[6] | 13/2 | 103751.660 | ($^3$H)4d $^4$H$_{11/2}$ | 5084.259 | −0.750 | | |
| | | | | 103832.050 | ($^3$H)4d $^4$K$_{13/2}$ | 5105.131 | −0.704 | | |
| | | | | 103973.780 | ($^3$H)4d $^4$K$_{11/2}$ | 5142.349 | −0.245 | 5142.35 | lab |
| | | | | 104064.670 | ($^3$H)4d $^4$I$_{13/2}$ | 5166.504 | −0.525 | | blend |
| | | | | 104174.270 | ($^3$H)4d $^4$I$_{11/2}$ | 5195.934 | +0.922 | 5195.942 | lab |
| | | | | 104315.370 | ($^3$H)4d $^2$K$_{13/2}$ | 5234.320 | −0.791 | | blend |
| | | | | 104622.300 | ($^3$H)4d $^2$I$_{13/2}$ | 5319.812 | −1.134 | | |
| | | | | 104765.450 | ($^3$H)4d $^2$I$_{11/2}$ | 5360.646 | −0.638 | 5360.65 | computed too weak |
| | | | | 105063.550 | ($^3$F)4d $^4$G$_{11/2}$ | 5447.727 | −1.416 | | |
| | | | | 105398.850 | ($^3$F)4d $^4$H$_{11/2}$ | 5549.118 | −1.185 | | |
| | | | | 106045.690 | ($^3$H)4d $^2$H$_{11/2}$ | 5755.774 | −1.242 | | |

**Table 7.** Fe II lines in the 3800-8000 Å region with log $gf \geq -1.5$ and $3d^6(^3H)4f$ energy levels as upper levels

| Upper level | | | | Lower level | | λ(calc) | log gf | λ(obs) | Notes |
|---|---|---|---|---|---|---|---|---|---|
| cm$^{-1}$ | | | J | cm$^{-1}$ | | Å | KUR | Å | |
| 123427.119 | ($^3$H)4f | 4[6] | 11/2 | 103771.320 | ($^3$H)4d $^4$G$_{9/2}$ | 5086.139 | −0.441 | 5086.15 | |
| | | | | 103874.260 | ($^3$H)4d $^4$H$_{9/2}$ | 5112.917 | −0.423 | | blend |
| | | | | 103973.780 | ($^3$H)4d $^4$K$_{11/2}$ | 5139.074 | +0.124 | 5139.10 | |
| | | | | 104192.480 | ($^3$H)4d $^4$I$_{9/2}$ | 5197.506 | +0.465 | 5197.56 | blend |
| | | | | 104315.370 | ($^3$H)4d $^2$K$_{13/2}$ | 5230.927 | −1.051 | | |
| | | | | 104622.300 | ($^3$H)4d $^2$I$_{13/2}$ | 5316.307 | −1.253 | | |
| | | | | 104765.450 | ($^3$H)4d $^2$I$_{11/2}$ | 5357.088 | +0.165 | 5357.10 | J78,lab |
| | | | | 104807.210 | ($^3$H)4d $^2$G$_{9/2}$ | 5369.102 | −1.260 | | |
| | | | | 105063.550 | ($^3$F)4d $^4$G$_{11/2}$ | 5444.051 | −0.902 | | |
| | | | | 105763.270 | ($^3$F)4d $^2$H$_{11/2}$ | 5659.712 | −0.911 | | |
| | | | | 106018.640 | ($^3$F)4d $^2$H$_{9/2}$ | 5742.735 | −0.704 | | computed too strong |
| | | | | 106045.690 | ($^3$H)4d $^2$H$_{11/2}$ | 5751.672 | −1.454 | | |
| | | | | 106097.520 | ($^3$H)4d $^2$H$_{9/2}$ | 5768.874 | −0.115 | 5768.90 | J78, computed too weak |
| | | | | 106722.170 | ($^3$F)4d $^4$F$_{9/2}$ | 5984.595 | −1.089 | | |
| | | | | 106924.430 | ($^3$F)4d $^2$G$_{9/2}$ | 6057.941 | −0.358 | 6057.92 | blend |
| | | | | 109625.200 | ($^3$G)4d $^2$G$_{9/2}$ | 7243.378 | −1.142 | | |
| | | | | 110008.300 | ($^3$G)4d $^2$H$_{9/2}$ | 7450.174 | −1.329 | | |
| 123441.100 | ($^3$H)4f | 4[5] | 11/2 | 103771.320 | ($^3$H)4d $^4$G$_{9/2}$ | 5082.524 | −0.439 | 5082.51 | computed too strong |
| | | | | 103874.260 | ($^3$H)4d $^4$H$_{9/2}$ | 5109.263 | +0.037 | 5109.29 | lab |
| | | | | 103973.780 | ($^3$H)4d $^4$K$_{11/2}$ | 5135.383 | −1.089 | | |
| | | | | 104174.270 | ($^3$H)4d $^4$I$_{11/2}$ | 5188.822 | +0.224 | 5188.831 | lab |
| | | | | 104192.480 | ($^3$H)4d $^4$I$_{9/2}$ | 5193.731 | +0.573 | 5193.74 | J78, lab |
| | | | | 104315.370 | ($^3$H)4d $^2$K$_{13/2}$ | 5227.103 | −1.390 | | |
| | | | | 104765.450 | ($^3$H)4d $^2$I$_{11/2}$ | 5353.077 | −0.299 | | blend |
| | | | | 105063.550 | ($^3$F)4d $^4$G$_{11/2}$ | 5439.910 | −1.230 | | |
| | | | | 105524.460 | ($^3$F)4d $^4$H$_{9/2}$ | 5579.854 | −1.306 | | |
| | | | | 106018.640 | ($^3$F)4d $^2$H$_{9/2}$ | 5738.126 | −1.011 | | computed too strong, not obs |
| | | | | 106097.520 | ($^3$H)4d $^2$H$_{9/2}$ | 5764.224 | −0.455 | 5764.20 | |
| | | | | 106722.170 | ($^3$F)4d $^4$F$_{9/2}$ | 5979.588 | −1.109 | | |
| | | | | 106924.430 | ($^3$F)4d $^2$G$_{9/2}$ | 6052.813 | −0.460 | 6052.8 | |
| | | | | 109625.200 | ($^3$G)4d $^2$G$_{9/2}$ | 7236.043 | −1.361 | | |
| 123435.468 | ($^3$H)4f | 4[5] | 9/2 | 103921.630 | ($^3$H)4d $^4$G$_{7/2}$ | 5123.141 | +0.119 | 5123.190 | blend |
| | | | | 103973.780 | ($^3$H)4d $^4$K$_{11/2}$ | 5136.869 | −0.836 | | blend |
| | | | | 103983.510 | ($^3$G)5s $^2$G$_{7/2}$ | 5139.439 | +0.314 | | blend |
| | | | | 103986.330 | ($^3$H)4d $^4$H$_{7/2}$ | 5140.184 | −0.208 | 5140.2 | lab |
| | | | | 104107.950 | ($^3$P)4d $^4$F$_{7/2}$ | 5172.529 | −1.242 | | |
| | | | | 104174.270 | ($^3$H)4d $^4$I$_{11/2}$ | 5190.340 | −1.319 | | |
| | | | | 104192.480 | ($^3$H)4d $^4$I$_{9/2}$ | 5195.251 | +0.450 | 5195.26 | lab |
| | | | | 105589.670 | ($^3$F)4d $^4$H$_{7/2}$ | 5602.005 | −1.242 | | |
| 123460.690 | ($^3$H)4f | 4[4] | 9/2 | 103191.917 | ($^3$P)4d $^2$F$_{7/2}$ | 4932.321 | −1.442 | | |
| | | | | 103771.320 | ($^3$H)4d $^4$G$_{9/2}$ | 5077.467 | −0.602 | 5077.5 | lab |
| | | | | 103874.260 | ($^3$H)4d $^4$H$_{9/2}$ | 5104.153 | −0.047 | 5104.15 | |
| | | | | 103921.630 | ($^3$H)4d $^4$G$_{7/2}$ | 5116.528 | −0.613 | 5116.52 | |
| | | | | 103973.780 | ($^3$H)4d $^4$K$_{11/2}$ | 5130.220 | −1.289 | | |
| | | | | 103983.510 | ($^3$G)5s $^2$G$_{7/2}$ | 5132.783 | −0.961 | | |
| | | | | 103986.330 | ($^3$H)4d $^4$H$_{7/2}$ | 5133.527 | −0.989 | | |
| | | | | 104174.27 | ($^3$H)4d $^4$I$_{11/2}$ | 5183.552 | −0.937 | | |
| | | | | 104481.590 | ($^3$H)4d $^2$F$_{7/2}$ | 5267.488 | −0.494 | 5267.47 | |
| | | | | 104765.450 | ($^3$H)4d $^2$I$_{11/2}$ | 5347.468 | −0.307 | 5347.45 | lab |
| | | | | 104807.210 | ($^3$H)4d $^2$G$_{9/2}$ | 5359.439 | −1.442 | | |
| | | | | 104993.860 | ($^3$F)4d $^4$D$_{7/2}$ | 5413.610 | −0.234 | 5413.60 | lab |
| | | | | 105063.550 | ($^3$F)4d $^4$G$_{11/2}$ | 5434.117 | −1.217 | | |
| | | | | 105123.000 | ($^3$H)4d $^2$G$_{7/2}$ | 5451.734 | −0.292 | 5451.72 | |
| | | | | 105220.600 | ($^3$H)4d $^4$F$_{7/2}$ | 5480.906 | −0.700 | | blend |
| | | | | 105291.010 | ($^3$F)4d $^4$G$_{7/2}$ | 5502.146 | −0.769 | | |
| | | | | 105449.540 | ($^5$D)5d $^4$G$_{7/2}$ | 5550.575 | −1.270 | | |

**Table 7.** Fe II lines in the 3800-8000 Å region with log $gf \geq -1.5$ and $3d^6(^3H)4f$ energy levels as upper levels

| Upper level | | | | Lower level | | $\lambda$(calc) | log $gf$ | $\lambda$(obs) | Notes |
|---|---|---|---|---|---|---|---|---|---|
| cm$^{-1}$ | | | J | cm$^{-1}$ | | Å | KUR | Å | |
| 123460.690 | cont. | | | 106018.640 | $(^3F)4d\ ^2H_{9/2}$ | 5731.681 | $-0.446$ | | wrong, not observed |
| | | | | 106097.520 | $(^3H)4d\ ^2H_{9/2}$ | 5757.720 | $+0.118$ | 5757.72 | J78, computed too low |
| | | | | 106722.170 | $(^3F)4d\ ^4F_{9/2}$ | 5972.589 | $-0.946$ | | |
| | | | | 106767.210 | $(^3F)4d\ ^4F_{7/2}$ | 5988.704 | $-1.212$ | | |
| | | | | 106900.370 | $(^3F)4d\ ^2G_{7/2}$ | 6036.859 | $-0.912$ | | |
| | | | | 106924.430 | $(^3F)4d\ ^2G_{9/2}$ | 6045.643 | $-0.124$ | 6045.65 | |
| | | | | 109625.200 | $(^3G)4d\ ^2G_{9/2}$ | 7225.797 | $-0.960$ | | |
| | | | | 110008.300 | $(^3G)4d\ ^2H_{9/2}$ | 7431.576 | $-1.109$ | | |
| 123435.277 | $(^3H)4f$ | 4[4] | 7/2 | 103921.630 | $(^3H)4d\ ^4G_{7/2}$ | 5123.191 | $-0.068$ | | blend |
| | | | | 103983.510 | $(^3G)5s\ ^2G_{7/2}$ | 5139.489 | $+0.217$ | 5139.45 | lab, blend |
| | | | | 103986.330 | $(^3H)4d\ ^4H_{7/2}$ | 5140.234 | $-0.435$ | 5140.20 | blend |
| | | | | 104023.910 | $(^3H)4d\ ^4G_{5/2}$ | 5150.186 | $+0.144$ | 5150.15 | lab |
| | | | | 104120.270 | $(^5D)5d\ ^6P_{5/2}$ | 5175.880 | $-1.206$ | | blend |
| | | | | 104192.480 | $(^3H)4d\ ^4I_{9/2}$ | 5195.303 | $-0.325$ | | blend |
| | | | | 104209.610 | $(^3H)4d\ ^2F_{5/2}$ | 5199.932 | $-1.066$ | 5199.95 | computed too weak |
| | | | | 104569.230 | $(^3P)4d\ ^4F_{5/2}$ | 5299.053 | $-0.753$ | | computed too strong |
| | | | | 105414.180 | $(^3F)4d\ ^4G_{5/2}$ | 5547.511 | $-1.009$ | | at the level of the noise |
| | | | | 105589.670 | $(^3F)4d\ ^4H_{7/2}$ | 5602.065 | $-1.328$ | | blend |
| | | | | 105630.750 | $(^5D)5d\ ^4G_{5/2}$ | 5614.990 | $-1.423$ | | at the continuum level |
| | | | | 107407.800 | $(^3F)4d\ ^2D_{5/2}$ | 6237.560 | $-1.471$ | | at the continuum level |
| 123451.449 | $(^3H)4f$ | 4[3] | 7/2 | 103191.917 | $(^3P)4d\ ^2F_{7/2}$ | 4934.571 | $-1.453$ | | |
| | | | | 103597.402 | $(^3P)4d\ ^2D_{5/2}$ | 5035.352 | $-0.856$ | | |
| | | | | 103771.320 | $(^3H)4d\ ^4G_{9/2}$ | 5079.851 | $-1.218$ | | |
| | | | | 103874.260 | $(^3H)4d\ ^4H_{9/2}$ | 5106.563 | $-0.583$ | 5106.55 | |
| | | | | 103921.630 | $(^3H)4d\ ^4G_{7/2}$ | 5118.949 | $-1.061$ | | |
| | | | | 103983.510 | $(^3G)5s\ ^2G_{7/2}$ | 5135.220 | $-0.335$ | | |
| | | | | 103986.330 | $(^3H)4d\ ^4H_{7/2}$ | 5135.964 | $-1.420$ | 5135.95 | |
| | | | | 104023.910 | $(^3H)4d\ ^4G_{5/2}$ | 5145.899 | $-0.764$ | | |
| | | | | 104107.950 | $(^3P)4d\ ^4F_{7/2}$ | 5168.256 | $-1.230$ | | |
| | | | | 104120.270 | $(^5D)5d\ ^6P_{5/2}$ | 5171.550 | $-1.408$ | | |
| | | | | 104481.590 | $(^3H)4d\ ^2F_{7/2}$ | 5270.054 | $-0.654$ | | blend |
| | | | | 104569.230 | $(^3P)4d\ ^4F_{5/2}$ | 5294.515 | $-1.314$ | | |
| | | | | 104993.860 | $(^3F)4d\ ^4D_{7/2}$ | 5416.320 | $-0.276$ | 5416.32 | lab |
| | | | | 105123.000 | $(^3H)4d\ ^2G_{7/2}$ | 5454.483 | $-0.324$ | 5454.50 | blend |
| | | | | 105220.600 | $(^3H)4d\ ^4F_{7/2}$ | 5483.684 | $-0.695$ | | |
| | | | | 105291.010 | $(^3F)4d\ ^4G_{7/2}$ | 5504.945 | $-0.792$ | 5504.95 | |
| | | | | 105449.540 | $(^5D)5d\ ^4G_{7/2}$ | 5553.424 | $-1.292$ | | |
| | | | | 106018.640 | $(^3F)4d\ ^2H_{9/2}$ | 5734.719 | $-1.053$ | | |
| | | | | 106097.520 | $(^3H)4d\ ^2H_{9/2}$ | 5760.786 | $-0.536$ | 5760.78 | computed too weak |
| | | | | 106767.210 | $(^3F)4d\ ^4F_{7/2}$ | 5992.021 | $-1.212$ | | |
| | | | | 106900.370 | $(^3F)4d\ ^2G_{7/2}$ | 6040.230 | $-1.110$ | | |
| | | | | 106924.430 | $(^3F)4d\ ^2G_{9/2}$ | 6049.023 | $-0.751$ | | |
| 123430.181 | $(^3H)4f$ | 4[3] | 5/2 | 103597.402 | $(^3P)4d\ ^2D_{5/2}$ | 5040.752 | $-1.238$ | | blend |
| | | | | 103921.630 | $(^3H)4d\ ^4G_{7/2}$ | 5124.529 | $-0.535$ | 5124.52 | |
| | | | | 103983.510 | $(^3G)5s\ ^2G_{7/2}$ | 5140.836 | $-0.648$ | 5140.83 | |
| | | | | 103986.330 | $(^3H)4d\ ^4H_{7/2}$ | 5141.582 | $-0.884$ | | blend |
| | | | | 104023.910 | $(^3H)4d\ ^4G_{5/2}$ | 5151.538 | $+0.030$ | 5151.52 | J78, lab |
| | | | | 104120.270 | $(^5D)5d\ ^6P_{5/2}$ | 5177.246 | $-0.906$ | | blend |
| | | | | 104209.610 | $(^3H)4d\ ^2F_{5/2}$ | 5201.311 | $-0.851$ | | blend, wrong ? |
| | | | | 104569.230 | $(^3P)4d\ ^4F_{5/2}$ | 5300.485 | $-0.786$ | | blend, computed too strong |
| | | | | 104572.920 | $(^3P)4d\ ^4F_{3/2}$ | 5301.522 | $-0.742$ | | wrong, not observed |
| | | | | 104993.860 | $(^3F)4d\ ^4D_{7/2}$ | 5422.568 | $-1.395$ | | at the continuum level |
| | | | | 105317.440 | $(^3P)4d\ ^2P_{3/2}$ | 5519.442 | $-1.271$ | 5519.43 | at the level of the noise |

**Table 7.** Fe II lines in the 3800-8000 Å region with log $gf \geq -1.5$ and $3d^6(^3H)4f$ energy levels as upper levels

| Upper level | | Lower level | | $\lambda$(calc) | log $gf$ | $\lambda$(obs) | Notes |
|---|---|---|---|---|---|---|---|
| cm$^{-1}$ | J | cm$^{-1}$ | | Å | KUR | Å | |
| 123430.181 | cont. | 105379.430 | $(^3F)4d\ ^4D_{5/2}$ | 5538.397 | −1.442 | | at the level of the noise |
| | | 105414.180 | $(^3F)4d\ ^4G_{5/2}$ | 5549.080 | −0.905 | | blend |
| | | 105630.750 | $(^5D)5d\ ^4G_{5/2}$ | 5616.598 | −1.451 | | blend |
| | | 106846.650 | $(^3F)4d\ ^4F_{3/2}$ | 6028.409 | −1.085 | 6028.40 | at the level of the noise |
| | | 106866.760 | $(^3F)4d\ ^4F_{5/2}$ | 6035.729 | −1.269 | | |
| | | 107407.800 | $(^3F)4d\ ^2D_{5/2}$ | 6239.544 | −1.446 | | |
| | | 110428.280 | $(^3G)4d\ ^4F_{5/2}$ | 7689.067 | −1.409 | | |
| | | 110609.540 | $(^3G)4d\ ^4F_{3/2}$ | 7797.776 | −1.406 | | |

**Table 8.** Fe II lines in the 3800-8000 Å region with log $gf \geq -1.5$ and $3d^6(^3F)4f$ energy levels as upper levels.

| Upper level | | | | Lower level | | $\lambda$(calc) | log $gf$ | $\lambda$(obs) | Notes |
|---|---|---|---|---|---|---|---|---|---|
| cm$^{-1}$ | | | J | cm$^{-1}$ | | Å | K09 | Å | |
| 124421.468 | ($^3$F)4f | 4[7] | 15/2 | 103617.580 | ($^3$H)4d $^4$H$_{13/2}$ | 4805.451 | −0.972 | 4805.42 | |
| | | | | 104064.670 | ($^3$H)4d $^4$I$_{13/2}$ | 4910.993 | −1.090 | | at the continuum level |
| | | | | 104119.710 | ($^3$H)4d $^2$K$_{15/2}$ | 4924.307 | −1.174 | | not obs |
| | | | | 104622.300 | ($^3$H)4d $^2$I$_{13/2}$ | 5049.309 | −1.258 | 5049.3 | very weak |
| | | | | 105288.847 | ($^3$F)4d $^4$H$_{13/2}$ | 5225.221 | +0.974 | 5225.229 | lab, J78 |
| 124436.436 | ($^3$F)4f | 4[7] | 13/2 | 103600.430 | ($^3$H)4d $^4$G$_{11/2}$ | 4798.043 | −1.190 | | at the continuum level |
| | | | | 103751.660 | ($^3$H)4d $^4$H$_{11/2}$ | 4833.123 | −1.441 | | |
| | | | | 104315.370 | ($^3$H)4d $^2$K$_{13/2}$ | 4968.529 | −1.078 | 4968.53 | very weak |
| | | | | 104765.450 | ($^3$H)4d $^2$I$_{11/2}$ | 5082.213 | −1.265 | | blend |
| | | | | 105063.550 | ($^3$F)4d $^4$G$_{11/2}$ | 5160.416 | −0.003 | 5160.4 | lab |
| | | | | 105288.847 | ($^3$F)4d $^4$H$_{13/2}$ | 5221.136 | −0.831 | | blend, weak component |
| | | | | 105398.852 | ($^3$F)4d $^4$H$_{11/2}$ | 5251.306 | +0.664 | 5251.321 | blend |
| | | | | 105763.270 | ($^3$F)4d $^2$H$_{11/2}$ | 5353.789 | +0.076 | 5353.80 | |
| | | | | 106045.690 | ($^3$H)4d $^2$H$_{11/2}$ | 5436.006 | −0.154 | 5436.12 | |
| | | | | 108630.429 | ($^1$I)5s e$^2$I$_{11/2}$ | 6324.960 | −1.433 | | at the continuum level |
| 124400.107 | ($^3$F)4f | 4[6] | 13/2 | 103600.430 | ($^3$H)4d $^4$G$_{11/2}$ | 4806.424 | −0.542 | 4806.4 | |
| | | | | 104174.270 | ($^3$H)4d $^4$I$_{11/2}$ | 4942.792 | −1.458 | | very weak |
| | | | | 104765.450 | ($^3$H)4d $^2$I$_{11/2}$ | 5091.616 | −0.517 | 5091.6 | |
| | | | | 105063.550 | ($^3$F)4d $^4$G$_{11/2}$ | 5170.111 | +0.742 | 5170.10 | J78,lab, blended |
| | | | | 105288.850 | ($^3$F)4d $^4$H$_{13/2}$ | 5231.062 | +0.278 | 5231.067 | lab |
| | | | | 105398.850 | ($^3$F)4d $^4$H$_{11/2}$ | 5261.345 | +0.080 | 5261.339 | shifted ? |
| | | | | 105763.270 | ($^3$F)4d $^2$H$_{11/2}$ | 5364.226 | −0.538 | 5364.22 | |
| | | | | 106045.690 | ($^3$H)4d $^2$H$_{11/2}$ | 5446.766 | −0.314 | 5446.75 | blend |
| 124402.557 | ($^3$F)4f | 4[6] | 11/2 | 103683.070 | ($^5$D)5d $^4$F$_{9/2}$ | 4825.028 | −1.407 | | |
| | | | | 104765.450 | ($^3$H)4d $^2$I$_{11/2}$ | 5090.983 | −1.256 | | blend |
| | | | | 104807.210 | ($^3$H)4d $^2$G$_{9/2}$ | 5101.830 | −1.382 | 5101.82 | |
| | | | | 104916.550 | ($^3$H)4d $^4$F$_{9/2}$ | 5130.460 | +0.158 | | |
| | | | | 105063.550 | ($^3$F)4d $^4$G$_{11/2}$ | 5169.456 | −0.871 | | computed too strong |
| | | | | 105155.090 | ($^3$F)4d $^4$G$_{9/2}$ | 5194.042 | −0.084 | 5194.047 | |
| | | | | 105211.062 | ($^5$D)5d $^4$G$_{9/2}$ | 5209.193 | −0.494 | 5209.199 | |
| | | | | 105398.852 | ($^3$F)4d $^4$H$_{11/2}$ | 5260.668 | −0.049 | 5260.682 | |
| | | | | 105524.461 | ($^3$F)4d $^4$H$_{9/2}$ | 5295.671 | −1.274 | 5295.662 | computed too weak |
| | | | | 105763.270 | ($^3$F)4d $^2$H$_{11/2}$ | 5363.520 | −0.269 | 5363.51 | |
| | | | | 106018.643 | ($^3$F)4d $^2$H$_{9/2}$ | 5438.027 | −0.914 | | blend |
| | | | | 106045.690 | ($^3$H)4d $^2$H$_{11/2}$ | 5446.039 | −0.626 | 5446.05 | |
| | | | | 106097.520 | ($^3$H)4d $^2$H$_{9/2}$ | 5461.459 | +0.179 | 5461.48 | |
| | | | | 106722.170 | ($^3$F)4d $^4$F$_{9/2}$ | 5654.418 | −0.044 | | computed too strong |
| | | | | 106924.430 | ($^3$F)4d $^2$G$_{9/2}$ | 5719.850 | +0.097 | 5719.85 | lab,J78 |
| | | | | 109925.200 | ($^3$G)4d $^2$G$_{9/2}$ | 6765.246 | −1.049 | | |
| | | | | 110008.300 | ($^3$G)4d $^2$H$_{9/2}$ | 6945.303 | −1.190 | | |
| 124388.840 | ($^3$F)4f | 4[5] | 11/2 | 103600.430 | ($^3$H)4d $^4$G$_{11/2}$ | 4809.029 | −0.852 | 4809.02 | |
| | | | | 103683.070 | ($^5$D)5d $^4$F$_{9/2}$ | 4828.222 | −0.829 | | |
| | | | | 103771.320 | ($^3$H)4d $^4$G$_{9/2}$ | 4848.889 | −0.699 | | weak, on the H$_\beta$ wing |
| | | | | 104765.450 | ($^3$H)4d $^2$I$_{11/2}$ | 5094.540 | −0.517 | 5094.55 | lab |
| | | | | 104807.210 | ($^3$H)4d $^2$G$_{9/2}$ | 5105.404 | +0.158 | 5105.4 | |
| | | | | 104868.500 | ($^5$D)5d $^6$G$_{9/2}$ | 5121.435 | −0.968 | 5121.45 | weak |
| | | | | 104916.550 | ($^3$H)4d $^4$F$_{9/2}$ | 5134.072 | −0.161 | | blend |
| | | | | 105063.550 | ($^3$F)4d $^4$G$_{11/2}$ | 5173.126 | +0.425 | 5173.12 | lab |
| | | | | 105155.090 | ($^3$F)4d $^4$G$_{9/2}$ | 5197.747 | −0.166 | 5197.756 | |
| | | | | 105211.062 | ($^5$D)5d $^4$G$_{9/2}$ | 5212.916 | −0.199 | | blend |
| | | | | 105288.847 | ($^3$F)4d $^4$H$_{13/2}$ | 5234.147 | −0.630 | | blend |
| | | | | 105398.852 | ($^3$F)4d $^4$H$_{11/2}$ | 5264.468 | −0.717 | 5264.45 | |
| | | | | 106045.690 | ($^3$H)4d $^2$H$_{11/2}$ | 5450.112 | −1.282 | | blend |
| | | | | 106722.170 | ($^3$F)4d $^4$F$_{9/2}$ | 5658.806 | −0.643 | | blend |
| | | | | 106924.430 | ($^3$F)4d $^2$G$_{9/2}$ | 5724.343 | −0.429 | | blend, computed too strong |
| | | | | 109811.920 | ($^3$G)4d $^4$F$_{9/2}$ | 6858.267. | −0.903 | | at the continuum level |

**Table 8.** Fe II lines in the 3800-8000 Å region with log $gf \geq -1.5$ and $3d^6(^3F)4f$ energy levels as upper levels.

| Upper level | | | | Lower level | | $\lambda$(calc) | log $gf$ | $\lambda$(obs) | Notes |
|---|---|---|---|---|---|---|---|---|---|
| cm$^{-1}$ | | | J | cm$^{-1}$ | | Å | K09 | Å | |
| 124385.706 | ($^3$F)4f | 4[5] | 9/2 | 103771.320 | ($^3$H)4d $^4$G$_{9/2}$ | 4849.626 | $-1.159$ | | H$_\beta$ wing, not obs. |
| | | | | 103986.330 | ($^3$H)4d $^4$H$_{7/2}$ | 4900.742 | $-1.404$ | | at the continuum level |
| | | | | 104807.210 | ($^3$H)4d $^2$G$_{9/2}$ | 5106.222 | $-0.305$ | | |
| | | | | 104993.860 | ($^3$F)4d $^4$D$_{7/2}$ | 5155.371 | $-0.195$ | 5155.37 | computed too strong |
| | | | | 105063.550 | ($^3$F)4d $^4$G$_{11/2}$ | 5173.965 | $-0.955$ | 5173.98 | computed too weak |
| | | | | 105123.000 | ($^3$H)4d $^2$G$_{7/2}$ | 5189.933 | $-0.112$ | | blend |
| | | | | 105155.090 | ($^3$F)4d $^4$G$_{9/2}$ | 5198.594 | $-0.154$ | 5198.596 | |
| | | | | 105211.062 | ($^5$D)5d $^4$G$_{9/2}$ | 5213.769 | $-0.389$ | 5213.78 | |
| | | | | 105220.600 | ($^3$H)4d $^4$F$_{7/2}$ | 5216.634 | $-1.420$ | | |
| | | | | 105291.010 | ($^3$F)4d $^4$G$_{7/2}$ | 5235.599 | $-0.769$ | | blend |
| | | | | 105398.852 | ($^3$F)4d $^4$H$_{11/2}$ | 5265.337 | $-0.986$ | 5265.323 | |
| | | | | 105775.491 | ($^3$F)4d $^2$F$_{7/2}$ | 5371.899 | $+0.199$ | 5371.90 | |
| | | | | 106018.640 | ($^3$F)4d $^2$H$_{9/2}$ | 5443.015 | $-1.240$ | | |
| | | | | 106097.520 | ($^3$H)4d $^2$H$_{9/2}$ | 5466.492 | $-0.492$ | 5466.49 | blend |
| | | | | 106722.170 | ($^3$F)4d $^4$F$_{9/2}$ | 5659.810 | $-1.436$ | | blend |
| | | | | 106767.210 | ($^3$F)4d $^4$F$_{7/2}$ | 5674.279 | $-1.037$ | 5674.30 | |
| | | | | 106900.370 | ($^3$F)4d $^2$G$_{7/2}$ | 5717.492 | $-1.080$ | | blend |
| | | | | 106924.430 | ($^3$F)4d $^2$G$_{9/2}$ | 5725.370 | $-0.147$ | 5725.35 | |
| | | | | 110167.280 | ($^3$G)4d $^4$F$_{7/2}$ | 7031.188 | $-1.480$ | | not observed |
| | | | | 110570.300 | ($^3$G)4d $^2$F$_{7/2}$ | 7236.302 | $-1.125$ | | not observed |
| 124401.939 | ($^3$F)4f | 4[4] | 9/2 | 103683.070 | ($^5$D)5d $^4$F$_{9/2}$ | 4825.170 | $-0.851$ | | |
| | | | | 103771.320 | ($^3$H)4d $^4$G$_{9/2}$ | 4845.810 | $-1.216$ | | on the H$_\beta$ wing |
| | | | | 104481.590 | ($^3$H)4d $^2$F$_{7/2}$ | 5018.593 | $-0.782$ | | blend Fe II 5018.440 |
| | | | | 104765.450 | ($^3$H)4d $^2$I$_{11/2}$ | 5091.141 | $-1.199$ | 5091.15 | |
| | | | | 104807.210 | ($^3$H)4d $^2$G$_{9/2}$ | 5101.991 | $-0.285$ | | wrong, not observed |
| | | | | 104868.500 | ($^5$D)5d $^6$G$_{9/2}$ | 5118.000 | $-0.871$ | 5117.98 | |
| | | | | 104916.550 | ($^3$H)4d $^4$F$_{9/2}$ | 5130.621 | $+0.114$ | 5130.60 | lab |
| | | | | 104993.860 | ($^3$F)4d $^4$D$_{7/2}$ | 5151.058 | $-0.280$ | 5151.07 | lab |
| | | | | 105063.550 | ($^3$F)4d $^4$G$_{11/2}$ | 5169.622 | $-0.361$ | 5169.6 | |
| | | | | 105155.090 | ($^3$F)4d $^4$G$_{9/2}$ | 5194.209 | $-1.245$ | | blend Fe III |
| | | | | 105211.062 | ($^5$D)5d $^4$G$_{9/2}$ | 5209.359 | $-1.260$ | | |
| | | | | 105220.600 | ($^3$H)4d $^4$F$_{7/2}$ | 5211.949 | $+0.055$ | 5211.953 | lab |
| | | | | 105291.010 | ($^3$F)4d $^4$G$_{7/2}$ | 5231.152 | $-0.836$ | | blend |
| | | | | 105763.270 | ($^3$F)4d $^2$H$_{11/2}$ | 5363.698 | $-1.391$ | | blend |
| | | | | 105775.491 | ($^3$F)4d $^2$F$_{7/2}$ | 5367.218 | $-0.182$ | 5367.22 | |
| | | | | 106097.520 | ($^3$H)4d $^2$H$_{9/2}$ | 5461.644 | $-0.455$ | 5461.65 | |
| | | | | 106722.170 | ($^3$F)4d $^4$F$_{9/2}$ | 5654.613 | $-0.197$ | 5654.62 | |
| | | | | 106900.370 | ($^3$F)4d $^2$G$_{7/2}$ | 5712.189 | $-1.361$ | | at the level of the noise |
| | | | | 109811.920 | ($^3$G)4d $^4$F$_{9/2}$ | 6852.110 | $-0.955$ | | at the level of the noise |
| 124385.010 | ($^3$F)4f | 4[4] | 7/2 | 103191.917 | ($^3$P)4d $^2$F$_{7/2}$ | 4717.199 | $-1.461$ | | |
| | | | | 103597.402 | ($^3$P)4d $^2$D$_{5/2}$ | 4809.214 | $-1.233$ | | |
| | | | | 104807.210 | ($^3$H)4d $^2$G$_{9/2}$ | 5106.403 | $-1.091$ | | |
| | | | | 104993.860 | ($^3$F)4d $^4$D$_{7/2}$ | 5155.556 | $-0.412$ | 5155.56 | |
| | | | | 105123.000 | ($^3$H)4d $^2$G$_{7/2}$ | 5190.121 | $-0.246$ | 5190.123 | |
| | | | | 105155.090 | ($^3$F)4d $^4$G$_{9/2}$ | 5198.782 | $-0.950$ | | blend |
| | | | | 105211.062 | ($^5$D)5d $^4$G$_{9/2}$ | 5213.958 | $-1.188$ | | blend |
| | | | | 105220.600 | ($^3$H)4d $^4$F$_{7/2}$ | 5216.553 | $-1.332$ | | blend |
| | | | | 105234.237 | ($^3$H)4d $^4$F$_{5/2}$ | 5220.268 | $-1.463$ | | |
| | | | | 105291.010 | ($^3$F)4d $^4$G$_{7/2}$ | 5235.790 | $-0.829$ | | blend |
| | | | | 105775.836 | ($^3$F)4d $^2$F$_{7/2}$ | 5372.100 | $+0.165$ | 5372.10 | lab |
| | | | | 106097.520 | ($^3$H)4d $^2$H$_{9/2}$ | 5466.700 | $-1.095$ | | at the level of the noise |
| | | | | 106208.560 | ($^3$F)4d $^2$F$_{5/2}$ | 5500.096 | $-0.922$ | | blend |
| | | | | 106767.210 | ($^3$F)4d $^4$F$_{7/2}$ | 5674.503 | $-1.298$ | 5674.50 | computed too weak |
| | | | | 106796.660 | ($^3$F)4d $^4$P$_{5/2}$ | 5684.004 | $-0.895$ | | |
| | | | | 106866.760 | ($^3$F)4d $^4$F$_{5/2}$ | 5706.743 | $-0.920$ | | |
| | | | | 106900.370 | ($^3$F)4d $^2$G$_{7/2}$ | 5717.719 | $-1.023$ | | not observed |

**Table 8.** Fe II lines in the 3800-8000 Å region with $\log gf \geq -1.5$ and $3d^6(^3F)4f$ energy levels as upper levels.

| Upper level | | | | Lower level | | $\lambda$(calc) | $\log gf$ | $\lambda$(obs) | Notes |
|---|---|---|---|---|---|---|---|---|---|
| cm$^{-1}$ | | | J | cm$^{-1}$ | | Å | K09 | Å | |
| 124385.010 | cont. | | | 106924.430 | ($^3$F)4d $^2$G$_{9/2}$ | 5725.598 | −0.824 | 5725.60 | |
| | | | | 107407.800 | ($^3$F)4d $^2$D$_{5/2}$ | 5888.617 | −0.044 | 5888.61 | |
| | | | | 110570.300 | ($^3$G)4d $^2$F$_{7/2}$ | 7236.667 | −1.221 | | at the level of the noise |
| 124416.110 | ($^3$F)4f | 4[3] | 7/2 | 103683.070 | ($^5$D)5d $^4$F$_{9/2}$ | 4821.172 | −1.273 | | |
| | | | | 104481.590 | ($^3$H)4d $^2$F$_{7/2}$ | 5015.025 | −0.607 | 5015.02 | |
| | | | | 104807.210 | ($^3$H)4d $^2$G$_{9/2}$ | 5098.304 | −0.623 | | |
| | | | | 104868.500 | ($^5$D)5d $^6$G$_{9/2}$ | 5114.290 | −1.355 | | computed too strong |
| | | | | 104916.550 | ($^3$H)4d $^4$F$_{9/2}$ | 5126.892 | −0.477 | 5126.84 | lab, blend |
| | | | | 104993.860 | ($^3$F)4d $^4$D$_{7/2}$ | 5147.300 | +0.051 | 5147.25 | blend,lab |
| | | | | 105123.000 | ($^3$H)4d $^2$G$_{7/2}$ | 5181.754 | −1.028 | 5181.75 | computed too weak |
| | | | | 105155.090 | ($^3$F)4d $^4$G$_{9/2}$ | 5190.388 | −1.077 | | blend |
| | | | | 105211.062 | ($^5$D)5d $^4$G$_{9/2}$ | 5205.515 | −1.184 | | blend |
| | | | | 105220.600 | ($^3$H)4d $^4$F$_{7/2}$ | 5208.101 | +0.031 | 5208.99 | |
| | | | | 105291.010 | ($^3$F)4d $^4$G$_{7/2}$ | 5227.276 | −1.201 | | blend |
| | | | | 105379.430 | ($^3$F)4d $^4$D$_{5/2}$ | 5251.555 | −1.289 | | at the continuum level |
| | | | | 105775.491 | ($^3$F)4d $^2$F$_{7/2}$ | 5363.137 | −0.687 | 5363.15 | |
| | | | | 106097.520 | ($^3$H)4d $^2$H$_{9/2}$ | 5457.419 | −1.335 | | blend |
| | | | | 106722.170 | ($^3$F)4d $^4$F$_{9/2}$ | 5650.084 | −0.819 | | blend |
| | | | | 106767.210 | ($^3$F)4d $^4$F$_{7/2}$ | 5664.504 | −1.029 | | at the level of the noise |
| | | | | 106796.660 | ($^3$F)4d $^4$P$_{5/2}$ | 5673.972 | −0.486 | 5673.93 | blend |
| | | | | 107407.800 | ($^3$F)4d $^2$D$_{5/2}$ | 5877.850 | −1.281 | | at the level of the noise |
| | | | | 109811.920 | ($^3$G)4d $^4$F$_{9/2}$ | 6845.461 | −1.364 | | not observed |
| 124403.474 | ($^3$F)4f | 4[3] | 5/2 | 103597.402 | ($^3$P)4d $^2$D$_{5/2}$ | 4804.946 | −1.146 | 4804.93 | computed too weak |
| | | | | 104993.860 | ($^3$F)4d $^4$D$_{7/2}$ | 5150.651 | −0.855 | | |
| | | | | 105123.000 | ($^3$H)4d $^2$G$_{7/2}$ | 5185.150 | −0.746 | 5185.141 | lab,blend |
| | | | | 105234.237 | ($^3$H)4d $^4$F$_{5/2}$ | 5215.240 | −1.455 | | blend |
| | | | | 105291.010 | ($^3$F)4d $^4$G$_{7/2}$ | 5230.732 | −1.416 | | blend |
| | | | | 105317.440 | ($^3$P)4d $^2$P$_{3/2}$ | 5237.975 | −1.304 | | blend |
| | | | | 105460.230 | ($^3$F)4d $^4$D$_{3/2}$ | 5277.458 | −0.778 | | wrong, not observed |
| | | | | 105518.140 | ($^3$H)4d $^4$F$_{3/2}$ | 5293.641 | −1.294 | 5293.627 | computed too low ? |
| | | | | 105775.491 | ($^3$F)4d $^2$F$_{7/2}$ | 5366.775 | −0.450 | 5366.78 | |
| | | | | 106208.560 | ($^3$F)4d $^2$F$_{5/2}$ | 5494.515 | −0.721 | 5494.51 | |
| | | | | 106796.660 | ($^3$F)4d $^4$P$_{5/2}$ | 5678.044 | −1.006 | | computed too strong |
| | | | | 106866.760 | ($^3$F)4d $^4$F$_{5/2}$ | 5700.741 | −0.790 | 5700.76 | |
| | | | | 107065.900 | ($^3$F)4d $^4$P$_{3/2}$ | 5766.220 | −1.192 | | at the level of the noise |
| | | | | 107407.800 | ($^3$F)4d $^2$D$_{5/2}$ | 5882.220 | −0.040 | 5882.22 | |
| | | | | 107430.250 | ($^3$F)4d $^2$D$_{3/2}$ | 5890.000 | −0.918 | | blend Na I |
| | | | | 108105.900 | ($^3$F)4d $^2$P$_{3/2}$ | 6134.185 | −0.702 | 6134.2 | |
| | | | | 110611.800 | ($^3$G)4d $^2$F$_{5/2}$ | 7248.754 | −1.434 | | blend with telluric |
| 124434.563 | ($^3$F)4f | 4[2] | 5/2 | 103597.402 | ($^3$P)4d $^2$D$_{5/2}$ | 4797.777 | −1.440 | | |
| | | | | 104120.270 | ($^5$D)5d $^6$P$_{5/2}$ | 4921.269 | −0.982 | | blend |
| | | | | 104209.610 | ($^3$H)4d $^2$F$_{5/2}$ | 4943.008 | −1.371 | 4943.0 | |
| | | | | 104481.590 | ($^3$H)4d $^2$F$_{7/2}$ | 5010.387 | −0.817 | 5010.4 | |
| | | | | 104993.860 | ($^3$F)4d $^4$D$_{7/2}$ | 5142.414 | −0.113 | 5142.42 | lab |
| | | | | 105213.000 | ($^3$H)4d $^2$G$_{7/2}$ | 5176.803 | −1.156 | | blend |
| | | | | 105127.770 | ($^5$D)5d $^4$D$_{5/2}$ | 5178.082 | −1.132 | 5178.08 | computed too weak |
| | | | | 105220.600 | ($^3$H)4d $^4$F$_{7/2}$ | 5203.100 | −0.191 | 5203.10 | |
| | | | | 105379.430 | ($^3$F)4d $^4$D$_{5/2}$ | 5246.469 | −0.830 | | at the noise level, computed too strong |
| | | | | 105775.491 | ($^3$F)4d $^2$F$_{7/2}$ | 5357.833 | −1.105 | | |
| | | | | 106208.560 | ($^3$F)4d $^2$F$_{5/2}$ | 5485.142 | −1.413 | | |
| | | | | 106767.210 | ($^3$F)4d $^4$F$_{7/2}$ | 5658.587 | −1.147 | | blend |
| | | | | 106796.660 | ($^3$F)4d $^4$P$_{5/2}$ | 5668.035 | −0.132 | 5668.05 | computed too strong |
| | | | | 106866.760 | ($^3$F)4d $^4$F$_{5/2}$ | 5690.652 | −1.300 | 5690.68 | computed too weak |
| | | | | 107407.800 | ($^3$F)4d $^2$D$_{5/2}$ | 5871.480 | −1.133 | | |

**Table 8.** Fe II lines in the 3800-8000 Å region with log $gf \geq -1.5$ and $3d^6(^3F)4f$ energy levels as upper levels.

| Upper level | | | | Lower level | | λ(calc) | log gf | λ(obs) | Notes |
|---|---|---|---|---|---|---|---|---|---|
| cm$^{-1}$ | | | J | cm$^{-1}$ | | Å | K09 | Å | |
| 124460.410 | ($^3$F)4f | 4[2] | 3/2 | 104120.270 | ($^5$D)5d $^6$P$_{5/2}$ | 4915.015 | −1.449 | | |
| | | | | 104189.380 | ($^5$D)5d $^4$P$_{3/2}$ | 4931.772 | −1.122 | | wrong, not observed |
| | | | | 105234.060 | ($^3$H)4d $^4$F$_{5/2}$ | 5199.747 | −1.496 | | |
| | | | | 105317.440 | ($^3$P)4d $^2$P$_{3/2}$ | 5222.396 | −0.923 | | blend |
| | | | | 105379.430 | ($^3$F)4d $^4$D$_{5/2}$ | 5239.362 | −1.350 | | blend |
| | | | | 105460.230 | ($^3$F)4d $^4$D$_{3/2}$ | 5261.644 | −0.436 | | wrong, not observed |
| | | | | 105518.140 | ($^3$H)4d $^4$F$_{3/2}$ | 5277.730 | −1.098 | | blend |
| | | | | 106208.560 | ($^3$F)4d $^2$F$_{5/2}$ | 5477.375 | −1.153 | | at the level of the noise |
| | | | | 106846.650 | ($^3$F)4d $^4$F$_{3/2}$ | 5675.805 | −1.332 | | at the level of the noise |
| | | | | 106866.760 | ($^3$F)4d $^4$F$_{5/2}$ | 5682.292 | −0.926 | | at the level of the noise |
| | | | | 107065.930 | ($^3$F)4d $^4$P$_{3/2}$ | 5747.356 | −0.824 | | at the level of the noise |
| | | | | 107407.800 | ($^3$F)4d $^2$D$_{5/2}$ | 5862.580 | −0.452 | 5862.58 | at the level of thec noise |
| | | | | 107430.250 | ($^3$F)4d $^2$D$_{3/2}$ | 5870.308 | −0.663 | 5870.30 | computed too weak |
| | | | | 108105.900 | ($^3$F)4d $^2$P$_{3/2}$ | 6112.829 | −0.452 | | EMISSION ? |
| 124661.274 | ($^3$F)4f | 3[6] | 13/2 | 103751.660 | ($^3$H)4d $^4$H$_{11/2}$ | 4781.152 | −1.241 | 4781.15 | computed too weak |
| | | | | 105063.550 | ($^3$F)4d $^4$G$_{11/2}$ | 5101.212 | −1.511 | 5101.2 | computed too weak |
| | | | | 105398.852 | ($^3$F)4d $^4$H$_{11/2}$ | 5190.010 | +0.482 | 5190.012 | |
| | | | | 105763.270 | ($^3$F)4d $^2$H$_{11/2}$ | 5290.092 | +0.589 | 5290.094 | |
| | | | | 106045.690 | ($^3$H)4d $^2$H$_{11/2}$ | 5370.350 | +0.111 | 5370.3 | Fe II,5270.284 main comp. |
| 124656.535 | ($^3$F)4f | 3[6] | 11/2 | 103874.260 | ($^3$H)4d $^4$H$_{9/2}$ | 4810.449 | −1.268 | 4810.45 | weak |
| | | | | 104192.480 | ($^3$H)4d $^4$I$_{9/2}$ | 4885.254 | −1.238 | | blend |
| | | | | 105155.090 | ($^3$F)4d $^4$G$_{9/2}$ | 5126.398 | −0.847 | | very weak |
| | | | | 105398.852 | ($^3$F)4d $^4$H$_{11/2}$ | 5191.288 | −1.025 | | blend |
| | | | | 105524.461 | ($^3$F)4d $^4$H$_{9/2}$ | 5225.371 | +0.768 | 5225.364 | lab + unid |
| | | | | 105763.270 | ($^3$F)4d $^2$H$_{11/2}$ | 5291.420 | −1.047 | | very weak |
| | | | | 106018.643 | ($^3$F)4d $^2$H$_{9/2}$ | 5363.923 | +0.201 | 5363.92 | lab |
| | | | | 106722.170 | ($^3$F)4d $^4$F$_{9/2}$ | 5574.341 | −1.111 | 5574.25 | |
| | | | | 106924.430 | ($^3$F)4d $^2$G$_{9/2}$ | 5637.925 | −0.160 | 5637.92 | |
| | | | | 109625.200 | ($^3$G)4d $^2$G$_{9/2}$ | 6650.935 | −1.387 | | blend |
| 124626.900 | ($^3$F)4f | 3[5] | 11/2 | 103683.070 | ($^5$D)5d $^4$F$_{9/2}$ | 4773.341 | −1.317 | | |
| | | | | 103771.320 | ($^3$H)4d $^4$G$_{9/2}$ | 4793.540 | −0.748 | 4793.55 | |
| | | | | 104807.210 | ($^3$H)4d $^2$G$_{9/2}$ | 5044.081 | −0.396 | | wrong, not observed |
| | | | | 104916.550 | ($^3$H)4d $^4$F$_{9/2}$ | 5072.063 | −0.515 | 5072.05 | |
| | | | | 105063.550 | ($^3$F)4d $^4$G$_{11/2}$ | 5110.175 | −1.355 | | blend |
| | | | | 105155.090 | ($^3$F)4d $^4$G$_{9/2}$ | 5134.199 | +0.353 | 5134.20 | blend |
| | | | | 105211.062 | ($^5$D)5d $^4$G$_{9/2}$ | 5149.000 | −0.004 | | blend |
| | | | | 105398.852 | ($^3$F)4d $^4$H$_{11/2}$ | 5199.288 | −0.178 | 5199.29 | |
| | | | | 105524.461 | ($^3$F)4d $^4$H$_{9/2}$ | 5233.477 | −0.662 | 5233.47 | computed too weak |
| | | | | 105763.270 | ($^3$F)4d $^2$H$_{11/2}$ | 5299.732 | −0.158 | 5299.717 | lab |
| | | | | 106018.643 | ($^3$F)4d $^2$H$_{9/2}$ | 5372.464 | −0.223 | | blend |
| | | | | 106045.690 | ($^3$H)4d $^2$H$_{11/2}$ | 5380.285 | −0.656 | 5380.29 | |
| | | | | 106097.520 | ($^3$H)4d $^2$H$_{9/2}$ | 5395.335 | +0.054 | 5395.32 | computed too strong |
| | | | | 106722.170 | ($^3$F)4d $^4$F$_{9/2}$ | 5583.566 | −1.347 | | |
| | | | | 106924.430 | ($^3$F)4d $^2$G$_{9/2}$ | 5647.362 | −0.074 | | blend |
| | | | | 109811.920 | ($^3$G)4d $^4$F$_{9/2}$ | 6748.062 | −1.222 | | at the level of the noise |
| 124636.116 | ($^3$F)4f | 3[5] | 9/2 | 103771.320 | ($^3$H)4d $^4$G$_{9/2}$ | 4791.423 | −1.349 | | at the level of the noise |
| | | | | 104107.950 | ($^3$P)4d $^4$F$_{7/2}$ | 4869.996 | −1.378 | | blend |
| | | | | 104481.590 | ($^3$H)4d $^2$F$_{7/2}$ | 4960.280 | −1.109 | 4960.28 | weak |
| | | | | 104807.210 | ($^3$H)4d $^2$G$_{9/2}$ | 5041.737 | −1.101 | | weak |
| | | | | 104873.230 | ($^5$D)5d $^4$D$_{7/2}$ | 5058.579 | −1.461 | | weak |
| | | | | 104916.550 | ($^3$H)4d $^4$F$_{9/2}$ | 5069.692 | −1.055 | | weak |

**Table 8.** Fe II lines in the 3800-8000 Å region with $\log gf \geq -1.5$ and $3d^6(^3F)4f$ energy levels as upper levels.

| Upper level | | | | Lower level | | λ(calc) | log gf | λ(obs) | Notes |
|---|---|---|---|---|---|---|---|---|---|
| cm$^{-1}$ | | | J | cm$^{-1}$ | | Å | K09 | Å | |
| 124636.116 | cont. | | | 104993.860 | $(^3F)4d\ ^4D_{7/2}$ | 5089.646 | −0.797 | | weak |
| | | | | 105123.000 | $(^3H)4d\ ^2G_{7/2}$ | 5123.331 | −1.032 | | |
| | | | | 105155.090 | $(^3F)4d\ ^4G_{9/2}$ | 5131.770 | −0.298 | | blend |
| | | | | 105211.062 | $(^5D)5d\ ^4G_{9/2}$ | 5146.557 | −0.622 | | blend |
| | | | | 105220.600 | $(^3H)4d\ ^4F_{7/2}$ | 5149.085 | +0.286 | 5149.1 | lab |
| | | | | 105291.010 | $(^3F)4d\ ^4G_{7/2}$ | 5167.827. | −0.884 | 5167.82 | computed too weak |
| | | | | 105398.852 | $(^3F)4d\ ^4H_{11/2}$ | 5196.797 | −1.467 | | at the level of the noise |
| | | | | 105524.461 | $(^3F)4d\ ^4H_{9/2}$ | 5230.953 | −0.507 | 5230.959 | computed too weak |
| | | | | 105589.670 | $(^3F)4d\ ^4H_{7/2}$ | 5248.862 | −0.754 | 5248.801 | blend |
| | | | | 105763.270 | $(^3F)4d\ ^2H_{11/2}$ | 5297.144 | −1.481 | | weak |
| | | | | 105775.491 | $(^3F)4d\ ^2F_{7/2}$ | 5300.576 | −0.373 | | weak |
| | | | | 106018.643 | $(^3F)4d\ ^2H_{9/2}$ | 5369.805 | −0.547 | 5369.81 | |
| | | | | 106097.520 | $(^3H)4d\ ^2H_{9/2}$ | 5392.652 | −0.592 | | not obs, wrong |
| | | | | 106767.210 | $(^3F)4d\ ^4F_{7/2}$ | 5594.760 | −0.050 | | not obs, wrong |
| | | | | 106900.370 | $(^3F)4d\ ^2G_{7/2}$ | 5636.766 | −0.061 | 5636.78 | computed too weak |
| | | | | 106924.430 | $(^3F)4d\ ^2G_{9/2}$ | 5644.423 | −0.918 | | blend |
| | | | | 109901.500 | $(^3G)4d\ ^2G_{7/2}$ | 6784.867 | −1.141 | | at the level of the noise |
| | | | | 110167.280 | $(^3G)4d\ ^4F_{7/2}$ | 6909.500 | −1.099 | | at the level of the noise |
| 124623.120 | $(^3F)4f$ | 3[4] | 9/2 | 103921.630 | $(^3H)4d\ ^4G_{7/2}$ | 4829.221 | −1.017 | 4829.25 | computed too weak |
| | | | | 103983.510 | $(^3G)5s\ ^2G_{7/2}$ | 4843.700 | −1.308 | | computed too strong, not obs |
| | | | | 103986.330 | $(^3H)4d\ ^4H_{7/2}$ | 4844.361 | −1.133 | | |
| | | | | 104916.550 | $(^3H)4d\ ^4F_{9/2}$ | 5073.036 | −1.028 | | |
| | | | | 104993.860 | $(^3F)4d\ ^4D_{7/2}$ | 5093.016 | −1.142 | 5093.01 | weak |
| | | | | 105123.000 | $(^3H)4d\ ^2G_{7/2}$ | 5126.745 | −0.382 | 5126.75 | lab, blend |
| | | | | 105155.090 | $(^3F)4d\ ^4G_{9/2}$ | 5135.196 | −0.318 | | blend |
| | | | | 105211.062 | $(^5D)5d\ ^4G_{9/2}$ | 5150.003 | −0.755 | 5150.02 | |
| | | | | 105220.600 | $(^3H)4d\ ^4F_{7/2}$ | 5152.534 | −1.333 | | blend |
| | | | | 105291.010 | $(^3F)4d\ ^4G_{7/2}$ | 5171.301 | +0.425 | 5171.305 | |
| | | | | 105398.852 | $(^3F)4d\ ^4H_{11/2}$ | 5200.310 | −1.359 | | blend |
| | | | | 105449.540 | $(^5D)5d\ ^4G_{7/2}$ | 5214.058 | −0.628 | | blend |
| | | | | 105524.461 | $(^3F)4d\ ^4H_{9/2}$ | 5234.513 | −0.157 | | blend |
| | | | | 105763.270 | $(^3F)4d\ ^2H_{11/2}$ | 5300.794 | −1.386 | | blend |
| | | | | 105775.491 | $(^3F)4d\ ^2F_{7/2}$ | 5304.231 | −0.076 | 5304.25 | blend |
| | | | | 106018.640 | $(^3F)4d\ ^2H_{9/2}$ | 5373.555 | −1.277 | | |
| | | | | 106097.520 | $(^3H)4d\ ^2H_{9/2}$ | 5396.435 | −0.899 | 5396.45 | computed too weak |
| | | | | 106900.370 | $(^3F)4d\ ^2G_{7/2}$ | 5640.900 | −0.389 | 5640.9 | computed too strong |
| | | | | 106924.430 | $(^3F)4d\ ^2G_{9/2}$ | 5648.568 | −0.369 | 5648.57 | blend |
| | | | | 110570.300 | $(^3G)4d\ ^2F_{7/2}$ | 7114.048 | −1.243 | | |
| 124620.914 | $(^3F)4f$ | 3[4] | 7/2 | 103921.630 | $(^3H)4d\ ^4G_{7/2}$ | 4829.735 | −1.435 | | |
| | | | | 104023.910 | $(^3H)4d\ ^4G_{5/2}$ | 4853.719 | −0.883 | | |
| | | | | 104569.230 | $(^3P)4d\ ^4F_{5/2}$ | 4985.721 | −0.873 | 4985.72 | weak |
| | | | | 104993.860 | $(^3F)4d\ ^4D_{7/2}$ | 5093.588 | −1.437 | | blend |
| | | | | 105123.000 | $(^3H)4d\ ^2G_{5/2}$ | 5127.325 | −0.784 | | blend |
| | | | | 105155.090 | $(^3F)4d\ ^4G_{9/2}$ | 5135.778 | −1.386 | | weak |
| | | | | 105234.237 | $(^3H)4d\ ^4F_{5/2}$ | 5156.745 | −0.254 | | blend |
| | | | | 105291.010 | $(^3F)4d\ ^4G_{7/2}$ | 5171.891 | +0.011 | 5171.9 | |
| | | | | 105379.430 | $(^3F)4d\ ^4D_{5/2}$ | 5195.658 | −0.478 | 5195.661 | lab |
| | | | | 105414.180 | $(^3F)4d\ ^4G_{5/2}$ | 5205.058 | −0.783 | | blend |
| | | | | 105449.540 | $(^5D)5d\ ^4G_{7/2}$ | 5214.658 | −1.042 | | weak |
| | | | | 105524.461 | $(^3F)4d\ ^4H_{9/2}$ | 5235.117 | −1.185 | | blend |
| | | | | 105711.730 | $(^5D)5d\ ^6S_{5/2}$ | 5286.964 | −0.934 | | blend |
| | | | | 105775.491 | $(^3F)4d\ ^2F_{7/2}$ | 5304.852 | −0.525 | 5304.87 | blend |
| | | | | 106208.560 | $(^3F)4d\ ^2F_{5/2}$ | 5429.627 | −0.531 | 5429.62 | computed too weak |
| | | | | 106866.760 | $(^3F)4d\ ^4F_{5/2}$ | 5630.922 | −1.421 | | weak |
| | | | | 106900.370 | $(^3F)4d\ ^2G_{7/2}$ | 5641.602 | −0.724 | 5641.61 | weak |
| | | | | 106924.430 | $(^3F)4d\ ^2G_{9/2}$ | 5649.272 | −1.404 | | not observed |
| | | | | 107407.800 | $(^3F)4d\ ^2D_{5/2}$ | 5807.914 | −0.295 | 5807.9 | blend |

**Table 8.** Fe II lines in the 3800-8000 Å region with log $gf \geq -1.5$ and $3d^6(^3F)4f$ energy levels as upper levels.

| Upper level | | | | Lower level | | $\lambda$(calc) | log $gf$ | $\lambda$(obs) | Notes |
|---|---|---|---|---|---|---|---|---|---|
| cm$^{-1}$ | | | J | cm$^{-1}$ | | Å | K09 | Å | |
| 124641.989 | ($^3$F)4f | 3[3] | 7/2 | 104107.950 | ($^3$P)4d $^4$F$_{7/2}$ | 4868.603 | −1.393 | | |
| | | | | 104120.270 | ($^5$D)5d $^6$P$_{5/2}$ | 4871.525 | −1.423 | | |
| | | | | 104481.590 | ($^3$H)4d $^2$F$_{7/2}$ | 4958.835 | −1.370 | | blend |
| | | | | 105123.000 | ($^3$H)4d $^2$G$_{7/2}$ | 5121.789 | −0.828 | | |
| | | | | 105155.090 | ($^3$F)4d $^4$G$_{9/2}$ | 5130.223 | −0.928 | | blend |
| | | | | 105211.062 | ($^5$D)5d $^4$G$_{9/2}$ | 5145.002 | −1.290 | | at the level of the noise |
| | | | | 105220.600 | ($^3$H)4d $^4$F$_{7/2}$ | 5147.528 | −0.014 | 5147.52 | |
| | | | | 105291.010 | ($^3$F)4d $^4$G$_{7/2}$ | 5166.258 | −1.096 | | weak |
| | | | | 105379.430 | ($^3$F)4d $^4$D$_{5/2}$ | 5189.973 | −0.210 | | blend |
| | | | | 105414.180 | ($^3$F)4d $^4$G$_{5/2}$ | 5199.353 | −1.041 | | blend |
| | | | | 105589.670 | ($^3$F)4d $^4$H$_{7/2}$ | 5247.244 | −0.996 | 5247.25 | weak |
| | | | | 105711.730 | ($^5$D)5d $^6$S$_{5/2}$ | 5281.078 | −0.874 | | not observed |
| | | | | 105775.491 | ($^3$F)4d $^2$F$_{7/2}$ | 5298.926 | −0.405 | | blend |
| | | | | 106097.520 | ($^3$H)4d $^2$H$_{9/2}$ | 5390.945 | −1.384 | | blend |
| | | | | 106208.560 | ($^3$F)4d $^2$F$_{5/2}$ | 5423.419 | −0.138 | 5423.41 | lab |
| | | | | 106767.210 | ($^3$F)4d $^4$F$_{7/2}$ | 5592.922 | −0.422 | | wrong |
| | | | | 106796.660 | ($^3$F)4d $^4$P$_{5/2}$ | 5602.152 | −0.795 | | blend |
| | | | | 106866.760 | ($^3$F)4d $^4$F$_{5/2}$ | 5624.245 | −1.195 | | blend |
| | | | | 106900.370 | ($^3$F)4d $^2$G$_{7/2}$ | 5634.900 | −0.588 | 5634.9 | computed too weak |
| | | | | 106924.430 | ($^3$F)4d $^2$G$_{9/2}$ | 5642.552 | −1.377 | | at the level of the noise |
| | | | | 107407.800 | ($^3$F)4d $^2$D$_{5/2}$ | 5800.811 | −0.993 | | at the level of the noise |
| | | | | 110167.280 | ($^3$G)4d $^4$F$_{7/2}$ | 6906.696 | −1.294 | | at the level of the noise |
| | | | | 110611.800 | ($^3$G)4d $^2$F$_{5/2}$ | 7125.523 | −1.233 | | at the level of the noise |
| 124653.022 | ($^3$F)4f | 3[3] | 5/2 | 104023.910 | ($^3$H)4d $^4$G$_{5/2}$ | 4846.164 | −1.115 | | weak |
| | | | | 104569.230 | ($^3$P)4d $^4$F$_{5/2}$ | 4977.751 | −0.819 | 4977.75 | computed too weak |
| | | | | 104839.998 | ($^3$P)4d $^2$D$_{3/2}$ | 5045.778 | −0.981 | 5045.79 | computed too weak |
| | | | | 105123.000 | ($^3$H)4d $^2$G$_{7/2}$ | 5118.896 | −1.484 | | |
| | | | | 105234.237 | ($^3$H)4d $^4$F$_{5/2}$ | 5148.219 | −0.286 | | computed too strong |
| | | | | 105291.010 | ($^3$F)4d $^4$G$_{7/2}$ | 5163.314 | −0.700 | 5163.29 | weak |
| | | | | 105317.440 | ($^3$P)4d $^2$P$_{3/2}$ | 5170.372 | −1.129 | | |
| | | | | 105379.430 | ($^3$F)4d $^4$D$_{5/2}$ | 5187.002 | −0.628 | 5187.0 | |
| | | | | 105414.180 | ($^3$F)4d $^4$G$_{5/2}$ | 5196.371 | −0.956 | | blend |
| | | | | 105460.230 | ($^3$F)4d $^4$D$_{3/2}$ | 5208.839 | −0.132 | 5208.862 | lab, computed too strong |
| | | | | 105711.730 | ($^5$D)5d $^6$S$_{5/2}$ | 5278.002 | −1.442 | | |
| | | | | 105775.491 | ($^3$F)4d $^2$F$_{7/2}$ | 5295.829 | −1.021 | | blend |
| | | | | 106208.560 | ($^3$F)4d $^2$F$_{5/2}$ | 5420.175 | −0.824 | 5420.2 | computed too weak |
| | | | | 106846.650 | ($^3$F)4d $^4$F$_{3/2}$ | 5614.409 | −0.773 | | computed too strong |
| | | | | 107065.930 | ($^3$F)4d $^4$P$_{3/2}$ | 5684.411 | −1.018 | | |
| | | | | 107407.800 | ($^3$F)4d $^2$D$_{5/2}$ | 5797.100 | −0.273 | 5797.1 | |
| | | | | 107430.250 | ($^3$F)4d $^2$D$_{3/2}$ | 5804.657 | −0.981 | | at the level of the noise |
| | | | | 108105.900 | ($^3$F)4d $^2$P$_{3/2}$ | 6041.674 | −0.519 | | |
| 124731.762 | ($^3$F)4f | 3[0] | 1/2 | 104189.380 | ($^5$D)5d $^4$P$_{3/2}$ | 4866.625 | −0.710 | | on the H$_\beta$ wing |
| | | | | 104588.710 | ($^5$D)5d $^6$D$_{3/2}$ | 4963.106 | −1.473 | | |
| | | | | 104736.460 | ($^3$P)4d $^2$P$_{1/2}$ | 4999.780 | −1.476 | | |
| | | | | 105460.230 | ($^3$F)4d $^4$D$_{3/2}$ | 5187.556 | −1.137 | | |
| | | | | 105477.920 | ($^3$F)4d $^4$D$_{1/2}$ | 5192.323 | −0.902 | | blend |
| | | | | 105518.140 | ($^3$H)4d $^4$F$_{3/2}$ | 5203.192 | −0.854 | | blend |
| | | | | 107065.930 | ($^3$F)4d $^4$P$_{3/2}$ | 5659.074 | −0.650 | 5659.05 | computed too weak |
| | | | | 107176.100 | ($^5$D)5d $^4$P$_{1/2}$ | 5694.588 | −0.810 | 5694.59 | good agreement |
| | | | | 107430.250 | ($^3$F)4d $^2$D$_{3/2}$ | 5778.239 | −0.939 | | blend |
| | | | | 108105.900 | ($^3$F)4d $^2$P$_{3/2}$ | 6013.060 | −1.184 | | |

**Table 8.** Fe II lines in the 3800-8000 Å region with log $gf \geq -1.5$ and $3d^6(^3F)4f$ energy levels as upper levels.

| Upper level | | | | Lower level | | $\lambda$(calc) | log $gf$ | $\lambda$(obs) | Notes |
|---|---|---|---|---|---|---|---|---|---|
| cm$^{-1}$ | | | J | cm$^{-1}$ | | Å | K09 | Å | |
| 124803.873 | ($^3$F)4f | 2[5] | 11/2 | 103771.320 | ($^3$H)4d $^4$G$_{9/2}$ | 4753.206 | −1.359 | | |
| | | | | 104807.210 | ($^3$H)4d $^2$G$_{9/2}$ | 4999.441 | −1.315 | | |
| | | | | 105524.461 | ($^3$F)4d $^4$H$_{9/2}$ | 5185.437 | +0.377 | 5185.422 | lab |
| | | | | 106018.643 | ($^3$F)4d $^2$H$_{9/2}$ | 5321.852 | +0.731 | 5321.83 | lab |
| | | | | 106097.520 | ($^3$H)4d $^2$H$_{9/2}$ | 5344.292 | −1.008 | 5344.28 | |
| | | | | 106924.430 | ($^3$F)4d $^2$G$_{9/2}$ | 5591.464 | −0.173 | | computed too strong |
| | | | | 109625.200 | ($^3$G)4d $^2$G$_{9/2}$ | 6586.373 | −1.344 | | not observed |
| 124809.727 | ($^3$F)4f | 2[5] | 9/2 | 103921.630 | ($^3$H)4d $^4$G$_{7/2}$ | 4786.078 | −1.434 | | |
| | | | | 103983.510 | ($^3$G)5s $^2$G$_{7/2}$ | 4800.298 | −1.342 | | |
| | | | | 105291.010 | ($^3$F)4d $^4$G$_{7/2}$ | 5121.860 | −1.107 | | blend |
| | | | | 105449.540 | ($^5$D)5d $^4$G$_{7/2}$ | 5163.801 | −1.335 | | blend |
| | | | | 105524.461 | ($^3$F)4d $^4$H$_{9/2}$ | 5183.862 | −1.227 | | blend |
| | | | | 105589.670 | ($^3$F)4d $^4$H$_{7/2}$ | 5201.450 | +0.802 | 5201.444 | lab |
| | | | | 105775.491 | ($^3$F)4d $^2$F$_{7/2}$ | 5252.229 | −1.121 | | |
| | | | | 106018.643 | ($^3$F)4d $^2$H$_{9/2}$ | 5320.193 | −0.866 | | blend |
| | | | | 106767.210 | ($^3$F)4d $^4$F$_{7/2}$ | 5540.925 | −1.367 | | |
| | | | | 106900.370 | ($^3$F)4d $^2$G$_{7/2}$ | 5582.123 | −0.405 | 5582.12 | |
| 124793.905 | ($^3$F)4f | 2[4] | 9/2 | 103921.630 | ($^3$H)4d $^4$G$_{7/2}$ | 4789.706 | −1.174 | 4789.7 | computed too weak |
| | | | | 103986.330 | ($^3$H)4d $^4$H$_{7/2}$ | 4804.599 | −1.426 | | blend |
| | | | | 104481.590 | ($^3$H)4d $^2$F$_{7/2}$ | 4921.748 | −1.081 | | blend |
| | | | | 105123.000 | ($^3$H)4d $^2$G$_{7/2}$ | 5082.234 | −0.341 | | blend |
| | | | | 105220.600 | ($^3$H)4d $^4$F$_{7/2}$ | 5107.576 | −0.574 | | blend |
| | | | | 105291.010 | ($^3$F)4d $^4$G$_{7/2}$ | 5126.016 | +0.065 | 5126.00 | lab. |
| | | | | 105449.540 | ($^5$D)5d $^4$G$_{7/2}$ | 5168.025 | −1.175 | | good agreement |
| | | | | 105524.460 | ($^3$F)4d $^4$H$_{9/2}$ | 5188.118 | −0.544 | 5188.12 | good agreement |
| | | | | 105589.670 | ($^3$F)4d $^4$H$_{7/2}$ | 5205.735 | −0.340 | | blend |
| | | | | 105775.491 | ($^3$F)4d $^2$F$_{7/2}$ | 5256.599 | −0.442 | 5256.599 | good agreement |
| | | | | 106018.640 | ($^3$F)4d $^2$H$_{9/2}$ | 5324.675 | −0.131 | 5234.68 | good agreementy= |
| | | | | 106900.370 | ($^3$F)4d $^2$G$_{7/2}$ | 5587.059 | +0.466 | | blend |
| | | | | 106924.430 | ($^3$F)4d $^2$G$_{9/2}$ | 5594.582 | −1.114 | | blend |
| | | | | 109901.500 | ($^3$G)4d $^2$G$_{7/2}$ | 6712.979 | −1.436 | | |
| | | | | 110167.280 | ($^3$G)4d $^4$F$_{7/2}$ | 6834.961 | −1.262 | | |
| | | | | 110570.300 | ($^3$G)4d $^2$F$_{7/2}$ | 7028.628 | −1.389 | | |
| 124783.748 | ($^3$F)4f | 2[4] | 7/2 | 104023.910 | ($^3$H)4d $^4$G$_{5/2}$ | 4815.647 | −0.780 | | not observed |
| | | | | 104120.270 | ($^5$D)5d $^6$P$_{5/2}$ | 4838.105 | −1.439 | | |
| | | | | 104209.610 | ($^3$H)4d $^2$F$_{5/2}$ | 4859.114 | −1.499 | | |
| | | | | 104569.230 | ($^3$P)4d $^4$F$_{5/2}$ | 4945.559 | −1.176 | | weak |
| | | | | 105123.000 | ($^3$H)4d $^2$G$_{7/2}$ | 5084.859 | −1.401 | | |
| | | | | 105291.010 | ($^3$F)4d $^4$G$_{7/2}$ | 5128.687 | −0.876 | | blend |
| | | | | 105414.180 | ($^3$F)4d $^4$G$_{5/2}$ | 5161.300 | +0.512 | 5161.3 | lab, computed too strong |
| | | | | 105589.670 | ($^3$F)4d $^4$H$_{7/2}$ | 5208.490 | −0.196 | 5208.501 | |
| | | | | 105630.750 | ($^5$D)5d $^4$G$_{5/2}$ | 5219.661 | −0.923 | | blend |
| | | | | 106018.640 | ($^3$F)4d $^2$H$_{9/2}$ | 5327.557 | −1.482 | | |
| | | | | 106208.560 | ($^3$F)4d $^2$F$_{5/2}$ | 5382.029 | −0.281 | 5382.12 | |
| | | | | 106900.370 | ($^3$F)4d $^2$G$_{7/2}$ | 5590.233 | −0.326 | 5590.22 | |
| | | | | 107407.800 | ($^3$F)4d $^2$D$_{5/2}$ | 5753.486 | −0.930 | | at the level of the noise |
| | | | | 110611.800 | ($^3$G)4d $^2$F$_{5/2}$ | 7054.248 | −1.377 | | at the level of the noise |

**Table 9.** Fe II lines in the 3800-8000 Å region with $\log gf \geq -1.5$ and $3d^6(^3G)4f$ energy levels as upper levels.

| Upper level | | | | Lower level | | λ(calc) | log gf | λ(obs) | Notes |
|---|---|---|---|---|---|---|---|---|---|
| cm$^{-1}$ | | | J | cm$^{-1}$ | | Å | KUR | Å | |
| 127507.241 | ($^3$G)4f | 5[8] | 17/2 | 103878.370 | ($^3$H)4d $^4$I$_{15/2}$ | 4230.919 | −1.017 | 4230.93 | |
| | | | | 108337.860 | ($^3$G)4d $^4$I$_{15/2}$ | 5215.200 | +1.119 | 5215.21 | |
| 127524.122 | ($^3$G)4f | 5[8] | 15/2 | 104064.670 | ($^3$H)4d $^4$I$_{13/2}$ | 4261.475 | −1.477 | | |
| | | | | 104622.300 | ($^3$H)4d $^2$I$_{13/2}$ | 4365.238 | −1.210 | | |
| | | | | 108133.440 | ($^3$G)4d $^4$H$_{13/2}$ | 5155.680 | −0.971 | | |
| | | | | 108463.910 | ($^3$G)4d $^4$I$_{13/2}$ | 5245.071 | +0.889 | 5245.073 | lab, J78 |
| | | | | 108648.695 | ($^1$I)5s e$^2$I$_{13/2}$ | 5296.420 | −0.047 | 5296.418 | |
| | | | | 109049.600 | ($^3$G)4d $^2$I$_{13/2}$ | 5411.356 | +0.449 | | blend |
| 127484.653 | ($^3$G)4f | 5[7] | 15/2 | 108133.440 | ($^3$G)4d $^4$H$_{13/2}$ | 5166.196 | +0.934 | 5166.2 | lab |
| | | | | 108337.860 | ($^3$G)4d $^4$I$_{15/2}$ | 5221.353 | +0.453 | 5221.335 | lab |
| | | | | 108463.910 | ($^3$G)4d $^4$I$_{13/2}$ | 5255.955 | −0.980 | | |
| | | | | 108648.695 | ($^1$I)5s e$^2$I$_{13/2}$ | 5307.518 | −0.940 | | |
| | | | | 109049.600 | ($^3$G)4d $^2$I$_{13/2}$ | 5422.941 | −1.415 | | |
| 127515.235 | ($^3$G)4f | 5[7] | 13/2 | 105763.270 | ($^3$F)4d $^2$H$_{11/2}$ | 4595.998 | −1.059 | | |
| | | | | 106045.690 | ($^3$H)4d $^2$H$_{11/2}$ | 4656.457 | −0.284 | | |
| | | | | 108133.440 | ($^3$G)4d $^4$H$_{13/2}$ | 5158.044 | −0.684 | | |
| | | | | 108181.550 | ($^3$G)4d $^4$G$_{11/2}$ | 5170.879 | −0.639 | | |
| | | | | 108387.920 | ($^3$G)4d $^4$H$_{11/2}$ | 5226.670 | +0.474 | 5226.686 | lab |
| | | | | 108463.910 | ($^3$G)4d $^4$I$_{13/2}$ | 5247.518 | +0.157 | 5247.536 | lab |
| | | | | 108648.695 | ($^1$I)5s e$^2$I$_{13/2}$ | 5298.915 | −1.299 | | |
| | | | | 108775.080 | ($^3$G)4d $^4$I$_{11/2}$ | 5334.651 | −0.859 | | |
| | | | | 109049.600 | ($^3$G)4d $^2$I$_{13/2}$ | 5413.960 | −0.246 | | |
| | | | | 109683.280 | ($^3$G)4d $^2$H$_{11/2}$ | 5606.354 | +0.514 | 5606.38 | |
| 127489.429 | ($^3$G)4f | 5[6] | 13/2 | 103600.430 | ($^3$H)4d $^4$G$_{11/2}$ | 4184.848 | −1.133 | | |
| | | | | 106045.690 | ($^3$H)4d $^2$H$_{11/2}$ | 4662.061 | −1.312 | | |
| | | | | 108133.440 | ($^3$G)4d $^4$H$_{13/2}$ | 5164.921 | +0.601 | 5164.9 | lab |
| | | | | 108181.550 | ($^3$G)4d $^4$G$_{11/2}$ | 5177.791 | +0.705 | 5177.77 | lab |
| | | | | 108337.860 | ($^3$G)4d $^4$I$_{15/2}$ | 5220.051 | −0.463 | | |
| | | | | 108387.920 | ($^3$G)4d $^4$H$_{11/2}$ | 5233.732 | −1.225 | | |
| | | | | 108463.910 | ($^3$G)4d $^4$I$_{13/2}$ | 5254.636 | −0.596 | | |
| | | | | 108648.695 | ($^1$I)5s e$^2$I$_{13/2}$ | 5306.173 | −0.818 | | |
| | | | | 109683.280 | ($^3$G)4d $^2$H$_{11/2}$ | 5614.479 | −0.728 | | |
| 127489.977 | ($^3$G)4f | 5[6] | 11/2 | 103600.430 | ($^3$H)4d $^4$G$_{11/2}$ | 4184.752 | −1.422 | | |
| | | | | 103683.070 | ($^5$D)5d $^4$F$_{9/2}$ | 4199.279 | −1.301 | | |
| | | | | 106045.690 | ($^3$H)4d $^2$H$_{11/2}$ | 4661.942 | −1.108 | | |
| | | | | 106722.170 | ($^3$F)4d $^4$F$_{9/2}$ | 4813.800 | −0.314 | 4813.8 | |
| | | | | 106924.430 | ($^3$F)4d $^2$G$_{9/2}$ | 4861.143 | −0.513 | | |
| | | | | 108133.440 | ($^3$G)4d $^4$H$_{13/2}$ | 5164.775 | −0.273 | 5164.77 | |
| | | | | 108181.550 | ($^3$G)4d $^4$G$_{11/2}$ | 5177.644 | +0.437 | 5177.64 | lab |
| | | | | 108387.920 | ($^3$G)4d $^4$H$_{11/2}$ | 5233.581 | −0.349 | 5233.58 | |
| | | | | 108391.500 | ($^3$G)4d $^4$G$_{9/2}$ | 5234.562 | −0.887 | | |
| | | | | 109049.600 | ($^3$G)4d $^2$I$_{13/2}$ | 5421.376 | −1.110 | | |
| | | | | 109625.200 | ($^3$G)4d $^2$G$_{9/2}$ | 5596.053 | −0.050 | | computed too strong |
| | | | | 109683.280 | ($^3$G)4d $^2$H$_{11/2}$ | 5614.306 | −0.230 | | |
| | | | | 109811.920 | ($^3$G)4d $^4$F$_{9/2}$ | 5655.161 | −0.047 | 5655.15 | |
| | | | | 110008.300 | ($^3$G)4d $^2$H$_{9/2}$ | 5718.689 | −0.545 | | |
| 127482.748 | ($^3$G)4f | 5[5] | 11/2 | 105763.270 | ($^3$F)4d $^2$H$_{11/2}$ | 4602.873 | −1.478 | | |
| | | | | 106045.690 | ($^3$H)4d $^2$H$_{11/2}$ | 4663.514 | −0.736 | | |
| | | | | 106722.170 | ($^3$F)4d $^4$F$_{9/2}$ | 4815.476 | −0.239 | | computed too strong |
| | | | | 108133.440 | ($^3$G)4d $^4$H$_{13/2}$ | 5166.704 | −0.401 | | computed too strong |
| | | | | 108181.550 | ($^3$G)4d $^4$G$_{11/2}$ | 5179.583 | +0.320 | | blend |
| | | | | 108387.920 | ($^3$G)4d $^4$H$_{11/2}$ | 5235.563 | −0.190 | 5235.585 | blend |
| | | | | 108391.500 | ($^3$G)4d $^4$G$_{9/2}$ | 5236.545 | +0.191 | | blend, computed too strong |

**Table 9.** Fe II lines in the 3800-8000 Å region with log $gf \geq -1.5$ and $3d^6(^3G)4f$ energy levels as upper levels.

| Upper level | | | | Lower level | | λ(calc) | log gf | λ(obs) | Notes |
|---|---|---|---|---|---|---|---|---|---|
| cm$^{-1}$ | | | J | cm$^{-1}$ | | Å | KUR | Å | |
| 127482.748 | cont. | | | 108463.910 | ($^3$G)4d $^4$I$_{13/2}$ | 5256.482 | −0.830 | 5256.5 | |
| | | | | 108648.695 | ($^1$I)5s e$^2$I$_{13/2}$ | 5308.055 | −1.341 | | |
| | | | | 108775.080 | ($^3$G)4d $^4$I$_{11/2}$ | 5343.915 | −1.043 | | |
| | | | | 109625.200 | ($^3$G)4d $^2$G$_{9/2}$ | 5598.319 | −0.100 | 5598.32 | computed too weak |
| | | | | 109683.280 | ($^3$G)4d $^2$H$_{11/2}$ | 5616.586 | −0.042 | 5616.6 | computed too weak |
| | | | | 109811.920 | ($^3$G)4d $^4$F$_{9/2}$ | 5657.474 | −0.662 | 5657.50 | computed too weak |
| | | | | 110008.300 | ($^3$G)4d $^2$H$_{9/2}$ | 5721.054 | −0.506 | | |
| 127485.362 | ($^3$G)4f | 5[4] | 9/2 | 104107.950 | ($^3$P)4d $^4$F$_{7/2}$ | 4276.430 | −1.168 | | |
| | | | | 104481.590 | ($^3$H)4d $^2$F$_{7/2}$ | 4345.891 | −1.316 | | |
| | | | | 105775.491 | ($^3$F)4d $^2$F$_{7/2}$ | 4604.910 | −1.176 | | |
| | | | | 106045.690 | ($^3$H)4d $^2$H$_{11/2}$ | 4662.945 | −1.404 | | |
| | | | | 106722.170 | ($^3$F)4d $^4$F$_{9/2}$ | 4814.870 | −0.945 | | |
| | | | | 106767.210 | ($^3$F)4d $^4$F$_{7/2}$ | 4825.337 | −1.318 | | |
| | | | | 106924.430 | ($^3$F)4d $^2$G$_{9/2}$ | 4862.235 | −0.425 | | |
| | | | | 108181.550 | ($^3$G)4d $^4$G$_{11/2}$ | 5178.882 | −0.635 | | |
| | | | | 108365.320 | ($^3$G)4d $^4$D$_{7/2}$ | 5228.658 | −0.224 | | blend |
| | | | | 108387.920 | ($^3$G)4d $^4$H$_{11/2}$ | 5234.846 | −0.695 | 5234.80 | |
| | | | | 108391.500 | ($^3$G)4d $^4$G$_{9/2}$ | 5235.828 | −0.195 | 5235.80 | blend |
| | | | | 108537.610 | ($^3$G)4d $^4$G$_{7/2}$ | 5276.203 | −1.169 | | |
| | | | | 108577.560 | ($^3$G)4d $^4$H$_{9/2}$ | 5287.351 | −1.391 | | |
| | | | | 109625.200 | ($^3$G)4d $^2$G$_{9/2}$ | 5597.499 | +0.251 | 5597.50 | computed too strong |
| | | | | 109683.280 | ($^3$G)4d $^2$H$_{11/2}$ | 5615.762 | −0.466 | 5615.75 | |
| | | | | 109811.920 | ($^3$G)4d $^4$F$_{9/2}$ | 5656.638 | −0.349 | 5656.55 | blend |
| | | | | 109901.500 | ($^3$G)4d $^2$G$_{7/2}$ | 5685.455 | −0.333 | 5685.45 | |
| | | | | 110008.300 | ($^3$G)4d $^2$H$_{9/2}$ | 5720.199 | −0.468 | 5720.20 | |
| | | | | 110167.280 | ($^3$G)4d $^4$F$_{7/2}$ | 5772.711 | −1.064 | | |
| | | | | 110570.300 | ($^3$G)4d $^2$F$_{7/2}$ | 5910.253 | −0.120 | | blend H2O |
| 127485.699 | ($^3$G)4f | 5[4] | 7/2 | 103683.070 | ($^5$D)5d $^4$F$_{9/2}$ | 4200.033 | −1.226 | | |
| | | | | 106722.170 | ($^3$F)4d $^4$F$_{9/2}$ | 4814.791 | +0.017 | 4814.8 | computed too strong |
| | | | | 106767.210 | ($^3$F)4d $^4$F$_{7/2}$ | 4825.259 | −0.375 | 4825.30 | blend |
| | | | | 106900.370 | ($^3$F)4d $^2$G$_{7/2}$ | 4856.472 | −1.384 | | |
| | | | | 106924.430 | ($^3$F)4d $^2$G$_{9/2}$ | 4862.155 | −0.753 | | |
| | | | | 108365.320 | ($^3$G)4d $^4$D$_{7/2}$ | 5228.566 | +0.266 | | blend |
| | | | | 108391.500 | ($^3$G)4d $^4$G$_{9/2}$ | 5235.735 | −0.618 | | blend |
| | | | | 108537.610 | ($^3$G)4d $^4$G$_{7/2}$ | 5276.109 | −0.999 | | |
| | | | | 109625.200 | ($^3$G)4d $^2$G$_{9/2}$ | 5597.394 | −1.025 | | |
| | | | | 109811.920 | ($^3$G)4d $^4$F$_{9/2}$ | 5656.530 | +0.034 | 5656.55 | |
| | | | | 110065.750 | ($^3$G)4d $^2$D$_{5/2}$ | 5738.953 | −1.494 | | |
| | | | | 110167.280 | ($^3$G)4d $^4$F$_{7/2}$ | 5772.598 | −0.676 | | |
| | | | | 110570.300 | ($^3$G)4d $^2$F$_{7/2}$ | 5910.135 | −1.369 | | |
| 127510.913 | ($^3$G)4f | 5[3] | 5/2 | 106767.210 | ($^3$F)4d $^4$F$_{7/2}$ | 4819.393 | −0.294 | 4819.40 | |
| | | | | 106900.370 | ($^3$F)4d $^2$G$_{7/2}$ | 4850.531 | −1.345 | | |
| | | | | 108365.320 | ($^3$G)4d $^4$D$_{7/2}$ | 5221.680 | +0.447 | 5221.68 | lab |
| | | | | 108537.610 | ($^3$G)4d $^4$G$_{7/2}$ | 5269.097 | −0.794 | 5369.12 | |
| | | | | 110065.750 | ($^3$G)4d $^2$D$_{5/2}$ | 5730.658 | −0.761 | | |
| | | | | 110167.280 | ($^3$G)4d $^4$F$_{7/2}$ | 5764.206 | −0.654 | | blend |
| | | | | 110570.300 | ($^3$G)4d $^2$F$_{7/2}$ | 5901.339 | −1.193 | | |
| 127487.681 | ($^3$G)4f | 5[2] | 3/2 | 106866.760 | ($^3$F)4d $^4$F$_{5/2}$ | 4848.090 | −0.945 | | |
| | | | | 108642.410 | ($^3$G)4d $^4$D$_{5/2}$ | 5304.895 | −0.425 | 5304.89 | blend |
| | | | | 110065.750 | ($^3$G)4d $^2$D$_{5/2}$ | 5738.300 | −0.104 | 5738.30 | |

**Table 9.** Fe II lines in the 3800-8000 Å region with log $gf \geq -1.5$ and $3d^6(^3G)4f$ energy levels as upper levels.

| Upper level | | | Lower level | | $\lambda$(calc) | log $gf$ | $\lambda$(obs) | Notes |
|---|---|---|---|---|---|---|---|---|
| cm$^{-1}$ | | J | cm$^{-1}$ | | Å | KUR | Å | |
| 127892.981 | ($^3$G)4f 4[7] | 15/2 | 104064.670 | ($^3$H)4d $^4$I$_{13/2}$ | 4195.506 | −1.455 | | |
| | | | 104622.300 | ($^3$H)4d $^2$I$_{13/2}$ | 4296.044 | −1.387 | | |
| | | | 108133.440 | ($^3$G)4d $^4$H$_{13/2}$ | 5059.436 | −0.484 | 5059.42 | lab |
| | | | 108463.910 | ($^3$G)4d $^4$I$_{13/2}$ | 5145.493 | −0.007 | 5145.5 | |
| | | | 108648.695 | ($^1$I)5s e$^2$I$_{13/2}$ | 5194.901 | +0.482 | | blend |
| | | | 109049.600 | ($^3$G)4d $^2$I$_{13/2}$ | 5305.427 | +0.862 | 5305.42 | lab |
| 127895.260 | ($^3$G)4f 4[7] | 13/2 | 104174.270 | ($^3$H)4d $^4$I$_{11/2}$ | 4214.489 | −1.351 | | |
| | | | 108387.920 | ($^3$G)4d $^4$H$_{11/2}$ | 5124.848 | −0.679 | | blend |
| | | | 108630.429 | ($^1$I)5s e$^2$I$_{11/2}$ | 5189.361 | −0.144 | 5189.371 | lab |
| | | | 108648.695 | ($^1$I)5s e$^2$I$_{13/2}$ | 5194.286 | −1.434 | | |
| | | | 108775.080 | ($^3$G)4d $^4$I$_{11/2}$ | 5228.621 | +0.896 | 5228.635 | lab |
| | | | 109389.880 | ($^3$G)4d $^2$I$_{11/2}$ | 5402.332 | +0.099 | 5402.32 | lab |
| 127875.000 | ($^3$G)4f 4[6] | 13/2 | 106045.690 | ($^3$H)4d $^2$H$_{11/2}$ | 4579.713 | −0.754 | | |
| | | | 108133.440 | ($^3$G)4d $^4$H$_{13/2}$ | 5064.044 | −1.045 | | |
| | | | 108387.920 | ($^3$G)4d $^4$H$_{11/2}$ | 5130.176 | +0.662 | 5130.18 | lab |
| | | | 108463.910 | ($^3$G)4d $^4$I$_{13/2}$ | 5150.259 | −0.700 | | |
| | | | 108648.695 | ($^1$I)5s e$^2$I$_{13/2}$ | 5199.759 | −0.190 | | blend |
| | | | 109049.600 | ($^3$G)4d 2I$_{13/2}$ | 5310.495 | +0.113 | 5310.5 | lab |
| | | | 109683.280 | ($^3$G)4d $^2$H$_{11/2}$ | 5495.480 | +0.481 | 5495.49 | lab, J78 |
| 127880.436 | ($^3$G)4f 4[6] | 11/2 | 106097.520 | ($^3$H)4d $^2$H$_{9/2}$ | 4589.468 | −0.765 | | |
| | | | 108387.920 | ($^3$G)4d $^4$H$_{11/2}$ | 5128.745 | −0.375 | | blend |
| | | | 108391.500 | ($^3$G)4d $^4$G$_{9/2}$ | 5129.687 | −1.085 | | |
| | | | 108577.560 | ($^3$G)4d $^4$H$_{9/2}$ | 5179.133 | +0.652 | 5179.14 | lab |
| | | | 108630.429 | ($^1$I)5s e$^2$I$_{11/2}$ | 5193.357 | −0.797 | | |
| | | | 108775.080 | ($^3$G)4d $^4$I$_{11/2}$ | 5232.678 | −0.047 | | blend |
| | | | 108929.040 | ($^3$G)4d $^4$I$_{9/2}$ | 5275.188 | −0.897 | | |
| | | | 109389.880 | ($^3$G)4d $^2$I$_{11/2}$ | 5406.663 | −0.491 | | |
| | | | 109625.200 | ($^3$G)4d $^2$G$_{9/2}$ | 5476.359 | −0.333 | 5476.38 | |
| | | | 109683.280 | ($^3$G)4d $^2$H$_{11/2}$ | 5493.838 | −1.052 | | |
| | | | 109811.920 | ($^3$G)4d $^4$F$_{9/2}$ | 5532.952 | −0.700 | | |
| | | | 110008.300 | ($^3$G)4d $^2$H$_{9/2}$ | 5593.749 | +0.039 | 5593.85 | |
| 127869.158 | ($^3$G)4f 4[5] | 11/2 | 106045.690 | ($^3$H)4d $^2$H$_{11/2}$ | 4580.939 | −1.153 | | |
| | | | 106722.170 | ($^3$F)4d $^4$F$_{9/2}$ | 4727.483 | −0.893 | | |
| | | | 108387.920 | ($^3$G)4d $^4$H$_{11/2}$ | 5131.714 | +0.220 | 5131.7 | lab |
| | | | 108391.500 | ($^3$G)4d $^4$G$_{9/2}$ | 5132.657 | +0.408 | | blend |
| | | | 108577.560 | ($^3$G)4d $^4$H$_{9/2}$ | 5182.161 | −0.938 | | |
| | | | 108648.695 | ($^1$I)5s e$^2$I$_{13/2}$ | 5201.340 | −1.171 | | |
| | | | 108775.080 | ($^3$G)4d $^4$I$_{11/2}$ | 5235.768 | −0.234 | | blend |
| | | | 108929.040 | ($^3$G)4d $^4$I$_{9/2}$ | 5278.329 | −1.413 | | |
| | | | 109049.600 | ($^3$G)4d $^2$I$_{13/2}$ | 5312.143 | −0.846 | | |
| | | | 109625.200 | ($^3$G)4d $^2$G$_{9/2}$ | 5479.744 | −0.089 | 5479.72 | lab |
| | | | 109683.280 | ($^3$G)4d $^2$H$_{11/2}$ | 5497.245 | +0.050 | 5497.25 | |
| | | | 109811.920 | ($^3$G)4d $^4$F$_{9/2}$ | 5536.408 | −0.555 | 5536.40 | |
| | | | 110008.300 | ($^3$G)4d $^2$H$_{9/2}$ | 5597.281 | −0.105 | 5597.30 | |
| 127855.952 | ($^3$G)4f 4[5] | 9/2 | 106722.170 | ($^3$F)4d $^4$F$_{9/2}$ | 4730.437 | −0.906 | | |
| | | | 106767.210 | ($^3$F)4d $^4$F$_{7/2}$ | 4740.541 | −0.409 | | |
| | | | 106900.370 | ($^3$F)4d $^2$G$_{7/2}$ | 4770.664 | −1.118 | | |
| | | | 108365.320 | ($^3$G)4d $^4$D$_{7/2}$ | 5129.241 | −0.301 | 5129.25 | |
| | | | 108387.920 | ($^3$G)4d $^4$H$_{11/2}$ | 5135.195 | −0.409 | | blend |
| | | | 108391.500 | ($^3$G)4d $^4$G$_{9/2}$ | 5136.140 | +0.294 | | blend |
| | | | 108577.560 | ($^3$G)4d $^4$H$_{9/2}$ | 5185.710 | −0.829 | | |
| | | | 108709.450 | ($^3$G)4d $^4$H$_{7/2}$ | 5221.432 | −1.407 | | |
| | | | 109625.200 | ($^3$G)4d $^2$G$_{9/2}$ | 5483.714 | +0.010 | 5483.70 | |

**Table 9.** Fe II lines in the 3800-8000 Å region with log $gf \geq -1.5$ and $3d^6(^3G)4f$ energy levels as upper levels.

| Upper level | | | Lower level | | $\lambda$(calc) | log $gf$ | $\lambda$(obs) | Notes |
|---|---|---|---|---|---|---|---|---|
| cm$^{-1}$ | | J | cm$^{-1}$ | | Å | KUR | Å | |
| 127855.952 | cont. | | 109683.280 | ($^3$G)4d $^2$H$_{11/2}$ | 5501.240 | −0.659 | | |
| | | | 109811.920 | ($^3$G)4d $^4$F$_{9/2}$ | 5540.460 | −0.431 | 5540.47 | |
| | | | 109901.500 | ($^3$G)4d $^2$G$_{7/2}$ | 5568.103 | −0.216 | 5568.10 | |
| | | | 110167.280 | ($^3$G)4d $^4$F$_{7/2}$ | 5651.767 | −0.160 | 5651.78 | computed too weak |
| | | | 110570.300 | ($^3$G)4d $^2$F$_{7/2}$ | 5783.541 | −0.854 | | |
| 127869.892 | ($^3$G)4f | 4[4] 9/2 | 106097.520 | ($^3$H)4d $^2$H$_{9/2}$ | 4591.690 | −1.043 | | no soectrum |
| | | | 106900.370 | ($^3$F)4d $^2$G$_{7/2}$ | 4767.493 | −1.141 | | no spectrum |
| | | | 108365.320 | ($^3$G)4d $^4$D$_{7/2}$ | 5125.575 | −1.117 | | weak |
| | | | 108391.500 | ($^3$G)4d $^4$G$_{9/2}$ | 5132.464 | −0.690 | | blend |
| | | | 108537.610 | ($^3$G)4d $^4$G$_{7/2}$ | 5171.255 | +0.332 | 5171.25 | lab, J78 |
| | | | 108577.560 | ($^3$G)4d $^4$H$_{9/2}$ | 5181.963 | +0.101 | 5181.97 | lab |
| | | | 108709.450 | ($^3$G)4d $^4$H$_{7/2}$ | 5217.634 | −1.196 | | weak |
| | | | 108775.080 | ($^3$G)4d $^4$I$_{11/2}$ | 5235.567 | −0.810 | | blend |
| | | | 108929.040 | ($^3$G)4d $^4$I$_{9/2}$ | 5278.125 | −0.704 | | blend |
| | | | 109389.880 | ($^3$G)4d $^2$I$_{11/2}$ | 5409.748 | −1.407 | | blend |
| | | | 109901.500 | ($^3$G)4d $^2$G$_{7/2}$ | 5563.783 | −0.269 | 5563.79 | |
| | | | 110008.300 | ($^3$G)4d $^2$H$_{9/2}$ | 5597.051 | +0.023 | 5597.05 | |
| | | | 110167.280 | ($^3$G)4d $^4$F$_{7/2}$ | 5647.317 | −0.723 | | blend |
| | | | 110570.300 | ($^3$G)4d $^2$F$_{7/2}$ | 5778.881 | −0.074 | 5778.88 | |
| 127874.745 | ($^3$G)4f | 4[3] 5/2 | 106767.210 | ($^3$F)4d $^4$F$_{7/2}$ | 4736.320 | −0.862 | | no spectrum |
| | | | 106796.660 | ($^3$F)4d $^4$P$_{5/2}$ | 4742.937 | −1.442 | | no spectrum |
| | | | 106866.760 | ($^3$F)4d $^4$F$_{5/2}$ | 4758.764 | −0.354 | | no spectrum |
| | | | 107407.800 | ($^3$F)4d $^2$D$_{5/2}$ | 4884.563 | −1.137 | | blend |
| | | | 108365.320 | ($^3$G)4d $^4$D$_{7/2}$ | 5124.300 | −0.351 | 5124.3 | |
| | | | 108537.610 | ($^3$G)4d $^4$G$_{7/2}$ | 5169.957 | −0.493 | 5169.95 | |
| | | | 108613.960 | ($^3$G)4d $^4$G$_{5/2}$ | 5190.451 | −1.336 | | blend |
| | | | 108642.410 | ($^3$G)4d $^4$D$_{5/2}$ | 5198.129 | −0.577 | 5198.12 | |
| | | | 108859.470 | ($^3$G)4d $^4$D$_{3/2}$ | 5257.467 | −1.074 | | weak |
| | | | 109901.500 | ($^3$G)4d $^2$G$_{7/2}$ | 5562.281 | −0.790 | | weak |
| | | | 110065.750 | ($^3$G)4d $^2$D$_{5/2}$ | 5613.582 | −0.302 | 5613.55 | blend |
| | | | 110167.280 | ($^3$G)4d $^4$F$_{7/2}$ | 5645.769 | −0.897 | | weak |
| | | | 110428.280 | ($^3$G)4d $^4$F$_{5/2}$ | 5730.231 | −0.236 | | blend |
| | | | 110570.300 | ($^3$G)4d $^2$F$_{7/2}$ | 5777.260 | −0.288 | 5777.73 | computed too weak |
| | | | 110611.800 | ($^3$G)4d $^2$F$_{5/2}$ | 5791.149 | −1.493 | | blend |
| 128110.214 | ($^3$G)4f | 3[6] 13/2 | 104765.450 | ($^3$H)4d $^2$I$_{11/2}$ | 4282.411 | −1.266 | | blend |
| | | | 108387.920 | ($^3$G)4d $^4$H$_{11/2}$ | 5068.991 | −0.821 | 5068.99 | |
| | | | 108630.429 | ($^1$I)5s e$^2$I$_{11/2}$ | 5132.097 | −0.929 | | blend |
| | | | 108775.080 | ($^3$G)4d $^4$I$_{11/2}$ | 5170.492 | +0.154 | 5170.5 | lab |
| | | | 109389.880 | ($^3$G)4d $^2$I$_{11/2}$ | 5340.300 | +0.922 | 5340.30 | lab, J78 |
| 128071.171 | ($^3$F)4f | 3[5] 11/2 | 106097.520 | ($^3$H)4d $^2$H$_{9/2}$ | 4549.630 | −0.731 | | no spectrum |
| | | | 106924.430 | ($^3$F)4d $^2$G$_{9/2}$ | 4727.539 | −0.926 | | no spectrum |
| | | | 108387.920 | ($^3$G)4d $^4$H$_{11/2}$ | 5079.046 | −1.376 | | blend |
| | | | 108391.500 | ($^3$G)4d $^4$G$_{9/2}$ | 5079.970 | −1.401 | | at the continuum level |
| | | | 108577.560 | ($^3$G)4d $^4$H$_{9/2}$ | 5128.457 | +0.377 | 5128.47 | lab |
| | | | 108775.080 | ($^3$G)4d $^4$I$_{11/2}$ | 5180.954 | −0.687 | | blend |
| | | | 108929.040 | ($^3$G)4d $^4$I$_{9/2}$ | 5222.625 | −0.245 | 5222.62 | computed too strong |
| | | | 109389.880 | ($^3$G)4d $^2$I$_{11/2}$ | 5351.461 | +0.043 | 5351.47 | |
| | | | 106925.200 | ($^3$G)4d $^2$G$_{9/2}$ | 5419.731 | −0.013 | 5419.73 | lab |
| | | | 110008.300 | ($^3$G)4d $^2$H$_{9/2}$ | 5534.681 | +0.459 | 5534.68 | |

**Table 9.** Fe II lines in the 3800-8000 Å region with log $gf \geq -1.5$ and $3d^6(^3G)4f$ energy levels as upper levels.

| Upper level | | | | Lower level | | $\lambda$(calc) | log $gf$ | $\lambda$(obs) | Notes |
|---|---|---|---|---|---|---|---|---|---|
| cm$^{-1}$ | | | J | cm$^{-1}$ | | Å | KUR | Å | |
| 128055.658 | ($^3$F)4f | 3[5] | 9/2 | 106097.520 | ($^3$H)4d $^2$H$_{9/2}$ | 4552.844 | −1.204 | | no spectrum |
| | | | | 106767.210 | ($^3$F)4d $^4$F$_{7/2}$ | 4696.069 | −0.812 | | no spectrum |
| | | | | 106924.430 | ($^3$F)4d $^2$G$_{9/2}$ | 4731.009 | −1.380 | | no spectrum |
| | | | | 108537.610 | ($^3$G)4d $^4$G$_{7/2}$ | 5122.036 | +0.148 | 5122.02 | lab |
| | | | | 108577.560 | ($^3$G)4d $^4$H$_{9/2}$ | 5132.541 | +0.038 | 5132.55 | lab |
| | | | | 108709.450 | ($^3$G)4d $^4$H$_{7/2}$ | 5167.532 | −0.521 | | blend |
| | | | | 108775.080 | ($^3$G)4d $^4$I$_{11/2}$ | 5185.122 | −1.448 | 5185.141 | blend |
| | | | | 109389.880 | ($^3$G)4d $^2$I$_{11/2}$ | 5355.908 | −0.925 | 5355.9 | weak |
| | | | | 106925.200 | ($^3$G)4d $^2$G$_{9/2}$ | 5424.293 | −0.649 | | blend |
| | | | | 109901.500 | ($^3$G)4d $^2$G$_{7/2}$ | 5506.850 | +0.159 | 5506.85 | |
| | | | | 110008.300 | ($^3$G)4d $^2$H$_{9/2}$ | 5539.439 | +0.045 | 5539.41 | |
| | | | | 110167.280 | ($^3$G)4d $^4$F$_{7/2}$ | 5588.670 | −0.697 | 5588.65 | |
| | | | | 110570.300 | ($^3$G)4d $^2$F$_{7/2}$ | 5717.485 | −0.176 | 5717.50 | |
| 128062.710 | ($^3$F)4f | 3[4] | 9/2 | 106900.370 | ($^3$F)4d $^2$G$_{7/2}$ | 4724.054 | −1.276 | | no spectrum |
| | | | | 108709.450 | ($^3$G)4d $^4$H$_{7/2}$ | 5165.649 | +0.734 | 5165.65 | lab |
| | | | | 108929.040 | ($^3$G)4d $^4$I$_{9/2}$ | 5224.934 | +0.139 | 5224.938 | |
| | | | | 109901.500 | ($^3$G)4d $^2$G$_{7/2}$ | 5504.712 | −0.840 | | not observed |
| | | | | 110008.300 | ($^3$G)4d $^2$H$_{9/2}$ | 5537.275 | −1.268 | | at the level of the noise |
| | | | | 110570.300 | ($^3$G)4d $^2$F$_{7/2}$ | 5715.180 | −1.173 | | at the level of the noise |
| 128066.823 | ($^3$F)4f | 3[4] | 7/2 | 104023.910 | ($^3$H)4d $^4$G$_{5/2}$ | 4158.057 | −1.351 | | not observed, wrong |
| | | | | 106208.560 | ($^3$F)4d $^2$F$_{5/2}$ | 4573.647 | −1.130 | | no spectrum |
| | | | | 106767.210 | ($^3$F)4d $^4$F$_{7/2}$ | 4693.607 | −1.067 | | no spectrum |
| | | | | 106900.370 | ($^3$F)4d $^2$G$_{7/2}$ | 4723.136 | −1.319 | | no spectrum |
| | | | | 108537.610 | ($^3$G)4d $^4$G$_{7/2}$ | 5119.108 | −0.444 | | computed too strong |
| | | | | 108577.560 | ($^3$G)4d $^4$H$_{9/2}$ | 5129.601 | −1.316 | | blend |
| | | | | 108613.960 | ($^3$G)4d $^4$G$_{5/2}$ | 5139.200 | +0.196 | 5139.20 | lab |
| | | | | 108709.450 | ($^3$G)4d $^4$H$_{7/2}$ | 5164.552 | −0.146 | 5164.52 | computed too weak |
| | | | | 108929.040 | ($^3$G)4d $^4$I$_{9/2}$ | 5223.811 | −0.993 | | blend |
| | | | | 109901.500 | ($^3$G)4d $^2$G$_{7/2}$ | 5503.465 | −0.078 | | blend |
| | | | | 110008.300 | ($^3$G)4d $^2$H$_{9/2}$ | 5536.014 | −0.751 | 5536.0 | |
| | | | | 110570.300 | ($^3$G)4d $^2$F$_{7/2}$ | 5713.836 | −0.308 | 5713.8 | |
| | | | | 110611.800 | ($^3$G)4d $^2$F$_{5/2}$ | 5727.421 | −0.043 | 5727.45 | |
| 128063.103 | ($^3$G)4f | 3[3] | 5/2 | 106864.650 | ($^3$G)4d $^4$F$_{3/2}$ | 4712.005 | −0.481 | | no spectrum |
| | | | | 106866.760 | ($^3$F)4d $^4$F$_{5/2}$ | 4716.475 | −1.431 | | no spectrum |
| | | | | 107430.250 | ($^3$F)4d $^2$D$_{3/2}$ | 4845.286 | −0.946 | | blend, computed too strong |
| | | | | 108613.960 | ($^3$G)4d $^4$G$_{5/2}$ | 5140.183 | +0.037 | 5140.19 | |
| | | | | 108642.410 | ($^3$G)4d $^4$D$_{5/2}$ | 5147.713 | −0.412 | 5147.71 | computed too weak |
| | | | | 108709.450 | ($^3$G)4d $^4$H$_{7/2}$ | 5165.544 | −0.693 | | blend |
| | | | | 108859.470 | ($^3$G)4d $^4$D$_{3/2}$ | 5205.898 | −0.225 | 5205.879 | |
| | | | | 109901.500 | ($^3$G)4d $^2$G$_{5/2}$ | 5504.593 | −1.414 | | at the continuum level |
| | | | | 110428.280 | ($^3$G)4d $^4$F$_{5/2}$ | 5669.025 | −0.651 | 5669.03 | |
| | | | | 110461.260 | ($^3$G)4d $^2$D$_{3/2}$ | 5679.647 | −1.133 | | at the level of the noise |
| | | | | 110609.540 | ($^3$G)4d $^4$F$_{3/2}$ | 5727.900 | −0.186 | 5727.90 | |
| | | | | 110611.800 | ($^3$G)4d $^2$F$_{5/2}$ | 5728.642 | −0.772 | | weak |
| 128089.313 | ($^3$G)4f | 3[2] | 5/2 | 106208.560 | ($^3$F)4d $^2$F$_{5/2}$ | 4568.946 | −1.396 | | no spectrum |
| | | | | 106747.210 | ($^5$D)5d $^4$F$_{7/2}$ | 4688.657 | −1.457 | | no spectrum |
| | | | | 106796.660 | ($^3$F)4d $^4$P$_{5/2}$ | 4695.142 | −1.393 | | no spectrum |
| | | | | 106866.760 | ($^3$F)4d $^4$F$_{5/2}$ | 4710.650 | −1.102 | | no spectrum |
| | | | | 108537.610 | ($^3$G)4d $^4$G$_{7/2}$ | 5113.219 | −1.022 | | at the continuum level |
| | | | | 108642.410 | ($^3$G)4d $^4$D$_{5/2}$ | 5140.775 | −0.580 | | blend |
| | | | | 108859.470 | ($^3$G)4d $^4$D$_{3/2}$ | 5198.803 | −0.577 | | blend |
| | | | | 109901.500 | ($^3$G)4d $^2$G$_{5/2}$ | 5496.660 | −0.747 | | blend |
| | | | | 110428.280 | ($^3$G)4d $^4$F$_{5/2}$ | 5660.612 | −0.985 | | blend |
| | | | | 110461.260 | ($^3$G)4d $^2$D$_{3/2}$ | 5671.202 | −0.429 | 5671.20 | |
| | | | | 110570.300 | ($^3$G)4d $^2$F$_{7/2}$ | 5706.501 | −0.913 | | at the level of the noise |
| | | | | 110611.800 | ($^3$G)4d $^2$F$_{5/2}$ | 5720.051 | +0.065 | 5720.05 | |